\documentclass{iopjournal}

\usepackage[T1]{fontenc}
\usepackage[utf8]{inputenc}
\usepackage{lmodern}
\usepackage[czech,english]{babel}

\usepackage{graphicx}
\usepackage[normalem]{ulem}
\usepackage{booktabs}
\usepackage{array}
\usepackage{tabularx}
\usepackage{float}

\usepackage{amssymb}
\usepackage{siunitx}
\DeclareSIUnit\hartree{\text{\ensuremath{E}}_{\mathrm{h}}}
\usepackage[version=3]{mhchem}

\usepackage{placeins}
\usepackage{listings}
\usepackage{tikz}
\usetikzlibrary{svg.path}
\usepackage{framed}
\usepackage{appendix}
\usepackage{subcaption} 

\usepackage[nameinlink,capitalize,noabbrev]{cleveref}

\usepackage[printonlyused, nolist]{acronym}
\begin{acronym}
    \acro{asf}[ASF]{Active Space Finder}
    \acro{bfgs}[BFGS]{Broyden-Fletcher-Goldfarb-Shanno Algorithm}
    \acro{bp}[BP]{barren plateau}
    \acro{ch2}[CH\textsubscript{2}]{methylene}
    \acro{cmaes}[CMA-ES]{Covariance Matrix Adaptation Evolution Strategy}
    \acro{cobyla}[COBYLA]{Constrained Optimization By Linear Approximations}
    \acro{dev2-svp}[dev2-SVP]{Development 2\textsuperscript{nd} generation - Split Valence Polarization}
    \acro{es}[ES]{Evolution Strategies}
    \acro{fe}[FE]{Function Evaluation}
    \acro{hea}[HEA]{Hardware-Efficient Ansatz}
    \acro{HF}[HF]{Hartree-Fock}
    \acro{h2}[H\textsubscript{2}]{hydrogen molecule}
    \acro{h4}[H\textsubscript{4}]{hydrogen chain}
    \acro{fci}[FCI]{full Configuration Interaction}
    \acro{gd}[GD]{Gradient Descent}
    \acro{hea}[HEA]{Hardware-Efficient Ansatz}
    \acro{hva}[HVA]{Hamiltonian Variational Ansatz}
    \acro{lih}[LiH]{lithium hydride}
    \acro{nelder}[Nelder-Mead]{Nelder-Mead Algorithm}
    \acro{nisq}[NISQ]{Noisy Intermediate-Scale Quantum}
    \acro{pso}[PSO]{Particle Swarm Optimization}
    \acro{pyscf}[PySCF]{Python-based Simulations of Chemistry Framework}
    \acro{qaoa}[QAOA]{Quantum Approximate Optimization Algorithm}
    \acro{qpe}[QPE]{Quantum Phase Estimation}
    \acro{scf}[SCF]{self-consistent field}
    \acro{slsqp}[SLSQP]{Sequential Least Squares Programming}
    \acro{spsa}[SPSA]{Simultaneous Perturbation Stochastic Approximation}
    \acro{sto3g}[STO-3G]{Slater-type orbitals with 3 Gaussian functions}
    \acro{tvha}[tVHA]{truncated Variational Hamiltonian Ansatz}
    \acro{ucc}[UCC]{Unitary Coupled Cluster}
    \acro{uccsd}[UCCSD]{Unitary Coupled Cluster with Single and Double excitations}
    \acro{vha}[VHA]{Variational Hamiltonian Ansatz}
    \acro{vqa}[VQA]{Variational Quantum Algorithm}
    \acro{vqe}[VQE]{Variational Quantum Eigensolver}
\end{acronym}

\begin{document}

\articletype{Paper} 

\title{Numerical Optimization Strategies for the Variational Hamiltonian Ansatz in Noisy Quantum Environments}

\author{Silvie Illésová$^{1,2,3}$\orcid{0009-0002-5231-3714}, 
Vojtěch Novák$^{3,4}$\orcid{0009-0001-4432-8070}, 
Tomáš Bezděk$^{2,5}$\orcid{0009-0006-3021-1855}, 
Clemens Possel$^{6,7}$\orcid{0000-0002-1859-7533}, 
Martin Beseda$^{8}$\orcid{0000-0001-5792-2872}}

\affil{$^1$Gran Sasso Science Institute, L’Aquila, Italy}

\affil{$2$Department of Applied Mathematics, Faculty of Electrical Engineering and Computer Science, VSB-Technical University of Ostrava, Ostrava, Czech Republic}

\affil{$^3$IT4Innovations National Supercomputing Center, VSB\textendash Technical University of Ostrava, 708\,00 Ostrava, Czech Republic}

\affil{$^4$Department of Computer Science, Faculty of Electrical Engineering and Computer Science, VSB\textendash Technical University of Ostrava, Ostrava, Czech Republic}

\affil{$^5$Department of Mathematics, TUM School of Computation, Information and Technology, Technical University of Munich, Garching bei München, Germany}

\affil{$^6$Fraunhofer Institute for Chemical Technology ICT, Joseph\textendash von\textendash Fraunhofer\textendash Str.\ 7, 76327 Pfinztal, Germany}

\affil{$^7$Department of Physics, Saarland University, 66123 Saarbrücken, Germany}

\affil{$^8$Department of Information Engineering, Computer Science and Mathematics, University of L'Aquila, via Vetoio (Coppito), 1 – 67100, L'Aquila, Italy}

\email{martin.beseda@univaq.it}


\begin{abstract}
The prevalence of variational methods in near-term quantum computing makes optimizer choice critical, yet selection is frequently intuition-based. We therefore present a systematic benchmark of eight classical optimization algorithms for variational quantum chemistry using the \ac{tvha}. Performance is evaluated on \ce{H2}, \ce{H4}, and \ce{LiH} in both full and active-space representations under noiseless and finite-shot sampling noise. Sampling noise substantially reshapes cost landscapes, induces wandering near minima, and flips optimizer rankings: gradient-based methods perform best in noiseless simulations, whereas population-based optimizers, particularly \ac{cmaes}, show greater robustness under finite-shot noise. Optimizer performance is strongly problem dependent: Hartree-Fock initialization aids small systems, but its advantage diminishes with system size. Also, we observe that finite shot sampling frequently violates the lower bound given by the variational principle, a principle that cannot be strictly held in the presence of noise. By exploiting the guaranteed convergence of Evolution Strategies to a steady state distribution defined by the noise floor, we utilize the symmetry of these violations to achieve energy estimation precision beyond the intrinsic sampling limit.

\end{abstract}

\section{Introduction}
Quantum computing has the potential to transform computational chemistry by enabling the simulation of strongly correlated electronic systems beyond the reach of classical algorithms \cite{Preskill2018}. To exploit this potential, it is crucial to develop an appropriate set of well-aligned algorithmic routines that account for both hardware constraints and application demands. Variational quantum algorithms \cite{cerezo2021variational} have gained significant attention in the \ac{nisq} era due to their compatibility with limited qubit counts \cite{niu2020hardware}, shallow circuit depths \cite{xu2024quantum,du2022quantum}, and hardware-specific constraints \cite{Peruzzo2014,callison2022hybrid,qi2024variational,stilck2021limitations}. The versatility of this variational framework has led to its adoption across a wide spectrum of disciplines beyond physics, ranging from machine learning, remote sensing \cite{gupta2022how,illesova2025classical} to software testing \cite{trovato2025preliminary} to general hybrid learning protocols \cite{illesova2025classical}. However, regardless of the domain, the combined effects of variational algorithms and optimization strategies in the presence of noise remain a critical hurdle, underscoring the need for systematic investigation \cite{novak2025optimization,illesova2025statistical}.

Among recent algorithmic innovations, the \ac{vha} has emerged as a promising---although still exploratory---approach for constructing compact, chemically motivated ansatz tailored to molecular Hamiltonians \cite{Wecker2015,anand2025hamiltonian}. The \ac{vha} leverages the structure of the electronic Hamiltonian by decomposing it into physically meaningful subcomponents, which are then mapped into a sequence of parametrized unitary transformations \cite{Wecker2015}. This design facilitates the incorporation of problem-specific knowledge while maintaining circuit expressibility under resource constraints. Designing efficient parameterized circuits is a shared challenge across the field, leading both to the efforts in benchmarking quantum circuits in quantum neural networks \cite{illesova2025qmetric} and the utilization of hybrid architectures for image classification \cite{illesova2025complementarity}. However, the practical deployment of the \ac{vha} on present-day quantum hardware is complicated by \textit{pervasive noise sources}, including gate infidelity, qubit decoherence, and stochastic measurement error \cite{Kandala2017,lewandowska2025benchmarking}. These imperfections manifest as \textit{sampling noise} in the cost function evaluation, severely distorting the optimization landscape and complicating parameter convergence \cite{McClean2016,Temme2017}.

In such stochastic regimes, \textit{optimization becomes the central bottleneck} for achieving reliable ground-state energy estimates \cite{ciaramelletti2025detecting}. Finite sampling introduces statistical fluctuations that can obscure true energy gradients, create false minima, and induce erratic convergence behavior \cite{novak2025reliable,bezdek2025classical}. These reliability issues are not unique to quantum chemistry; similar stochastic difficulties arise when applying quantum models to sensitive biomedical data \cite{novak2025predicting,novak2025quantum}. In all these cases, noise can be particularly detrimental in high-dimensional parameter spaces, where flat or rugged cost surfaces---exacerbated by the so-called \textit{barren plateau} phenomenon---may prevent even well-designed circuits from reaching their expressive potential \cite{mcclean2018barren,Cerezo2021}. As a result, understanding the fundamental properties of these learning models \cite{illesova2025importance} and tuning the classical optimization routine is critical to unlock the practical utility of the \ac{vha} on \ac{nisq} devices \cite{Stokes2020}. The \ac{vha} approach is compatible with a wide range of variational quantum eigensolver variants, including State-Averaged Orbital-Optimized VQE \cite{beseda2024state, illesova2025transformation}, ADAPT-VQE \cite{tang2021qubit}, and Subspace-Search VQE \cite{nakanishi2019subspace}, allowing for flexible integration across different quantum chemistry and optimization pipelines.

To explore this interplay between algorithm design and quantum noise, we conduct a \textit{comparative study of eight classical optimization strategies} applied to the \ac{vha} in noisy quantum simulations. The set includes both gradient-based and gradient-free methods, encompassing diverse optimization philosophies. \ac{gd} and \ac{bfgs} represent classical gradient-based approaches, with \ac{bfgs} leveraging approximate second-order information for rapid convergence in smooth landscapes~\cite{liu1989,dai2002,morales2002}. \ac{spsa}, a stochastic method specifically designed for noisy, high-dimensional optimization, requires only two function evaluations per iteration and is known for its sampling efficiency~\cite{spall1992,spall2002,maryak1999}. Among derivative-free techniques, \ac{cobyla} and \ac{slsqp} approximate the objective locally using trust-region models and are well suited for constrained problems~\cite{powell1994,powell1998,powell2007,kraft1988,boggs1995}. \ac{nelder}, a simplex-based heuristic, explores the landscape through geometric operations~\cite{nelder1965,lagarias1998}, while \ac{cmaes} adapts a multivariate Gaussian over candidate solutions to guide search in complex, non-convex terrains~\cite{hansen2003}. Finally, \ac{pso} employs a population of interacting solutions that update their positions based on both individual experience and global information, drawing inspiration from collective behavior in biological systems~\cite{eberhart1995,shi2001,jain2022}. This breadth of methods allows us to rigorously evaluate optimizer performance across multiple axes: noise resilience, convergence efficiency, and final energy accuracy.

This study relies on a Python-based simulation stack combining \textit{Qiskit}~\cite{qiskit2024} and \textit{PySCF}~\cite{pyscf2018} for quantum circuit construction and molecular integral computation, respectively. These tools enable the simulation of variational circuits under both ideal and noisy conditions, offering insights into the effects of noise on optimization dynamics. In this work, we focus on ideal (noiseless) and sampling-noise-based simulations, investigating the effects of sampling noise on the optimizers and the cost function landscape, aiming to find an efficient optimization approach, while understanding the underlying work the optimizer is performing in detail.

The paper is structured as follows. In the following Section~\ref{sec:vha}, we provide an overview of the \ac{vha}, its formulation, and its potential advantages for quantum chemistry simulations. Subsequently, in Section~\ref{sec:noise_landscape} we describe the effects of sampling noise on different cost functions corresponding to selected molecular systems. In this section, we are illustrating, which number of shots is sufficient to ``see'' the landscape clearly, while explaining this behavior statistically. Section~\ref{sec:noise_floor} furthermore quantifies the numerical properties of the sampling noise via a \textit{noise floor} together with a description of a robust estimation of expectation values. Section~\ref{sec:setup} outlines the whole simulation setup including the computational infrastructure, the software packages, and the effectiveness tweaks adopted for faster computation. Section~\ref{sec:results} presents the comparative results across both idealized and noisy scenarios, highlighting key trends and trade-offs. We conclude the discussion in Section~\ref{sec:conclusion} by summarizing the implications of our findings for future applications of the \ac{vha} and offering guidelines for optimizer selection in \ac{nisq}-era quantum chemistry. The following Section~\ref{sec:availability} contains the details about additional data and the software implementation. Additional details are provided in the appendices: \cref{sec:optimizers} describes the optimization algorithms in depth; \cref{sec:convergence_runs} includes convergence trajectories for individual optimization runs; and \cref{sec:pop_mean} analyzes the impact of population size on noise suppression in optimization.

\section{Overview of the Variational Hamiltonian Ansatz Framework}\label{sec:vha}
The \ac{vha}, along with its improved version \ac{tvha}, is a novel framework designed to enhance quantum computing applications in quantum chemistry, particularly when executed on \ac{nisq} devices~\cite{possel2025truncatedvariationalhamiltonianansatz}.
Based on the principles of the adiabatic theorem, \ac{vha} effectively addresses the challenges associated with simulating quantum systems, especially those exhibiting strong electron correlations.
At the heart of \ac{vha} is the adiabatic theorem, which posits that a quantum system remains in its instantaneous eigenstate when subjected to a sufficiently slow transformation between an initial Hamiltonian and a final Hamiltonian.
This principle serves as the foundation for determining the ground state of complex molecular systems.
\ac{vha} utilizes a linear interpolation of the Hamiltonian for state evolution, variationally ensuring that errors are suppressed that arise from discretization, Trotterization, and, in case of usage of \ac{tvha}, truncation of non-Coulomb two-body terms.
\ac{vha} uses the \ac{HF} state as a classically precomputed starting point, simplifying the Hamiltonian into a manageable form by applying a mean-field approximation, using unitary gate-based time evolution to reach the ground state of the final Hamiltonian with all its electron correlations. 
\ac{vha} is compatible with active-space calculations, allowing larger molecules to be executed.
By selecting a subset of molecular orbitals deemed crucial for accurately capturing electron correlations, the complexity of quantum circuits is efficiently reduced.

\ac{tvha} stands apart from traditional approaches such as \ac{uccsd} and \ac{hea} by minimizing the parameter count while retaining the capability to construct circuits of comparable size.
This innovative truncation scheme optimizes the operators involved in circuit design, ultimately leading to a more efficient quantum computing framework.
By balancing accuracy and efficiency, \ac{tvha} enables the exploration of more complex molecular systems on \ac{nisq} devices, paving the way for future advancements in both quantum chemistry and material science computations.

\ac{tvha} addresses three critical challenges in \ac{nisq} implementations of VQE: preservation of molecular symmetries, mitigation of barren plateau landscapes, and systematic construction of chemically relevant parameterized states. This approach combines adiabatic state preparation concepts with variational optimization, directly encoding electronic structure into the ansatz architecture.

\begin{figure}[htbp]
\includegraphics[width=\textwidth]{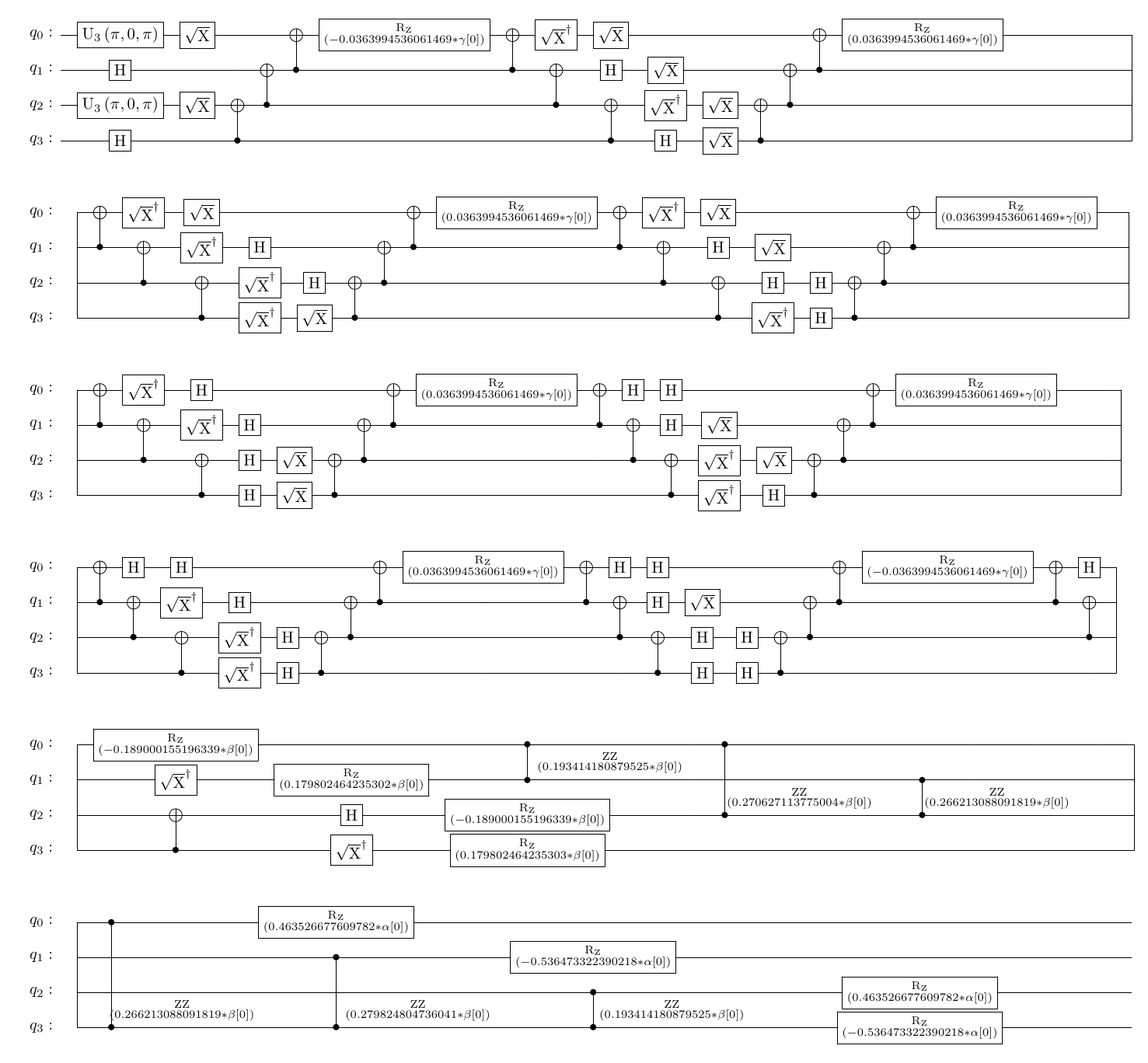}
    \caption{The figure illustrates a fully decomposed 4-qubit \ac{tvha} with 3 trainable parameters, generated for finding the ground state of the \ce{H2} molecule. The circuit is constructed using a gate set of \{\textbf{CX} (48), \textbf{H} (32), $\sqrt{\textbf{X}}$ (16), \textbf{RZ} (16), $\sqrt{\textbf{X}}^{\dagger}$ (16), \textbf{RZZ} (6), \textbf{U3} (2)\}. This gate configuration effectively balances chemical accuracy with \ac{nisq}-era hardware limitations.}
    \label{fig:circuit} 
\end{figure}

Traditional quantum chemistry ansatz face a fundamental tension between physical interpretability and \ac{nisq} feasibility. While \ac{ucc} methods provide chemically meaningful parameterizations, they often suffer from deep circuits exceeding coherence times, non-commuting Trotter steps complicating optimization, and exponential parameter growth with system size--unless the ansatz is restricted to a subset of excitations, typically single and double excitations, leading to \ac{uccsd} ansatz with polynomial parameter growth. \ac{tvha} circumvents these limitations through systematic construction from the molecular Hamiltonian 
\begin{align}
    H & = H_\alpha + H_\beta + H_\gamma \\
    & = \sum_{ij} h_{ij} a_i^\dagger a_j 
    + \frac{1}{2}  \sum_{ij} g_{ijk\ell} a_i^\dagger a_j^\dagger a_j a_i
    + \frac{1}{2}  \sum_{\substack{ijk\ell \\ i\neq k \\ j\neq \ell}} g_{ijk\ell} a_i^\dagger a_j^\dagger a_k a_\ell,
\end{align}
where $\alpha$ denotes the one-body terms, $\beta$ the Coulomb two-body terms, and $\gamma$ the non-Coulomb two-body terms.
The truncation scheme is applied to the non-Coulomb two-body terms using a truncation threshold $p$ such that 
\begin{equation}
    p = \frac{1}{ \sum_s \left| g^\gamma_s \right| } \sum_{s=1}^{s_{cut}} \left| g^\gamma_s \right|, \label{eq:truncation_threshold}
\end{equation}
where the index $s$ contracts the four indices $i$, $j$, $k$, $\ell$ in sorted descending order.
In the following, the truncated Hamiltonian is used.
The terms of the molecular Hamiltonian are transformed to spin operators, i.e. Pauli terms $H_\alpha =\sum_\alpha c_\alpha P_\alpha$, $H_\beta=\sum_\beta c_\beta P_\beta$, and $H_\gamma=\sum_\gamma c_\gamma P_\gamma$ using Jordan-Wigner transformation. Other transformations such as Bravyi-Kitaev transformation are in principle also feasible and might be preferable for large molecular systems but don't allow for an straight-forward interpretation of each qubit's state as the occupation number of a spin orbital.
The Hamiltonians are grouped into commuting Hamiltonian fragments, where $\mathcal{G}^\alpha$, $\mathcal{G}^\beta$, and $\mathcal{G}^\gamma$ represent commuting groups within the three fragments of the Hamiltonian, respectively.
With this, the ansatz can be written as a unitary transformation
\begin{equation}
    U(\boldsymbol{\alpha}, \boldsymbol{\beta}, \boldsymbol{\gamma}) 
    = \prod_{d=1}^D 
    \left[ \prod_{G \in \mathcal{G^\alpha}} e^{i \alpha_d c_{\alpha,G} P_\alpha} \right]
    \left[ \prod_{G \in \mathcal{G^\beta}} e^{i \beta_d c_{\beta,G} P_\beta} \right]
    \left[ \prod_{G \in \mathcal{G^\gamma}} e^{i \gamma_d c_{\gamma,G} P_\gamma} \right]
\end{equation}
where $D$ represents the number of discretization steps (linearly connected to the circuit depth) used to mimic adiabatic evolution and $\boldsymbol{\alpha}=\{\alpha_d\}$, $\boldsymbol{\beta}=\{\beta_d\}$, and $\boldsymbol{\gamma}=\{\gamma_d\}$ are free variational parameters (for better readability summarized as parameters $\boldsymbol{\theta}$).
Here, Suzuki-Trotter expansion of first order, also known as Lie-Trotter expansion, is applied to approximate the exponential of non-commuting groups.

As shown in \cref{fig:circuit}, the ansatz employs \textbf{CX} and \textbf{RZZ} gates to capture electronic correlations, while \textbf{RZ} and \textbf{U3} gates enable precise parameter optimization. \textbf{Hadamard} and $\sqrt{\textbf{X}}$/$\sqrt{\textbf{X}}^{\dagger}$ gates prepare orbital superpositions, with $\sqrt{\textbf{X}}^{\dagger}$ ensuring efficient decompositions. The high \textbf{CX} count (48) reflects strong electron-electron interactions, while minimal \textbf{U3} gates (2) provide targeted rotations. This structure preserves three crucial molecular symmetries: particle-number conservation through $[H_G,\hat{N}] = 0$, spin symmetry via $[H_G, S^2] = 0$, and point group symmetry through term selection.

Key implementation features include term grouping via graph coloring for partitioning Pauli terms into commuting sets $\mathcal{G}$, gate sequencing with diagonal terms implemented via Z-rotations and off-diagonal terms via Pauli-gadget synthesis (demonstrated in \cref{fig:circuit}), and symmetry locking through qubit tapering to remove conserved degrees of freedom. \ac{tvha}'s architecture provides distinct advantages: robustness to barren plateaus, chemical interpretability with parameters $\alpha_d$, $\beta_d$, and $\gamma_d$ directly correlating to Hamiltonian term contributions, depth efficiency (4-6 layers sufficient for chemical accuracy), and measurement reduction through parallel measurement of Pauli terms within commuting groups.
The resistance to barren plateaus arises from two factors. First, under certain parameter constraints, the initialization near the Hartree-Fock state ensures that gradient magnitudes decay only polynomially with respect to the number of qubits $n$, i.e., $\mathcal{O}(1/\text{poly}(n))$~\cite{Park2024hamiltonian}, avoiding the exponential decay typical for barren plateaus \cite{mcclean2018barren,Cerezo2021}.
Second, unlike highly over-parametrized ansatz such as \ac{ucc} or \ac{hea}, \ac{tvha} uses a compact, physically motivated parameterization \cite{possel2025truncatedvariationalhamiltonianansatz}. By dramatically reducing the number of free parameters while retaining expressivity, this design lowers the likelihood of large, flat regions in the high-dimensional energy landscape and therefore mitigates the emergence of barren plateaus.

Optimization landscape exploration depends critically on initial parameter selection. For adiabatic initialization with Hartree-Fock initial state, parameters emulate Trotterized adiabatic evolution
\begin{equation}
U_{\text{ad}} = \prod_{d=1}^D e^{-i \frac{\tau}{D} H_0} e^{-i \frac{\tau}{D} \frac{d}{D} V},
\end{equation}
where $H_0$ and $V$ represent non-interacting and interacting Hamiltonian components, respectively. For sub-operators $H_\alpha$ (representing $H_0$), initial parameters are set to $\alpha_{d}^{(0)} \frac{\tau}{D} \langle H_\alpha \rangle_{\text{HF}}$; for $H_\beta$ and $H_\gamma$ (representing $V$), $\beta_{d}^{(0)} = \gamma_{d}^{(0)} = \frac{\tau}{D} \langle V \rangle_{\text{HF}} \frac{d}{D}$. Additional sub-operators follow analogous adiabatic evolution time dependence. Random initialization uses $\theta_{i}^{(0)} \sim \mathcal{U}(0, 1)$ for the exploration of unbiased parameter space.

The constrained entanglement growth of \ac{tvha}, as visualized in the circuit diagram (\cref{fig:circuit}), enables efficient classical optimization while maintaining sufficient expressibility to capture multi-reference effects, striking a balance between computational tractability and physical accuracy.

\section{Sampling Noise Distortions in the Optimization Landscape}
\label{sec:noise_landscape}

The variational principle \cite{tilly2022variational} guarantees that exact energies $E(\boldsymbol{\theta}) = \langle \psi(\boldsymbol{\theta})|H|\psi(\boldsymbol{\theta})\rangle$ obey $E(\boldsymbol{\theta}) \geq E_0$. Under finite sampling, however, the estimator $\hat{E}(\boldsymbol{\theta})$ becomes a random variable with variance $\sigma^2 \propto 1/N_{\text{shots}}$. This stochasticity distorts the optimization landscape, allowing spurious minima below $E_0$ and reversing gradient directions.

\begin{figure*}[htbp]
\centering
\begin{subfigure}{1\textwidth}
\includegraphics[width=1\linewidth]{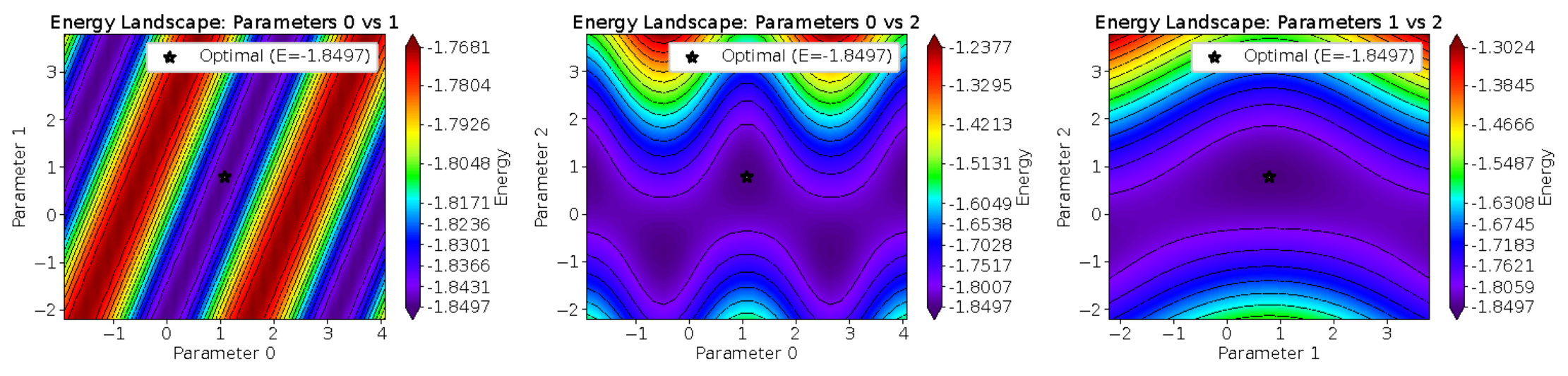}
\caption{Energy landscape for \ce{H2} molecule without sampling noise.}
\label{fig:landscapeno}
\end{subfigure}
\hfill
\begin{subfigure}{1\textwidth}
\includegraphics[width=1\linewidth]{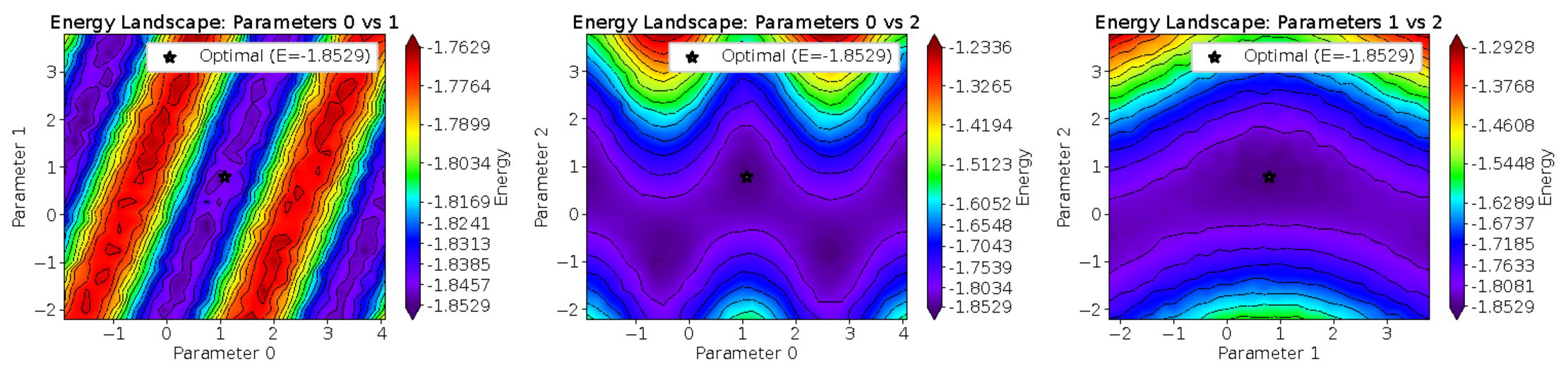}
\caption{Energy landscape for \ce{H2} with moderate sampling noise ($6 \times 1024$ shots).}
\label{fig:landscape1}
\end{subfigure}
\hfill
\begin{subfigure}{1\textwidth}
\includegraphics[width=1\linewidth]{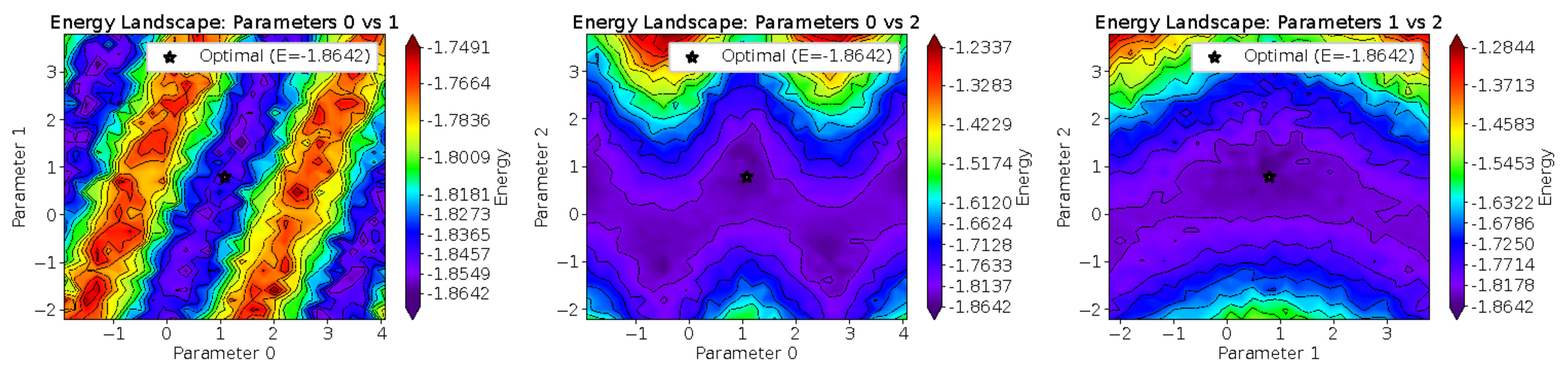}
\caption{Energy landscape for \ce{H2} with significant sampling noise (512 shots).}
\label{fig:landscape2}
\end{subfigure}
\caption{Energy landscapes for the \ce{H2} molecule under varying levels of sampling noise. (a) shows the smooth contours characteristic of exact statevector simulation, revealing quasi-degenerate valleys. (b) reveals emerging distortions, particularly in contour line warping, with moderate noise. (c) highlights key phenomena under significant noise: false minima (blue/purple regions below $E_0$), gradient reversals, and noise induced ruggedness within quasidegenerate valleys, where the sampling noise creates artificial local minima specifically along the flat valley floor.}
\label{fig:combined_landscape}
\end{figure*}

To characterize these distortions, we employ a parameter space slicing methodology. Reference parameters $\boldsymbol{\theta}^{\ast}$ are first obtained from high-precision sampling ($N_{\text{shots}} \ge 10^{6}$). For selected parameter pairs $(i,j)$ we then compute $\hat{E}(\theta_i + \delta_i,\ \theta_j + \delta_j,\ \boldsymbol{\theta}_{k\neq i,j}^{\ast})$ across a displacement grid $\delta_i,\delta_j \in [-\Delta,\Delta]$, fixing the other parameters at $\boldsymbol{\theta}_{k\neq i,j}^{\ast}$. Repeating this process for different shot counts quantifies the impact of finite sampling. As displayed in \cref{fig:combined_landscape}, three main effects emerge, the first one consists of the false minima below $E_0$, the second one is that the slopes of the gradients are inverted and lastly there is a noise induced roughness within the quasidegenerate valleys. This occurs because the signal to noise ratio varies drastically across the landscape: while the steep walls of the valley dominate the noise, the gradient along the flat valley floor is overwhelmed by sampling fluctuations, turning the path toward the minimum into a stochastic trap.

Flat areas of near-degeneracy are most susceptible. Without noise, such valleys direct deterministic descent. Under finite sampling they are statistical conduction bands: broad intervals where fluctuations $\eta(\boldsymbol{\theta})$ are equal to or larger than the true gradient. Within the bands, optimization is controlled by stochastic drift instead of convergence. The associated dynamics are equivalent to an assortment of random walks in parameter space, whose trajectories are aligned with noise instead of the true gradient.

A corollary phenomenon is overfitting by noise. Optimizers like PSO initially converge to $E_0$ but then stray off by getting attracted to transient minima formed by statistical artifacts. The result is non-monotonic convergence curves, where error reduction is initially followed by systematic divergence. Optimizers with large candidate pools are particularly vulnerable, as wider exploration widens the probability of exploiting noise artifacts.

In high-dimensional ansatz for the \ce{LiH} or \ce{H4} chain, the effects are cumulated. Quasi-degenerate submanifolds of parameter space allow noise-driven manifold diffusion: the parameters meander stochastically along nearly flat modes, similar to diffusion processes in physics. Optimization trajectories then fold up into narrow tubes acutely sensitive to noise perturbations, making convergence even harder for well-conditioned systems.

The \ac{bp} framework provides a complementary perspective \cite{BPreview}. A \ac{bp} occurs when the loss function or its gradient concentrates exponentially around the mean with qubit number $n$. This follows from the exponential growth of Hilbert space dimension $\dim \mathcal{H} = 2^n$ and operator space dimension $\dim \mathcal{B}(\mathcal{H}) = 4^n$. In this setting, the loss
\begin{equation}
\ell_\theta = \langle \rho(\theta), O \rangle, \qquad \rho(\theta)\in \mathcal{B}(\mathcal{H})
\end{equation}
is the Hilbert–Schmidt inner product between exponentially large vectors. As $n$ increases,
\begin{equation}
\mathrm{Var}_\theta(\ell_\theta) \in \mathcal{O}(1/4^n),
\end{equation}
suggesting exponentially disappearing gradients. Optimization thus becomes anti-aligning high-dimensional vectors whose overlap is concentrated around zero. Probabilistic concentration leads to flat but locally fertile areas (narrow gorges), while deterministic concentration generates uniformly flat domains. Both phenomena sharpen noise sensitiveness, as statistical fluctuations mask any residual gradient signal.

The dimensional scaling highlights this challenge. The operator space of \ac{fci} grows exponentially as $\mathcal{O}(4^n)$; even for a relatively small molecule like \ce{LiH} (12 qubits), this space already reaches $1.6 \times 10^7$ dimensions. This exponential growth makes barren plateau effects severe, amplifying the stochastic phenomena described above. This requires the use of ansatz with limited expressivity that grows polynomially, such as the \ac{vha} discussed earlier.

However, a trade-off for the inclusion of physical symmetries and complexity of the \ac{vha} is an increase in the number of gates and overall ansatz depth. We therefore employ the \ac{tvha}, a truncated version designed to produce less deep quantum circuits with fewer gates, making it more feasible on current hardware \cite{possel2025truncatedvariationalhamiltonianansatz}. This truncation introduces a \textit{reachability deficit} \cite{reach}. By using fewer gates, it is entirely possible that the set of parameters obtained from the optimization does not correspond to a sufficiently accurate solution. The ansatz may not be able to explore the full space required to find the lowest possible energy (the \ac{fci} energy), creating a gap between the attainable \ac{vqe} result and the true ground state~\cite{possel2025truncatedvariationalhamiltonianansatz}.

Noise-induced distortions and barren plateau concentration are thus coupled phenomena. Both suppress informative gradient signals and induce stochastic drift, narrowing optimization pathways into fragile, noise-dominated channels. Mitigating these effects requires ansatz design strategies that reduce expressiveness without sacrificing accuracy, together with sampling strategies that keep $P(\hat{E} < E_0)$ below practical thresholds.

\section{Violation of the Variational Principle Bound and the Sampling Noise Floor}
\label{sec:noise_floor}

The idea of a noise floor stems from signal processing, where it represents the lowest background level that obscures measurable precision. In classical systems, this floor results from thermal or environmental fluctuations and can often be lowered through improved hardware. In quantum algorithms, however, the noise floor is fundamentally different. It arises from the inherent randomness of quantum measurements: repeated sampling distributes the expectation values around the true mean, establishing a boundary below which improved solutions cannot be distinguished from noise~\cite{zeng2021simulating,huggins2021efficient}. This quantum sampling noise floor lacks a classical reference point and can even result in apparent violations of the variational principle, where finite sampling yields an energy estimate $\hat{C}(\theta) < E_0$~\cite{noise_impact_vqe_2022}. In this context, reducing the floor requires expensive additional circuit repetitions (shots), imposing a direct tradeoff between computational cost and precision.

While the variational principle strictly lower bounds the true energy expectation value such that $E(\theta) \ge E_0$~\cite{tilly2022variational}, our finite shot results demonstrate that the estimated energy $\hat{E}$ frequently violates this bound. Rather than discarding these violations as unphysical artifacts, we argue that they provide critical statistical information. The presence of estimates both above and below $E_0$ confirms the approximately symmetric nature of the sampling noise in the vicinity of the optimum, allowing us to leverage population averaging to enhance precision.

Foundational works of \ac{vqe} largely operate under the assumption that the variational principle serves as a hard lower bound ($E(\theta) \ge E_0$) \cite{kandala2017hardware, peruzzo2014variational, mcclean2018barren}, where hardware noise degrades accuracy by elevating energy estimates rather than violating the ground state limit. The original \ac{vqe} proposal by Peruzzo et al. \cite{peruzzo2014variational} states that the algorithm simply finds the optimal parameters for a noisy state, effectively maintaining the lower bound property. This perspective was experimentally reinforced by Kandala et al. \cite{kandala2017hardware}, where decoherence was observed to strictly shift energy curves upwards, away from chemical accuracy but respecting the variational limit. Similarly, seminal works on error mitigation by Temme et al. \cite{temme2017error}, and Li and Benjamin \cite{li2017efficient} frame noise primarily as a bias that produces mixed states with higher energies, treating the unmitigated bound as valid but practically limiting. Even regarding optimization dynamics, McClean et al. \cite{mcclean2018barren} identify noise as a source of vanishing gradients and barren plateaus that hinder trainability, rather than a mechanism that generates unphysical violations of the lower bound. 

However, recent studies suggest this boundary is more permeable than originally thought. Saib et al. \cite{saib2021effect} show that quantum hardware noise elevates energy estimates, while other literature identifies clear conditions where the variational lower bound is violated \cite{sagastizabal2019experimental, oliv2022evaluating}. One primary driver of these violations is error mitigation: methods like Probabilistic Error Cancellation can amplify variance to produce unphysical energies, as shown by Chi et al. \cite{chi2026variational}, while independent Pauli estimation in symmetry verification can yield non-positive density matrices that break the bound, as noted by Sagastizabal et al. \cite{sagastizabal2019experimental}. Beyond mitigation, Oliv et al. \cite{oliv2022evaluating} demonstrate that finite sampling noise alone is sufficient to generate statistical violations. This challenges the perspective of Cai et al. \cite{cai2023quantum} that treats sampling merely as a precision limit, supporting our finding that the bound is permeable in finite shot regimes.

We leverage the theoretical guarantee established in the literature~\cite{beyer2001theory} that Evolution Strategies, specifically \ac{cmaes}, reliably converge to a steady-state distribution determined by the noise floor $\sigma_{\text{noise}}$. While this ensures algorithmic stability, it does not guarantee convergence to the global minimum. Recent high detail exploration of the energy landscape for the 12 qubit LiH system~\cite{boy2025energy} analogous to the largest full space system in our benchmark reveals a ``glassy'' topography populated by numerous low lying local minima whose energy spread can exceed chemical accuracy.

Consequently, the optimizer may stabilize in one of these near optimal local traps rather than the true global ground state. We therefore only observe energies below the exact ground state (variational bound violations) if the optimizer successfully navigates this rugged landscape to a minimum where the remaining energy gap is smaller than the width of the sampling noise distribution.

At this noise-saturated stage, the algorithmic challenge shifts from finding better parameters to determining the energy with maximal precision. Given the high sensitivity of \ac{vqe} to energy precision, we identify three distinct strategies to minimize the estimation error at this steady state.

The first strategy is direct shot escalation, where one reduces the noise floor itself by increasing the number of shots $N_{\text{shots}}$ for every evaluation. While effective, this scales the computational cost linearly for every individual in the population, quickly becoming prohibitively expensive as the number of samples required grows quadratically to reduce the standard deviation \cite{hansen2008method}. Furthermore, implementations of dynamic shot schedules (such as three-stage telescoping) have been observed to yield mixed results \cite{bonet2023performance}. Optimization traces reported by Bonet-Monroig et al. \cite{bonet2023performance} (e.g., Fig. 5, p.11) reveal that the discrete transition between shot count stages frequently triggers significant error spikes. These discontinuities occur because the optimizer, having previously converged to an artificially low value driven by high-variance sampling noise (``good outliers''), is suddenly corrected when the shot budget increases and the variance tightens. This effective shift in the cost landscape invalidates the optimizer's internal model, forcing a disruptive re-convergence process.

The second strategy is selective reevaluation. One may attempt to identify the best individuals, those appearing to yield the lowest energies, and reevaluate them with a high shot count \cite{hansen2008method}. However, in a high noise environment, this approach risks selecting individuals that are merely lucky (beneficiaries of favorable noise tails or "good outliers") rather than those with superior parameters \cite{hansen2008method}. This strategy essentially amounts to a stochastic gamble: betting that a variational violation represents a true ground state proximity rather than a noise artifact.

The third strategy, which forms the basis of our method, is to increase the population size $\lambda$ \cite{hellwig2016evolution}. By aggregating information across a larger population, we utilize the collective statistics of the steady state distribution, which has been shown to be preferable to resampling for converging to the true, noise-free optimum in noisy environments \cite{hellwig2016evolution, hansen2008method}.

Our work focuses on this third option. While the analysis in this section highlights the robustness of the population mean against varying noise levels, we provide complementary evidence in \cref{sec:pop_mean} showing that the estimation error scales inversely with the square root of the population size, $\propto \sigma_{\text{noise}}/\sqrt{\lambda}$. This scaling allows us to implicitly resample the solution space and smooth out the symmetric noise responsible for the variational violations. Consequently, we achieve high precision energy estimates without the specific risks of selection bias or the massive overhead of uniformly high shot counts.

To quantify the noise floor limits discussed above, we model the noisy estimates formally. Variational quantum algorithms optimize a cost function $C(\boldsymbol{\theta})$ using noisy estimates $\bar{C}(\boldsymbol{\theta})$ derived from finite measurements $N_{\text{shots}}$:
\begin{equation}
\bar{C}(\boldsymbol{\theta}) = C(\boldsymbol{\theta}) + \epsilon_{\text{sampling}},
\end{equation}
where $\epsilon_{\text{sampling}}$ is a zero mean random variable with variance $\mathrm{Var}[C(\boldsymbol{\theta})]/N_{\text{shots}}$. For a Hamiltonian $\hat{H} = \sum_k c_k P_k$, the variance follows
\begin{equation}
\mathrm{Var}[C(\boldsymbol{\theta})] = \langle \hat{H}^2 \rangle - \langle \hat{H} \rangle^2,
\end{equation}
with contributions arising from Pauli term shot noise and covariance within commuting groups, as the partitioning into Abelian groups is applied to decrease the number of necessary measurements. The effective sampling noise floor $\sigma_{\text{noise}}$ is estimated by averaging the variance across the optimization trajectory as
\begin{equation}
\sigma_{\text{noise}} \approx \sqrt{\frac{1}{N} \sum_{i=1}^N \frac{\mathrm{Var}[C(\boldsymbol{\theta}_i)]}{N_{\text{shots}}}}.
\end{equation}
As this limit is approached, the previously described statistical artifacts appear, where downward fluctuations yield energies ostensibly below the true ground state.

\begin{figure}[htpb]
  \centering
  \includegraphics[width=0.75\textwidth]{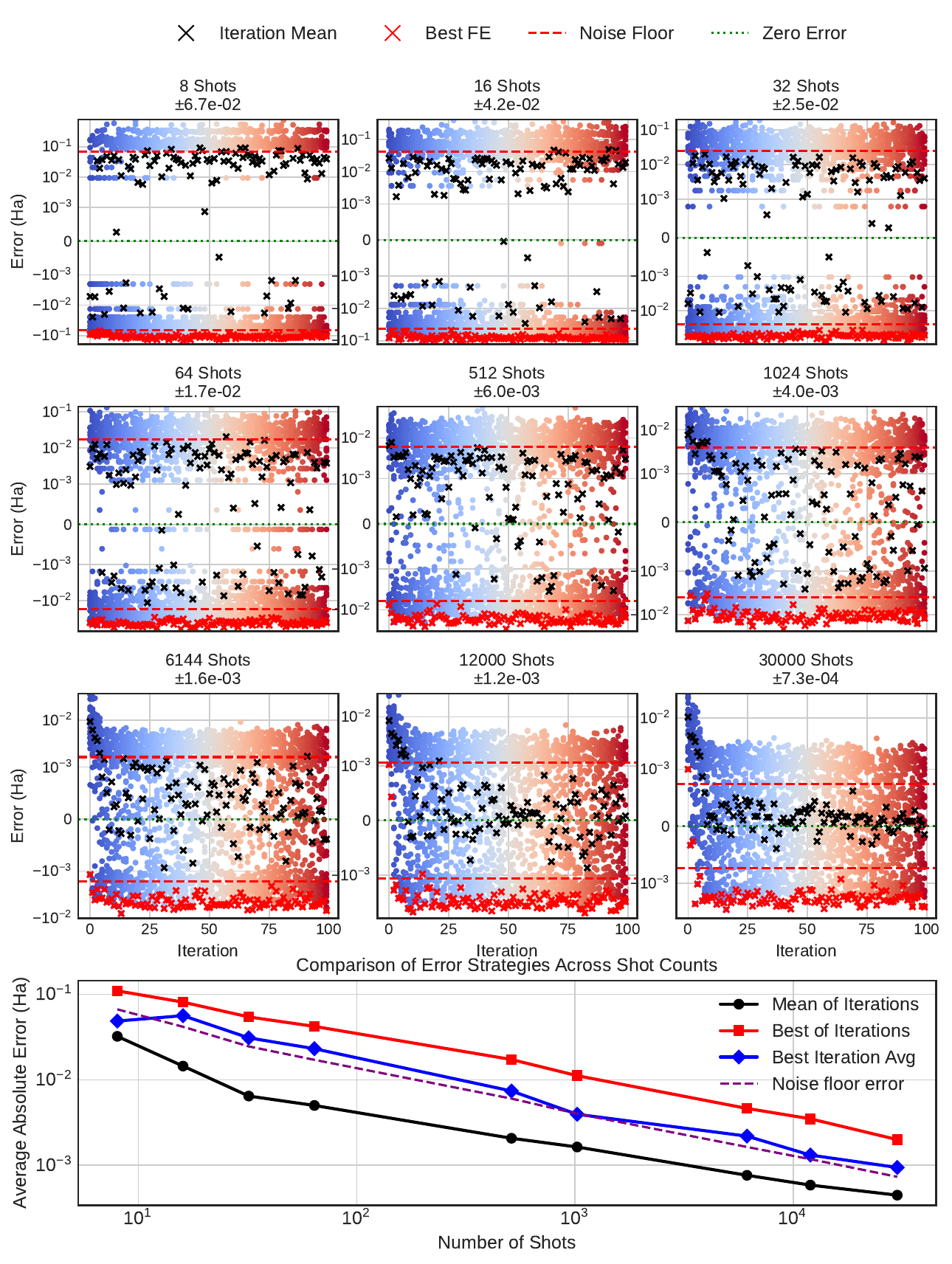}
  \footnotesize
  \caption{Energy error progression for \ce{H2} using \ac{tvha} with \ac{cmaes} population size 25. Top: \acp{fe} of all individuals (colored points), average of \acp{fe} in iteration (black crosses), best (lowest) \ac{fe} in iteration (red crosses) and noise floor (red dashed lines). Bottom: Aggregated absolute errors for the mean of \acp{fe} in iteration (black line), aggregated absolute errors for the best \ac{fe} in iteration (red line), and the best iteration average (blue) approaches compared to the computed noise floor error (purple).}
  \label{fig:error_analysis}
\end{figure}

We empirically validate these bounds in \cref{fig:error_analysis} using the \ac{cmaes} optimizer with a population size of $\lambda = 25$ under different shot budgets. The upper panel contains nine subplots, each tracking 100 iterations at shot counts ranging from 8 to 30,000, corresponding to empirical noise floors from 0.0667 down to 0.0007 Ha. The noise floor is computed as $\sigma_{\text{noise}} = \sqrt{\bar{\sigma}^2/N_{\text{shots}}}$, where $\bar{\sigma}^2$ is the mean variance of all evaluations. Each subplot tracks relative errors $\epsilon_{t,j} = E_{t,j} - E_{\text{true}}$ across individuals. Red crosses mark the lowest energy individuals, while black crosses indicate the population mean. In the vicinity of the optimum, the optimizer wanders within the noise floor, with energies scattered both above and below the ground state due to sampling noise. The low shot cases show little clustering near the zero error line, and many red crosses fall below it, illustrating noise driven overfitting. In contrast, black crosses cluster symmetrically around the optimum, reflecting the stability of mean estimates. The dashed red lines indicate the empirical noise floors $\pm\sigma_{\text{noise}}$.

The lower panel of \cref{fig:error_analysis} aggregates these results across shot counts to compare three distinct error metrics. The black line represents the average magnitude of the population mean error across all iterations. The red line tracks the average error of the single best individual per iteration, highlighting the bias toward minimal values. Finally, the blue line serves as a control, plotting the population mean specifically for the single iteration where the global best solution was recorded. The collective evidence from both panels demonstrates that population means naturally filter measurement noise, while best value selection amplifies it. This confirms that in noisy quantum environments, traditional optimization approaches relying on the ``best'' individual do not merely struggle with precision; they actively mislead by chasing statistical fluctuations rather than true physical minima. This discrepancy highlights our key finding that population means resist overfitting by averaging out statistical fluctuations, thereby providing a more robust and physically meaningful estimate of progress.

The statistical advantage of the population mean arises directly from the scaling law where, for a population size $\lambda$, the error scales as $\text{Error} \propto \sigma_{\text{noise}}/\sqrt{\lambda}$. This phenomenon is supported by recent theoretical and empirical studies. Huggins~\cite{huggins2021efficient} demonstrated that strategic averaging provides noise resilient measurements, while Barron~\cite{nature_comp_sci_2024} formally proved that conditional averaging yields provable bounds surpassing individual measurement precision. Our findings are also consistent with recent work by Mohammad~\cite{hopso_vqe_2025}, showing that population based optimizers exhibit superior noise resilience through inherent averaging mechanisms. Ultimately, while the sampling noise floor sets a precision limit that cannot be bypassed, robust estimation strategies such as population means allow variational algorithms to operate reliably within this boundary.

\FloatBarrier
\FloatBarrier

\section{Experimental Setup}
\label{sec:setup}

We conducted extensive simulations of \ac{tvha} on classical computing infrastructure to evaluate its performance for quantum chemistry applications. Our study encompasses four molecular systems, \ce{H2}, \ce{LiH} (full space), \ce{LiH} (active space), and \ce{H4} chain, with eight different optimizers (\ac{bfgs}, \ac{cmaes}, \ac{cobyla}, \ac{gd}, \ac{nelder}, \ac{pso}, \ac{slsqp}, and \ac{spsa}) tested for each molecule. The molecular configurations are shown in detail in \Cref{app:molecular_config}.

For each optimizer-molecule combination, we performed 40 independent runs (10 runs per configuration) across four distinct scenarios, Hartree-Fock initialization with statevector simulation, Hartree-Fock initialization with sampling noise (shot-based simulation), random parameter initialization with statevector simulation, and random parameter initialization with sampling noise. This resulted in 1,280 independent simulations (4 molecules $\times$ 8 optimizers $\times$ 4 configurations $\times$ 10 runs), with each simulation running up to  optimization iterations using optimizer-specific convergence criteria. We employed a fixed shot count of 6144 ($6 \times 1024$ standard batches) per evaluation, a relatively high budget given the extensive simulation sweep to guarantee that the statistical noise floor remains sufficiently suppressed to distinguish chemically relevant energy differences from sampling artifacts.

For random initialization, we drew parameters independently for each run from continuous uniform distribution over $\langle0, 1)$ via \texttt{numpy.random.rand}\footnote{\url{https://numpy.org/doc/2.1/reference/random/generated/numpy.random.rand.html}}. The starting parameters were not shared across optimizers, reflecting method differences; for example, population-based methods such as \ac{pso} cannot mirror single-point starts used by \ac{gd} or \ac{bfgs}. The pseudo-random seeds were not fixed, so that each trial is an independent stochastic draw with respect to the starting points, internal optimizer steps and shot sampling. Hyperparameters were taken from widely used defaults of the respective implementations rather than case-by-case tuning, which improves comparability and generalizability of conclusions across methods. The detailed convergence criteria can be found in \cref{app:convergence}.

To obtain expectation values, the Qiskit Aer \texttt{Estimator} primitive\footnote{\url{https://qiskit.github.io/qiskit-aer/stubs/qiskit_aer.primitives.Estimator.html}} was used for shot-based noisy simulations and the Qiskit \texttt{StatevectorEstimator}\footnote{\url{https://quantum.cloud.ibm.com/docs/en/api/qiskit/qiskit.primitives.StatevectorEstimator}} for exact reference simulations. Both estimators implement the same operator-expectation framework, but while the  \texttt{StatevectorEstimator} evaluates expectation values deterministically, the Aer \texttt{Estimator} draws samples according to Born’s rule, introducing binomial shot noise with variance scaling as $1/N_{\mathrm{shots}}$.

Given the computational intensity of classical quantum circuit simulations, we leveraged the Barbora supercomputer, a Bull Sequana X cluster featuring 192 standard nodes (2$\times$18-core Intel Xeon, 192 GB RAM), 8 GPU nodes (2$\times$12-core Intel Xeon, 4$\times$NVIDIA V100), 1 fat node (8$\times$16-core Intel Xeon, 6 TB RAM), Infiniband HDR interconnect (200 Gb/s), and 310 TB SCRATCH storage with 28 GB/s throughput.

Our parallelization strategy employed an embarrassingly parallel approach with job-level distribution. Each independent run was submitted as a separate Slurm job, eliminating inter-process communication overhead. For molecules requiring longer simulations (\ce{LiH}, \ce{H4}), we managed groups of 5--10 concurrent runs using Slurm job arrays, allocating each task to a dedicated CPU node with 1 core per job to maximize throughput. Simulation times scaled with molecular complexity, \ce{H2} statevector runs completed in hours (tens of hours with sampling noise), while \ce{LiH}/\ce{H4} statevector simulations took roughly one day (extending to more than one week per run with sampling noise). The complete study consumed $\sim${4500} node hours, efficiently utilizing the cluster's capacity for long-running, independent tasks.

The simulations were implemented using a Python-based quantum chemistry stack combining several specialized libraries with strict version control, as listed in \Cref{tab:software_dependencies}.

\begin{table}[htbp]
    \centering
    \caption{Key Software Dependencies and Versions}
    \label{tab:software_dependencies}
    \begin{tabular}{|l|l|}
        \hline
        \textbf{Python Library} & \textbf{{Version}} \\
        \hline
        \texttt{qiskit} \cite{qiskit} & {1.0.2} \\
        \texttt{qiskit\_algorithms} \cite{qiskit} & {0.3.1} \\
        \texttt{qiskit\_nature} \cite{qiskit} & {0.7.2} \\
        \texttt{qiskit\_aer} \cite{qiskit} &  {0.14.2} \\
        \texttt{pyscf} \cite{sun2018pyscf} &  {2.6.0} \\
        \texttt{scipy} \cite{2020SciPy-NMeth} & {1.15.2} \\
        \texttt{cma} \cite{hansen2019pycma} &  {3.3.0} \\
        \hline
    \end{tabular}
\end{table}

The computational workflow proceeded through several stages, constructing the second-quantized Hamiltonian via {\texttt{pyscf}} in \ac{sto3g} basis, applying Jordan-Wigner mapping to obtain the qubit Hamiltonian, pruning small terms ($|\gamma_i| < \mathrm{threshold}$). The pruning threshold was set according to \cref{eq:truncation_threshold} in {\cref{sec:vha}, using $p=0.999$ to retain the most significant Hamiltonian terms while reducing circuit complexity. This corresponds to keeping terms whose cumulative contribution accounts for 99.9\% of the total non-Coulomb two-body interaction strength. 
The workflow continued by building the \ac{tvha} ansatz with Trotterized time evolution operators, optimizing parameters using classical optimizers, and repeating the process for all molecule-optimizer-initialization combinations.
The implementation used {\texttt{qiskit\_nature}}'s operator formalism for efficient Pauli string manipulation, with custom modifications for variational Hamiltonian approximation terms. All simulations recorded complete optimization trajectories including energy evaluations, parameter updates, and convergence metrics.

\section{Benchmark Results}
\label{sec:results}

Our extensive benchmarking of eight optimization algorithms across four molecular systems using the \ac{tvha} framework reveals fundamental insights into the performance of the variational quantum eigensolver. \cref{fig:h2_mean_conv,fig:lih_as_mean_conv} show convergence of optimization algorithms for each molecule on statevector and sampling simulations. The plot shows the mean energy over 10 independent runs as a function of function evaluations (log-log scale). In these plots, we can observe the discrepancy between exact simulation (statevector) compared to noisy optimization (sampling noise) and also comparing optimization starting from Hartree-Fock initial points (\ac{HF} Init) and random initial points (Random Init). The first plotted energy in these plots is the energy after the first iteration of the optimization method, which is the reason, why the convergence curves have different starting values even in the case of Hartree-Fock initialization. In all plots, we do not show results for \ac{slsqp} under sampling noise due to large errors and divergence of the \ac{slsqp} optimizer. For more detail, see \cref{sec:convergence_runs} where we show plots of each optimizer individually with every single run displayed. In \cref{fig:all_errorbars} we show the final errors in more detail. These figures with numerical results being discussed in more detail in this section should provide more insight into three critical performance axes, initialization sensitivity, noise resilience, and molecular complexity scaling, directly tied to the \ac{tvha} architecture described in \cref{sec:vha}.

\begin{figure}[htpb]
\centering
\includegraphics[width=\linewidth]{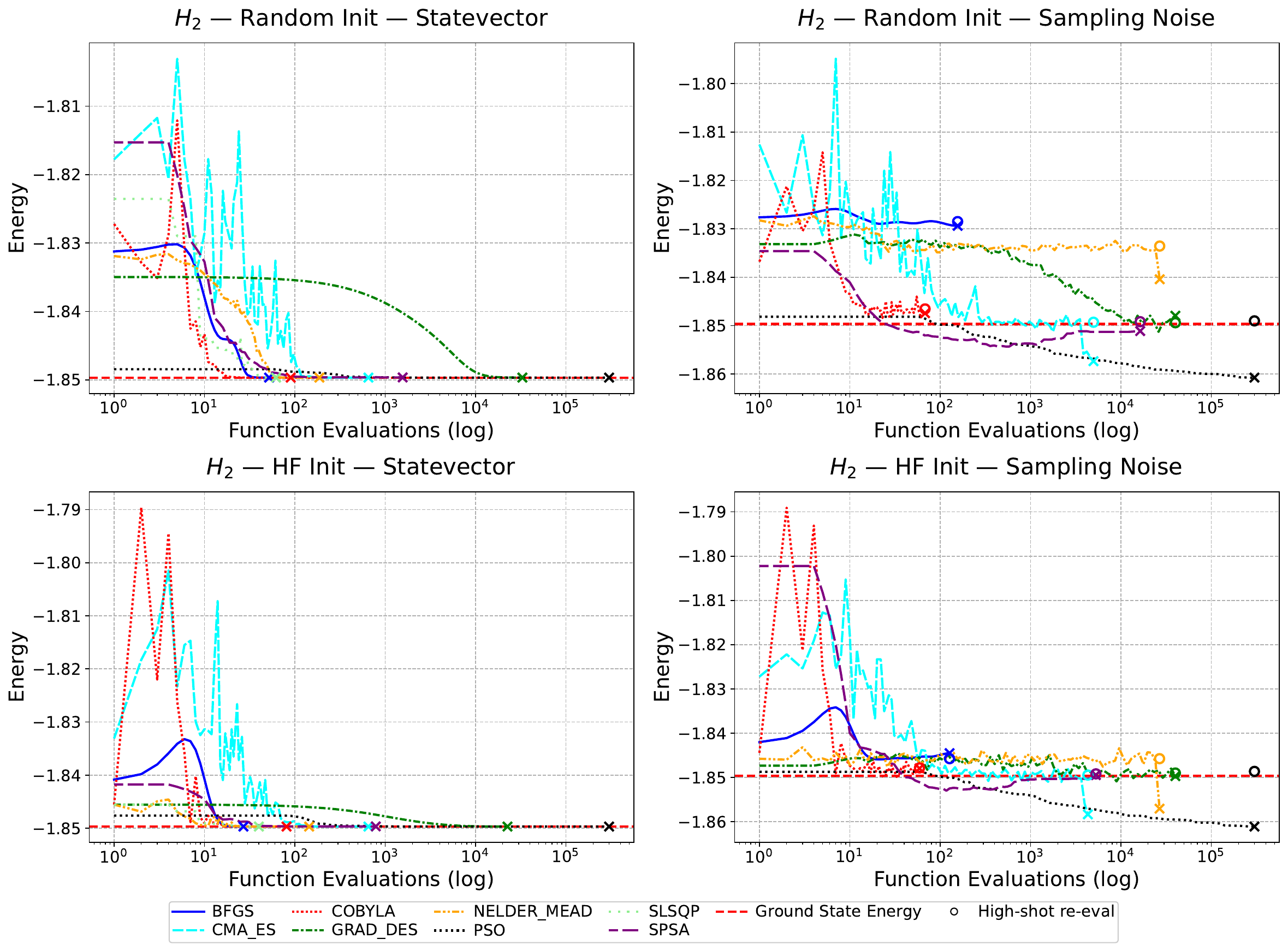}
\caption{Convergence of optimization algorithms for \ce{H2} energy on statevector and sampling simulations. The plot shows the mean energy over 10 independent runs as a function of function evaluations (log-log scale). Crosses mark the lowest energies encountered during optimization, while circles denote high-shot ($10^{5}$) reevaluations of the corresponding parameters.}
\label{fig:h2_mean_conv}
\end{figure}

For \ce{H2}, the simplest molecule, statevector simulations establish the expected baseline. All optimizers apart from \ac{gd} converge essentially to machine precision. \ac{bfgs}, \ac{cmaes}, \ac{cobyla}, \ac{nelder}, and \ac{pso} reach errors below $10^{-13}$–$10^{-14}$ $\mathrm{Ha}$. \ac{gd} converges more slowly, but still achieves micro-Hartree accuracy ($\sim 10^{-6}$ $\mathrm{Ha}$ with HF). \ac{spsa} performs similarly, with final errors around $3\times 10^{-5}$ {$\mathrm{Ha}$.} These results show that, for the noiseless case, differences are mainly in convergence speed, not in final accuracy.

The introduction of sampling noise reveals much stronger differentiation. Without correction, optimizers report apparent errors ranging from a few millihartree (\ac{cobyla}, \ac{spsa}) up to tens of millihartree (\ac{bfgs}, \ac{nelder}). Many runs even appear below the true ground state due to shot-induced fluctuations, an artifact confirmed by the crosses in \cref{fig:h2_mean_conv}. High-shot reevaluation exposes the actual optimizer hierarchy. The lowest corrected mean errors come from \ac{gd} with \ac{HF} initialization ($6.6\times 10^{-4}$ Ha), followed by \ac{spsa} with \ac{HF} ($4.1\times 10^{-4}$ Ha) and \ac{cmaes} ($4.5\times 10^{-4}$ Ha). These three optimizers consistently remain within sub-millihartree accuracy once statistical artifacts are filtered.

\ac{slsqp} fails to converge even for the most simple system in the presence of sampling noise even when initialized near minima. This instability shows that the local curvature of the cost landscape becomes comparable to or smaller than the sampling noise amplitude, leading to inconsistent gradient and Hessian estimates and unstable line searches. The result confirms that in noisy regimes the effective landscape becomes weakly curved and stochastic rather than deterministic, rendering curvature-based optimization unreliable. The collapse of \ac{slsqp} illustrates the limit of deterministic descent: when noise dominates curvature, optimization transitions from contraction toward a minimum to random diffusion across a fluctuating surface. Due to its complete failure in noisy environments for all molecular systems, \ac{slsqp} is not further discussed in subsequent sections as the result remains unchanged. See also \cref{sec:convergence_runs} for further details.

The results for \ac{bfgs} differ highly from the other optimization methods. While it is the fastest one in terms of convergence in the noiseless environment, it does degrade the most when noise is considered, resulting in errors close to $4\times 10^{-3}$ $\mathrm{Ha}$ even after the correction is applied. In constrast to the completely failing \ac{slsqp} method, \ac{bfgs}, though built on the same quasi-Newton principle, degrades only slightly because its unconstrained updates average out noise instead of amplifying it through nested constraint subproblems.
\ac{nelder} optimization methods also performs poorly with Hartree-Fock initial state with errors of $3.9\times 10^{-3}$ {$\mathrm{Ha}$.} This shows that this method is strongly influenced by stochastic ruggedness inside quasidegenerate valleys, which characterizes the noisy landscape. The next optimization method, \ac{cobyla}, places in the middle of all of the tested methods, with error rates of $2.4\times 10^{-3}$ $\mathrm{Ha}$.

Initialization effects remain strong. \ac{HF} starts consistently reduce error variance and mean error for most optimizers, but not universally. \ac{pso} is a counterexample: random initialization yields lower error under noise ($6.5\times 10^{-4}$ Ha) compared to \ac{HF} ($1.0\times 10^{-3}$ Ha). For \ac{cmaes} and \ac{spsa}, both \ac{HF} and random starts succeed, but \ac{HF} provides the tighter error distribution. This shows that while \ac{HF} initialization is generally stabilizing, population-based optimizers can sometimes benefit from broader random exploration.

In general, the \ce{H2} results highlight three principles. First, statevector simulations mask differences in optimizer quality since nearly all methods reach chemical accuracy effortlessly. Second, in noisy simulations, gradient methods like \ac{bfgs} lose their advantage, and robustness is instead found in stochastic and population-based approaches (\ac{spsa}, \ac{cmaes}, \ac{pso}, and even \ac{gd} after reevaluation). Third, high-shot validation is critical, as the apparent best results from noisy runs are often misleading. The corrected ranking shows that only a subset of optimizers can reliably reach sub-millihartree accuracy under realistic sampling conditions.

\begin{figure}[htpb]
\centering
\includegraphics[width=\linewidth]{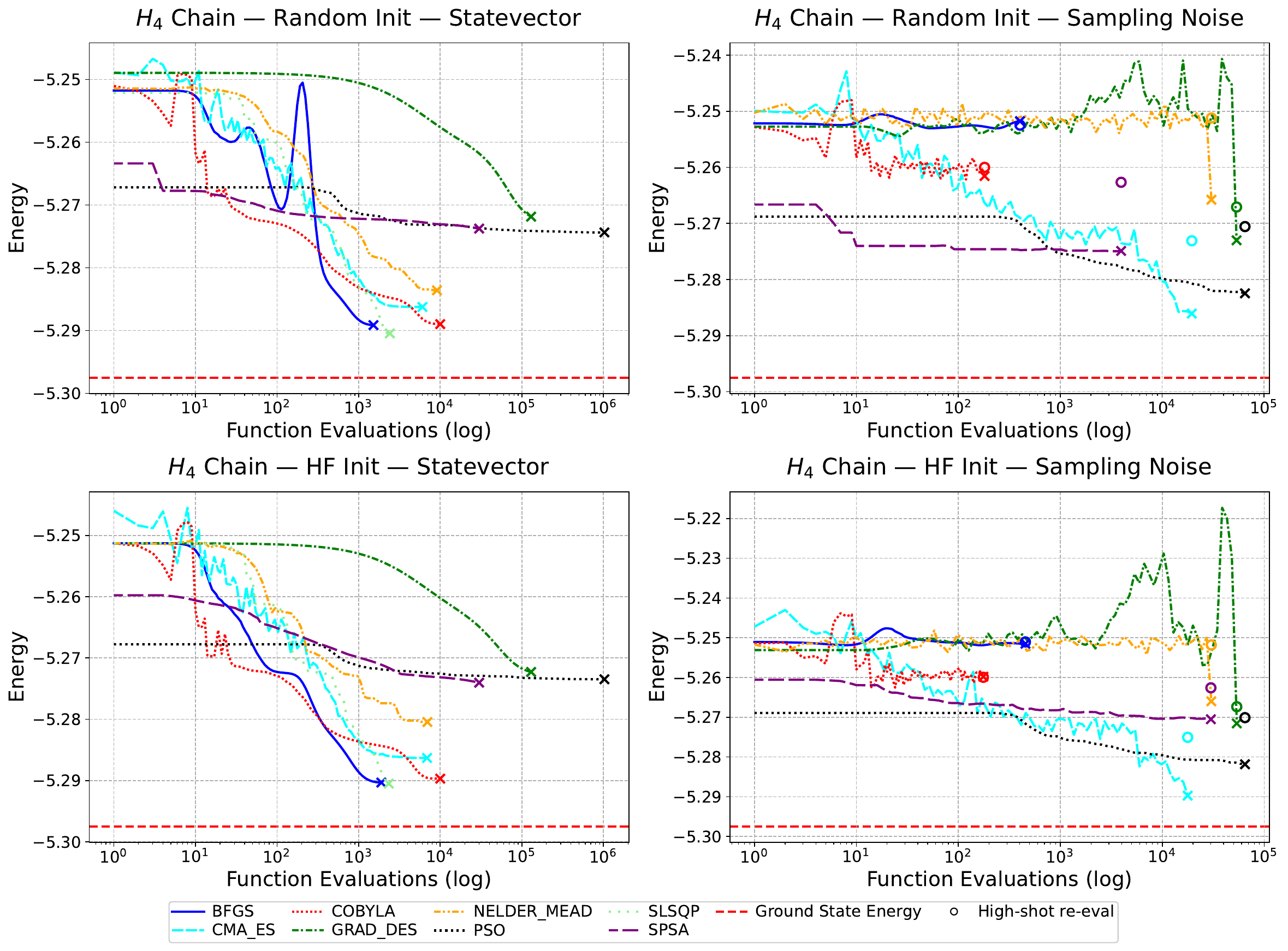}
\caption{Convergence of optimization algorithms for \ce{H4} energy on statevector and sampling simulations. The plot shows the mean energy over 10 independent runs as a function of function evaluations (log-log scale). Crosses mark the lowest energies encountered during optimization, while circles denote high-shot ($10^{5}$) reevaluations of the corresponding parameters.}
\label{fig:h4_mean_conv}
\end{figure}

The \ce{H4} system (\cref{fig:h4_mean_conv}) increases the Hamiltonian complexity and makes the optimizer differences more visible. In statevector runs, \ac{bfgs}, \ac{cobyla}, and \ac{slsqp} remain efficient, all reaching errors on the order of $7\times 10^{-3}$ {$\mathrm{Ha}$.} \ac{cmaes} and \ac{nelder} converge more slowly, while \ac{gd} is consistently less accurate ($\sim 2.5\times 10^{-2}$ Ha). As in \ce{H2}, most methods succeed in the noiseless case, but their separation increases with molecular size.

Under sampling noise, the raw optimization traces show large spreads. High-shot reevaluation clarifies the hierarchy: \ac{cmaes} with \ac{HF} initialization achieves the lowest corrected mean error ($2.25\times 10^{-2}$ Ha), closely followed by \ac{gd} ($3.01\times 10^{-2}$ Ha) and \ac{pso} ($2.7\times 10^{-2}$ Ha). By contrast, \ac{bfgs} and \ac{nelder} degrade strongly, both above $4.5\times 10^{-2}$ {$\mathrm{Ha}$ }after correction, confirming their high sensitivity to noisy landscapes. \ac{spsa} and \ac{cobyla} fall in the middle, around $3.5$ -$3.8\times 10^{-2}$ {$\mathrm{Ha}$.}

Initialization effects are weaker than in \ce{H2}. For \ac{bfgs} and \ac{cmaes}, \ac{HF} provides only improvement of few percents under noise, and for \ac{pso} the two starting strategies yield nearly the same error. This indicates that as system complexity increases, the stabilizing advantage of \ac{HF} initialization is partially lost.

In general, \ce{H4} shows that the increase in molecular size amplifies the differences between the optimizers under noise. Gradient-based methods lose their edge, with \ac{bfgs} dropping from top performer in statevector to one of the weakest under sampling. Population-based approaches (\ac{cmaes}, \ac{pso}) and even simple \ac{gd} become comparatively more reliable, though still limited to tens of millihartree accuracy. High-shot reevaluation is again critical, as raw noisy outputs exaggerate both under- and overestimates of the ground state.

\begin{figure}[htpb]
\centering
\includegraphics[width=\linewidth]{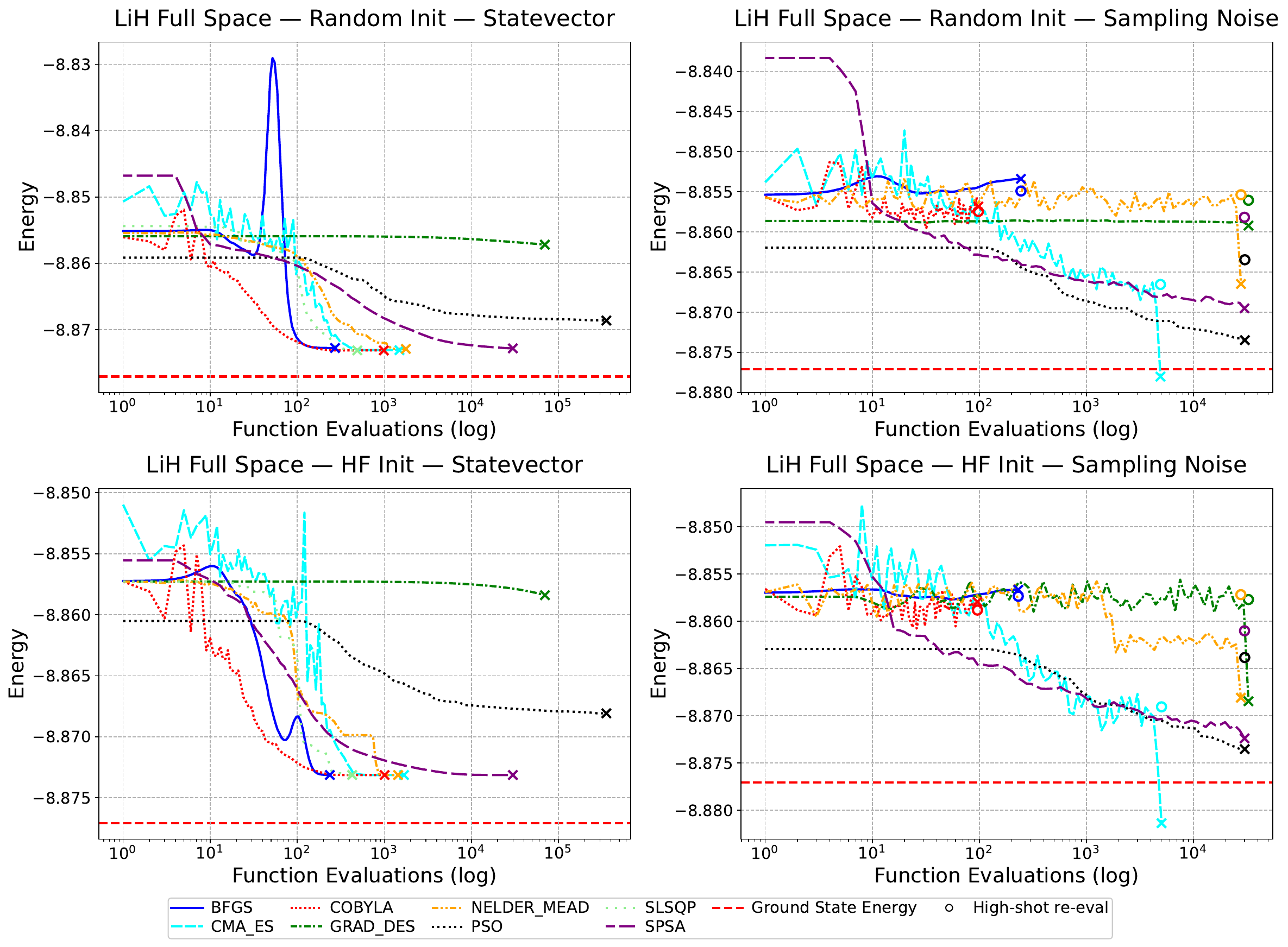}
\caption{Convergence of optimization algorithms for \ce{LiH} energy on statevector and sampling simulations. The plot shows the mean energy over 10 independent runs as a function of function evaluations (log-log scale). Crosses mark the lowest energies encountered during optimization, while circles denote high-shot ($10^{5}$) reevaluations of the corresponding parameters.}
\label{fig:lih_mean_conv}
\end{figure}

The \ce{LiH} system (\cref{fig:lih_mean_conv}) highlights the challenges of more complex ansatz. In statevector simulations most optimizers reach sub-millihartree precision, with \ac{bfgs}, \ac{cmaes}, \ac{cobyla}, \ac{nelder}, and \ac{spsa} all converging near $4\times 10^{-3}$ {$\mathrm{Ha}$.} \ac{gd} and \ac{pso} lag behind at $\sim 2\times 10^{-2}$ and $9\times 10^{-3}$ {$\mathrm{Ha}$ }respectively, reflecting scaling difficulties compared to \ce{H2} and \ce{H4}.

Noise reshapes the ranking. After high-shot reevaluation, \ac{cmaes} with \ac{HF} initialization emerges as the most accurate, with mean error $\sim 8\times 10^{-3}$ {$\mathrm{Ha}$.} \ac{pso} follows closely at $1.3\times 10^{-2}$ Ha, showing that population-based search retains robustness in the larger parameter space. \ac{spsa} also performs well, correcting to $1.6\times 10^{-2}$ {$\mathrm{Ha}$ }with HF, outperforming all deterministic gradient-based methods. By contrast, \ac{bfgs}, \ac{cobyla}, and \ac{nelder} all rise to $\sim 2\times 10^{-2}$ Ha, confirming strong sensitivity to stochastic noise in larger systems. \ac{gd} degrades similarly ($\sim 1.9\times 10^{-2}$ Ha), consistent with its poor scaling behavior.

Initialization effects are still visible but weaker in smaller molecules. For \ac{cmaes}, \ac{HF} provides a clear improvement (8.0 vs 10.5 mHa), while \ac{pso} shows almost no dependence (13.2 vs 13.6 mHa). \ac{bfgs} benefits moderately (19.7 vs 22.2 mHa), but for most optimizers \ac{HF} no longer guarantees dominance, reflecting that random exploration can sometimes avoid noise-amplified traps in the high-dimensional landscape.

Overall, \ce{LiH} illustrates that as molecular complexity increases, only population-based and stochastic optimizers (\ac{cmaes}, \ac{pso}, \ac{spsa}) maintain accuracy below 20 {$\mathrm{mHa}$ } under noise, while gradient-based methods lose their advantage entirely. High-shot reevaluation is again essential, revealing that raw noisy outputs, sometimes appearing better than the true ground state, are misleading without statistical correction.

\begin{figure}[htpb]
\centering
\includegraphics[width=\linewidth]{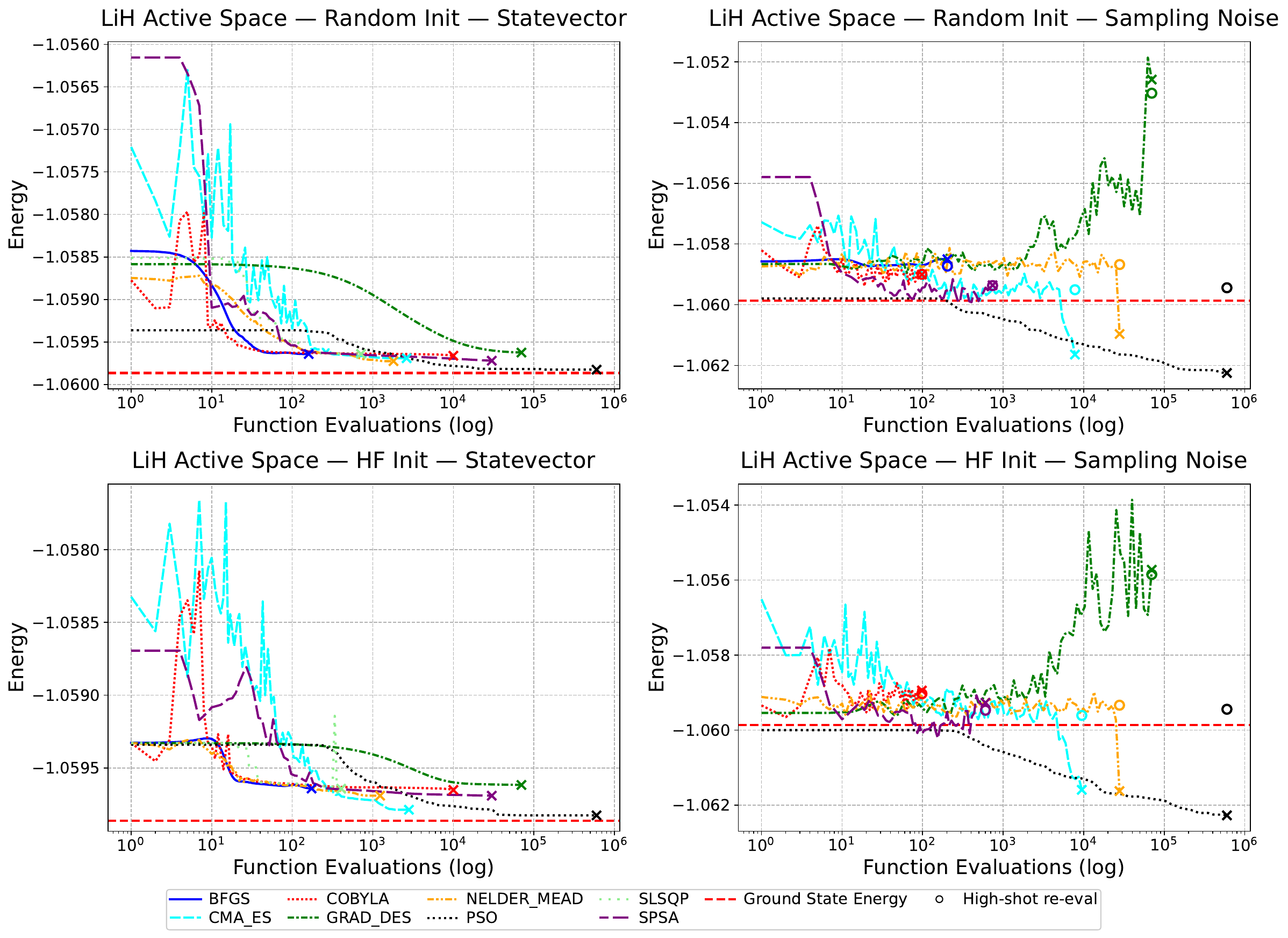}
\caption{Convergence of optimization algorithms for \ce{LiH} active space energy on statevector and sampling simulations. The plot shows the mean energy over 10 independent runs as a function of function evaluations (log-log scale). Crosses mark the lowest energies encountered during optimization, while circles denote high-shot ($10^{5}$) reevaluations of the corresponding parameters.}
\label{fig:lih_as_mean_conv}
\end{figure}

The \ce{LiH} active space (\cref{fig:lih_as_mean_conv}) illustrates the benefit of orbital truncation, which reduces both optimizer cost and noise sensitivity relative to the full molecule. In statevector runs all optimizers converge to sub-millihartree accuracy, with \ac{cmaes}, \ac{pso}, \ac{nelder}, and \ac{spsa} reaching errors below $2\times 10^{-4}$ {$\mathrm{Ha}$.} \ac{bfgs} and \ac{cobyla} also perform well ($\sim 2\times 10^{-4}$ Ha), confirming that in the simplified parameter space the choice of optimizer is less critical under noiseless conditions.

The presence of sampling noise, however, separates the methods. The high-shot reevaluation shows that \ac{cmaes} with \ac{HF} initialization is the best performer ($2.6\times 10^{-4}$ Ha), followed by \ac{spsa} ($3.9\times 10^{-4}$ Ha) and \ac{pso} ($4.2\times 10^{-4}$ Ha). \ac{cobyla} and \ac{bfgs} achieve $\sim 8\times 10^{-4}$ to $1.1\times 10^{-3}$ Ha, while \ac{nelder} lies in between ($5.3\times 10^{-4}$ {$\mathrm{Ha}$ }with HF). \ac{gd} fails to adapt, producing errors of several millihartree even after reevaluation. \ac{slsqp}, though accurate under statevector, completely diverges under noise with errors exceeding $3\times 10^{-1}$ Ha, marking the worst case across all molecules.

Initialization plays a more nuanced role than in full \ce{LiH}. \ac{cmaes} and \ac{spsa} gain modest advantage from \ac{HF} starts, while \ac{pso} performs nearly identically for \ac{HF} and random. For \ac{nelder}, \ac{HF} improves stability, cutting the error from $1.2\times 10^{-3}$ Ha (random) to $5.3\times 10^{-4}$ {$\mathrm{Ha}$.} \ac{bfgs} and \ac{cobyla} also benefit slightly, but not decisively.

Active space reduction lowers optimizer demands and improves noise robustness by roughly an order of magnitude compared to full \ce{LiH}. The ranking under noise shifts toward stochastic and population-based methods (\ac{cmaes}, \ac{spsa}, \ac{pso}), which reliably reach sub-millihartree accuracy. Gradient-based optimizers, once dominant in \ce{H2}, collapse in performance for larger systems and especially under noise, while \ac{slsqp} proves unusable.

In general, the optimizer hierarchy evolves systematically with system size and sampling noise. For the smallest system, \ce{H2}, deterministic gradient-based methods such as \ac{bfgs} achieve the fastest and most accurate convergence under noiseless conditions, while stochastic and population-based approaches offer no clear benefit. As molecular complexity increases in \ce{H4} and full \ce{LiH}, the situation reverses: gradient-based optimizers lose their effectiveness, and stochastic methods, particularly \ac{cmaes}, \ac{pso}, and \ac{spsa}, provide the most consistent convergence under realistic sampling noise. This transition marks a shift from curvature-driven optimization in smooth landscapes to noise-resilient exploration in high-dimensional, stochastic ones.

Under sampling noise, gradient-based optimizers exhibit a complete breakdown of convergence reliability. \ac{slsqp} consistently diverges, confirming that curvature and constraint estimates become dominated by stochastic variance. \ac{bfgs} shows no measurable improvement once noise is introduced and often worsens, with its quasi-Newton updates amplifying fluctuations instead of guiding descent. \ac{gd} remains numerically stable but quickly plateaus far above the ground state energy, revealing that gradients vanish into noise and the landscape becomes effectively flat. These behaviors indicate that the optimization surface under noise is weakly curved and stochastic, where gradient estimates fluctuate symmetrically around zero, eliminating any usable descent information.

For larger systems, \ac{nelder} traces suggest apparent convergence that vanishes after high-shot reevaluation, showing that its progress is driven almost entirely by noise rather than meaningful parameter updates. In contrast, stochastic and population-based optimizers such as \ac{cmaes}, \ac{pso}, and \ac{spsa} maintain steady but slower improvement through averaging effects that filter random fluctuations. The overall pattern across all systems demonstrates that as the parameter space grows, deterministic gradient-based methods lose their functional signal, and optimization transitions from gradient-driven contraction to diffusion dominated by sampling noise.

\begin{framed}
Sampling noise reverses optimizer performance hierarchy. Gradient-based methods lose reliability as gradients vanish into noise, while population-based and stochastic optimizers (\ac{cmaes}, \ac{pso}, \ac{spsa}) maintain stable progress through averaging and exploration, offering the highest robustness across system sizes.
\end{framed}

Initialization effects follow a similar trend. \ac{HF} starts offer clear efficiency gains in \ce{H2}, often cutting errors by more than half, but their influence weakens with system size. For \ce{LiH}, random starts perform comparably for \ac{pso} and even outperform in some noisy cases, suggesting that the structured advantage of \ac{HF} initialization is gradually lost in higher-dimensional landscapes and increasing electron correlation.

Active space reduction in \ce{LiH} provides a distinct mechanism for managing optimization difficulty. Truncation not only reduces the computational burden but also improves noise tolerance by more than an order of magnitude. Under this simplification, optimizers that otherwise struggle, such as \ac{cobyla} or \ac{spsa}, become competitive and achieve sub-millihartree accuracy at moderate cost.

A cross-figure comparison of the final error distributions in \cref{fig:all_errorbars} reveals several additional patterns from our results. First, the gap between noiseless and noisy runs systematically widens from \ce{H2} to \ce{LiH} full space, showing that noise amplification scales with molecular complexity and ansatz size. High-shot reevaluation reduces these discrepancies, but its effect is optimizer-dependent: for \ac{cmaes} and \ac{pso} the reevaluated bars collapse close to zero, while gradient-based methods such as \ac{bfgs} or \ac{nelder} remain elevated, indicating that their errors stem from structural mis-optimization rather than statistical fluctuations. Initialization advantages also vary non-monotonically: Hartree–Fock starts give a clear benefit for \ce{H2}, nearly vanish in \ce{H4}, and partially reappear in the \ce{LiH} active space, suggesting that contracted parameter spaces restore the usefulness of chemically informed priors. \ac{spsa} shows a consistent upward trajectory across system size, moving from mid-tier in \ce{H2} to near top-tier in \ce{LiH} active space, which highlights its scaling advantages compared to deterministic gradient-based schemes.

Another notable feature arises in the \ce{LiH} full space convergence curves, displayed in \cref{fig:lih_mean_conv}. Under noisy conditions, the best candidates from \ac{cmaes} achieve lower apparent errors than the statevector baseline, while \ac{spsa} and \ac{pso} perform comparably. However, once the same parameters are reevaluated at high shot counts, the errors increase, confirming that the originally lower values were artifacts of sampling fluctuations. This behavior reflects the fact that none of the optimizers converged tightly to the true ground state in the noisy landscape; instead, transient noise-induced underestimations created the illusion of superior performance. The reevaluation thus reveals that sampling noise can occasionally mask systematic convergence difficulties by artificially lowering measured energies, a phenomenon most visible in larger systems where optimizers struggle with high-dimensional rugged landscapes.

\begin{figure}[htpb]
    \centering
    \includegraphics[width=1\linewidth]{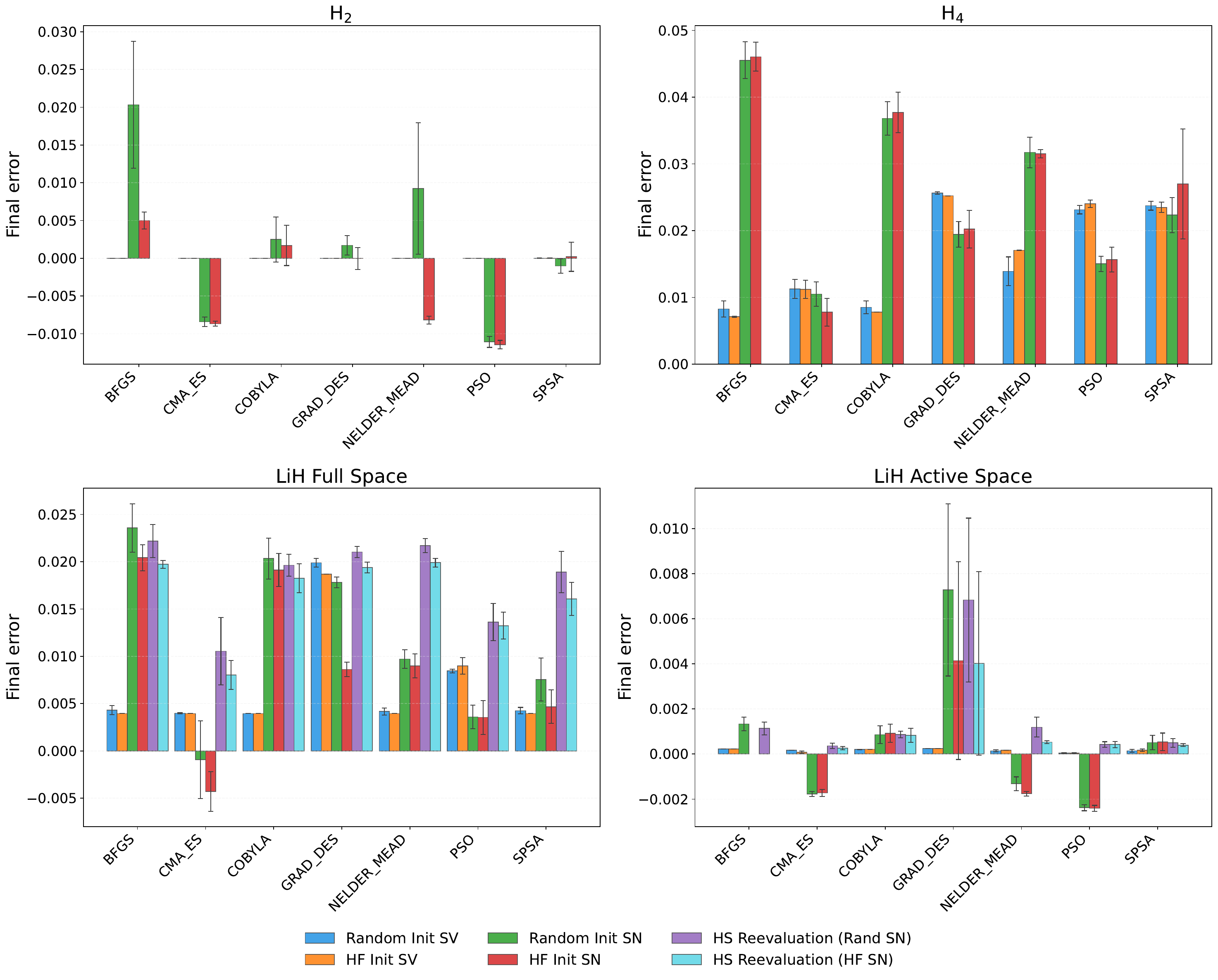}
    \caption{Final error bars for all four systems: \ce{H2}, \ce{H4} chain, \ce{LiH} Full Space, and \ce{LiH} Active Space. Grouped bars per optimizer show the mean final error with 95\% confidence intervals. Bars correspond to: Random Init SV (randomly initialized parameters with statevector simulation), \ac{HF} Init SV (Hartree–Fock initialization with statevector simulation), Random Init SN (randomly initialized parameters with sampling noise), and \ac{HF} Init SN (Hartree–Fock initialization with sampling noise). Two additional bars per optimizer represent High Shot (HS) reevaluation under sampling noise conditions.}
    \label{fig:all_errorbars}
\end{figure}

\cref{tab:optimizer_comparison} provides a concise summary of the optimizers utilized in our study. It outlines their respective strengths, weaknesses, and the types of problems for which they are best suited based on our experimental findings. This comparison offers insights into the trade-offs associated with each optimization method in the context of variational quantum simulations.

\begin{table}[htpb]
\centering
\scriptsize
\setlength{\tabcolsep}{2pt} 
\renewcommand{\arraystretch}{0.9} 
\begin{tabular}{|p{1.6cm}|p{4.1cm}|p{3.8cm}|p{3.6cm}|}
\hline
\textbf{Optimizer} & \textbf{Strengths} & \textbf{Weaknesses} & \textbf{Best for} \\
\hline
\ac{bfgs} & Fast noiseless convergence; effective with \ac{HF} initialization & Noise-sensitive; initialization-dependent & Clean small systems needing precision \\
\hline
\ac{cmaes} & Noise-robust; consistent across molecules & High function evaluation cost; slow to high precision & Noisy/high-dim problems or hardware \\
\hline
\ac{cobyla} & Effective in low-noise conditions; good for small-mid systems & Noise-sensitive; struggles with rugged landscapes & Weakly noisy gradient-free cases \\
\hline
\ac{gd} & Simple; works well in smooth, noiseless cases & Very noise-sensitive; poor for complex cases & Idealized examples/small spaces \\
\hline
\ac{nelder} & Derivative-free; simple to implement & Slow; poor scaling; fails in noisy or high-dim problems & Toy problems/prototyping \\
\hline
\ac{pso} & Gradient-free; noise-stable; easy & Slow; inconsistent across runs & Noisy tasks needing exploration \\
\hline
\ac{slsqp} & Strong clean simulations with good initialization & Fails with noise/poor starts & Small low-noise gradient problems \\
\hline
\ac{spsa} & Noise-tolerant; sampling-efficient & Slower convergence; tuning required & Hardware noisy optimization with budget limits \\
\hline
\end{tabular}
\caption{Comparison of optimization algorithms for variational quantum simulations.}
\label{tab:optimizer_comparison}
\end{table}

Our findings on optimizer performance in VQE using the \ac{tvha} framework demonstrate several alignments with recent literature, while also introducing new insights. Lavrijsen et al.~\cite{lavrijsen2020classopt} reported that gradient-based methods (particularly \ac{bfgs} and L-BFGS-B) outperform other approaches in noiseless settings, consistent with our observation that \ac{bfgs} achieves chemical accuracy in just 28 function evaluations with Hartree-Fock initialization for \ce{H2} under statevector simulations. Regarding noise sensitivity, we can observe a hierarchy, where the gradient-based methods are dominating in the noiseless settings, but are overtaken by population-based methods, most often by \ac{cmaes}, in the sampling noise cases. These results match those of Nannicini~\cite{nannicini2019performance}, who also described that gradient-free approaches are more resilient to sampling noise. The identification of a practical ``noise floor'' limiting accuracy regardless of optimizer selection corresponds with Kandala et al.~\cite{kandala2017hardware}, and we further quantify diminishing returns beyond $\sim$1000 shots, offering more concrete guidance on resource allocation.

Our work extends prior results by quantitatively examining optimizer behavior and noise resilience in the \ac{tvha} across increasing molecular complexity. In contrast to Tang~et~al.~\cite{tang2021qubit}, who reported minimal advantage from chemically informed initialization, our results show that Hartree–Fock initialization reduces total evaluations by 10–27\% and lowers final energy errors by up to 75\% for small systems, owing to \ac{tvha}’s adiabatic parameterization that better preserves molecular symmetries. Gradient-based optimizers perform efficiently in noiseless \ce{H2} but progressively fail as system size and sampling noise increase: \ac{slsqp} diverges entirely, while \ac{bfgs} becomes marginally stable as curvature information is overwhelmed by stochastic variance. This transition marks the onset of the noise-dominated regime, where curvature-driven methods lose reliability. When we consider the stochastic and the population-based optimization methods, in our study represented by \ac{cmaes}, \ac{pso} and \ac{spsa}, we observe a stable convergence across all setting, namely \ce{H2}, \ce{H4}, and \ce{LiH}, as the averaging over the candidate population work well, and effectively filters out fluctuations caused by sampling noise, thus sustaining stable optimization progression. In the active-space simulations, \ac{spsa} achieves an error of $6.5\times10^{-4}$~{$\mathrm{Ha}$ }for \ce{LiH} with 486 function evaluations under \ac{HF} initialization, outperforming the poor \ac{spsa} performance reported by Arrasmith~et~al.~\cite{arrasmith2020operator}. This improvement arises from the reduced parameter space and smoother energy landscape characteristic of \ac{tvha}. Our findings are consistent with recent benchmarks of variational quantum eigensolvers for the Kitaev model~\cite{BenchmarkKitaev}, which identified optimizer choice and ansatz structure as key determinants of convergence under noise, and complement the newly introduced HOPSO optimizer~\cite{ahamed2025hopso}, whose population-based robustness aligns with the noise-resilient behavior observed in \ac{tvha}.

Synthesizing our findings with existing literature refines practical guidelines. While Zhu et al.~\cite{zhu2020training} recommended gradient-based methods, our results suggest they are optimal only for small, low-noise problems; population-based optimizers (\ac{cmaes}, \ac{pso}, \ac{spsa}) become essential in noisy or higher-dimensional regimes. Active space reduction not only improves efficiency but also enhances noise robustness, confirming Smart and Mazziotti~\cite{smart2021efficient} while extending the discussion to optimization stability. For instance, we find a 95.2\% error reduction in \ce{LiH} \ac{cobyla} sampling through active-space truncation. Finally, our results align with Yuan et al.~\cite{yuan2024quantifying}, who also observed that optimizer choice and noise critically shape convergence and that initial state preparation influences performance.

\section{Conclusion}
\label{sec:conclusion}

This work benchmarks eight optimizers for the \ac{tvha} ansatz on \ce{H2}, \ce{H4}, and \ce{LiH} under both ideal statevector and finite-shot sampling. The study combines structured ansatz design, explicit visualization of noisy energy landscapes, and extensive simulation to expose the interplay among noise, initialization, and dimensionality.

Under Gaussian sampling noise, methods based on a quasi-Newton principle, namely \ac{slsqp} and \ac{bfgs}, fail to converge reliably. The divergence of \ac{slsqp} while \ac{bfgs} remains marginally stable indicates that the true Hessian signal is already drowned in sampling noise, and that additional constraint enforcement transforms noisy curvature estimates into instability. This marks a practical threshold beyond which second-order and line-search–based methods lose physical meaning, while stochastic or population-based strategies continue to function through implicit noise averaging.

The optimizer ranking is noise dependent. In statevector (exact) simulations on small systems, gradient-based methods converge the quickest and reach the lowest energies. Under finite-shot sampling, population-based and stochastic methods maintain accuracy with increasing problem size, whereas deterministic gradient methods decline. High-shot reevaluation distinguishes genuine convergence from low-shot-spurious improvement and shows that some seeming improvement is an artifact of estimator noise instead of further optimization advance.

The initialization and model size are inherently coupled. \ac{HF} initializations reduce function evaluations and stabilize trajectories in small systems, but this benefit diminishes with dimensionality. For certain population methods, random starts equal or exceed \ac{HF} as the search enjoys wider exploration. Active-space reductions narrow the search region and reduce estimator's variance downward, regaining sub-millihartree accuracy for several optimizers.

A notable artifact appears in the biggest system studied: \ce{LiH} under sampling noise, where \ac{cmaes} was the only optimizer that reached energies that appear to be lower than ground state energy. High-shot confirmation removes this advantage, identifying it as a sampling effect rather than a genuine noise-assisted convergence. The result cautions against interpreting low-shot improvements as evidence that noise is beneficial; without correction, the combination of estimator variance and broad exploration can transiently bias best-observed values. However, after high-shot reevaluation \ac{cmaes} still achieved the lowest error.

Finally, our work recontextualizes the violation of the variational lower bound in finite shot regimes. Leveraging the theoretical guarantee that \ac{cmaes} converges to a steady state distribution determined by the sampling noise floor, we recognize this limit not as a barrier, but as a transition point where the algorithmic challenge shifts from optimization to high precision estimation. While frequent bound violations in this regime are often dismissed as artifacts, we demonstrate that they signal a symmetric noise environment which can be exploited. By utilizing the population mean, we effectively perform an implicit resampling that smooths out these fluctuations. This strategy allows us to penetrate the noise floor and secure precise energy estimates without the prohibitive cost of shot escalation or the statistical risks of selective reevaluation, effectively converting the violation of the variational principle from a liability into a distinct statistical advantage.

Also, this study isolates finite shot sampling noise without including coherent or stochastic gate errors or readout bias. Although hardware noise typically imposes errors of higher magnitude, we demonstrate that sampling noise alone renders deterministic methods (such as \ac{gd}, \ac{bfgs}, and \ac{slsqp}) unusable and significantly degrades the performance of \ac{cobyla} and \ac{nelder}. Consequently, while optimizer rankings may shift under full device noise, the fundamental instability caused by sampling uncertainty remains a critical bottleneck.

Practical guidance follows from these findings. On noisy simulators or hardware, favor \ac{cmaes}, \ac{pso}, or \ac{spsa} for robustness. Use \ac{HF} initialization when chemically available but allow random starts when exploration is advantageous. Monitor population means rather than best value traces. Reserve high-shot budgets for late-stage confirmation to establish statistically reliable ordering. Future work should pair fast gradients with population search using noise-adaptive switching, and develop shot-allocation and stopping rules that track the measured noise floor to minimize total evaluation cost while preserving correctness.
Incorporating realistic channels, such as thermal relaxation, amplitude damping, readout assignment, and hardware backed calibration will serve as tests for the persistence of these trends found in the regime of sampling noise. We emphasize that sampling noise remains a first order concern even as quantum error correction reduces physical error rates, particularly because many error mitigation methods increase estimator variance, making robustness to sampling noise central to practical quantum computing.

\section*{Data Availability}
\label{sec:availability}
To support our findings and ensure reproducibility, we have published the dataset containing both the computed results and the scripts used to generate them.
The dataset serves as a companion resource to this work and facilitates further research in hybrid quantum-classical architectures.
It is publicly available on Zenodo \cite{beseda2025results}.

\section*{Conflict of Interest}
The authors declare that they have no known competing financial interests or personal relationships that could have appeared to influence the work reported in this paper.

\section*{Ethical Statement}
Ethical approval was not required for this study as the research did not involve human participants, human data, or animal experimentation.

\section*{Acknowledgements}

Clemens Possel thanks the Ministerium für Wirtschaft, Arbeit und Tourismus Baden-Württemberg (Ministry of Economic Affairs, Labour and Tourism of Baden-Württemberg) for their support through the projects QC-4-BW, QC-4-BW II, and KQCBW24. Tomáš Bezděk is supported by Grant of SGS No. SP2025/049, VŠB - Technical University of Ostrava, Czech Republic. Vojtěch Novák is supported by Grant of SGS No. SP2025/072, VSB-Technical University of Ostrava, Czech Republic. This work was supported by the Ministry of Education, Youth and Sports of the Czech Republic through the e-INFRA CZ (ID:90254 ).

\newpage
\appendix
\section{Details on optimization algorithms}

\label{sec:optimizers}
In this section, we provide an overview of all optimization methods that were compared in this paper.

In the realm of optimization algorithms, \ac{bfgs} is a quasi-Newton optimization method used to solve unconstrained nonlinear optimization problems \cite{liu1989limited}. It approximates the inverse Hessian matrix to improve search efficiency without requiring second derivatives. The algorithm iteratively updates the Hessian estimate using gradient information, ensuring rapid convergence for smooth, well-behaved objective functions. \ac{bfgs} belongs to the family of variable metric methods and is known for its robustness in solving medium-scale problems \cite{dai2002convergence, morales2002numerical}.

Another powerful method is \ac{cmaes}, a derivative-free, evolutionary optimization algorithm that belongs to the family of evolution strategies \cite{hansen2003reducing}. It maintains a population of candidate solutions and adapts a multivariate normal distribution's covariance matrix to guide the search process. The method is particularly effective for high-dimensional, non-convex, and multimodal optimization problems. By learning correlations between variables and adjusting step sizes dynamically, \ac{cmaes} improves convergence speed and robustness compared to simpler evolutionary approaches \cite{varelas2018comparative}.

For constrained nonlinear problems, \ac{cobyla} is a derivative-free optimization algorithm designed for constrained nonlinear optimization problems \cite{powell1994direct, powell1998direct, powell2007view}. It employs a simplex-based approach to construct linear approximations of the objective function and constraints. The algorithm iteratively refines the solution by adjusting a trust region radius that controls step sizes. Unlike gradient-based methods, \ac{cobyla} can handle noisy, discontinuous, or expensive-to-evaluate objective functions, making it suitable for applications where derivatives are unavailable. However, it may struggle with large-scale problems due to its reliance on local linear models.

Among the simplest and most widely applied techniques, \ac{gd} is a first-order optimization algorithm that minimizes differentiable functions by iteratively updating parameters in the direction of the negative gradient \cite{ruder2016overview, amari1993backpropagation}. Using a fixed step size, it continues until convergence criteria are met. Proper step size tuning is crucial, as overly large steps can cause divergence, while small steps slow progress.

For derivative-free optimization of non-smooth functions, the \ac{nelder} method is a direct search optimization technique that does not require derivatives \cite{nelder1965simplex}. It operates on a simplex of points, updating vertices through reflection, expansion, contraction, and shrinkage operations. The algorithm is well suited for optimizing discontinuous, noisy, or non-smooth objective functions, where gradient-based methods fail. However, it does not guarantee convergence to a global minimum and can be inefficient in high-dimensional spaces due to its reliance on heuristics rather than gradient information \cite{lagarias1998convergence}.

Inspired by natural phenomena, \ac{pso} is a population-based optimization algorithm inspired by swarm intelligence and collective behavior in nature \cite{eberhart1995particle, shi2001particle}. Each candidate solution, or particle, moves through the search space by updating its velocity based on its own best-known position and the best-known position of the entire swarm. The method balances exploration and exploitation through inertia weight and acceleration coefficients, allowing it to efficiently search complex, multimodal landscapes. \ac{pso} has been widely applied to global optimization problems, particularly in scenarios where gradients are unavailable or unreliable \cite{jain2022overview}.

For constrained problems where gradient-based methods are feasible, \ac{slsqp} is a constrained optimization algorithm that combines sequential quadratic programming with least-squares minimization techniques \cite{kraft1988software}. It constructs a quadratic approximation of the objective function and iteratively solves subproblems to update the solution. The method effectively handles both equality and inequality constraints and is well-suited for smooth, differentiable problems where gradient and Hessian information can be efficiently computed \cite{boggs1995sequential}.

Finally, for high-dimensional optimization without gradients, \ac{spsa} is a gradient-free optimization method that estimates gradients using only two function evaluations per iteration, making it particularly useful for high-dimensional problems \cite{spall1992multivariate, spall2002implementation}. Unlike traditional finite-difference approximations, which require a separate evaluation for each parameter, \ac{spsa} perturbs all parameters simultaneously, leading to significant efficiency gains. The method is widely used in noisy or stochastic environments, where exact gradients are either expensive or impossible to compute. Its convergence properties are well studied, and it performs well for large-scale optimization problems where function evaluations are costly \cite{maryak1999efficient}.

\begin{table}[htpb]
\centering
\begin{tabular}{@{}ll@{}}
\toprule
\textbf{Optimizer} & \textbf{Category}                        \\ \midrule
\ac{bfgs}               & Gradient-Based Method                    \\
\ac{cmaes}             & Metaheuristic Method (Evolutionary based)                    \\
\ac{cobyla}             & Gradient-Free Method                     \\
\ac{gd}   & Gradient-Based Method                    \\
\ac{nelder}        & Gradient-Free Method                     \\
\ac{pso}                & Metaheuristic Method (Swarm based)                     \\
\ac{slsqp}              & Gradient-Based Method                    \\
\ac{spsa}               & Gradient-Free Method                     \\ \bottomrule
\end{tabular}
\caption{Classification of Optimizers by Principle}
\label{tab:optimizer_classification}
\end{table}

\section{Optimization Convergence Traces of Individual Runs}
\label{sec:convergence_runs}
This appendix presents detailed convergence trajectories for each of the eight optimization algorithms across the four molecular systems studied. For each optimizer-molecule combination, we include plots visualizing the energy error evolution over function evaluations for all 10 independent runs per configuration (Hartree-Fock and random initialization, under both statevector and sampling noise conditions). These plots provide a granular view of optimization behavior, highlighting the variability and consistency of convergence paths. As discussed in \cref{sec:results}, the trajectories reveal critical insights into optimizer-specific responses to noise-induced landscape distortions (\cref{sec:noise_landscape}) and the benefits of Hartree-Fock initialization (\cref{sec:vha}). For instance, the plots illustrate how population-based methods like \ac{cmaes} exhibit robust convergence under sampling noise, while gradient-based methods like \ac{bfgs} show rapid convergence in noiseless settings but greater variability with noise. These detailed traces complement the mean convergence plots in \cref{fig:h2_mean_conv}--\cref{fig:lih_as_mean_conv}, offering a deeper understanding of optimizer stability and performance across diverse scenarios.

\begin{figure}[htbp]
\centering
\vspace*{-9.5mm}
\mbox{%
\begin{subfigure}[b]{0.49\linewidth}
\includegraphics[width=\linewidth]{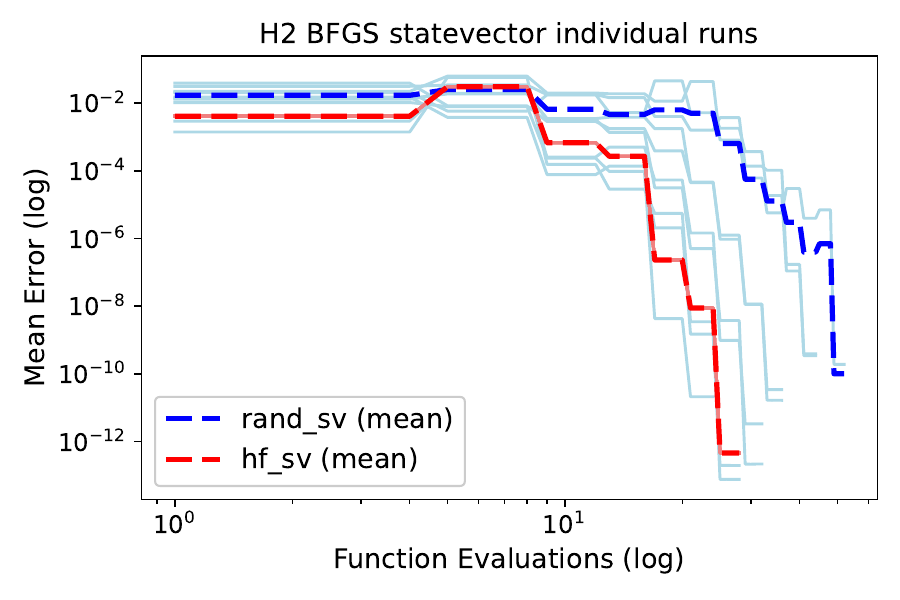}
\label{fig:bfgs_sv}
\end{subfigure}%
\begin{subfigure}[b]{0.49\linewidth}
\includegraphics[width=\linewidth]{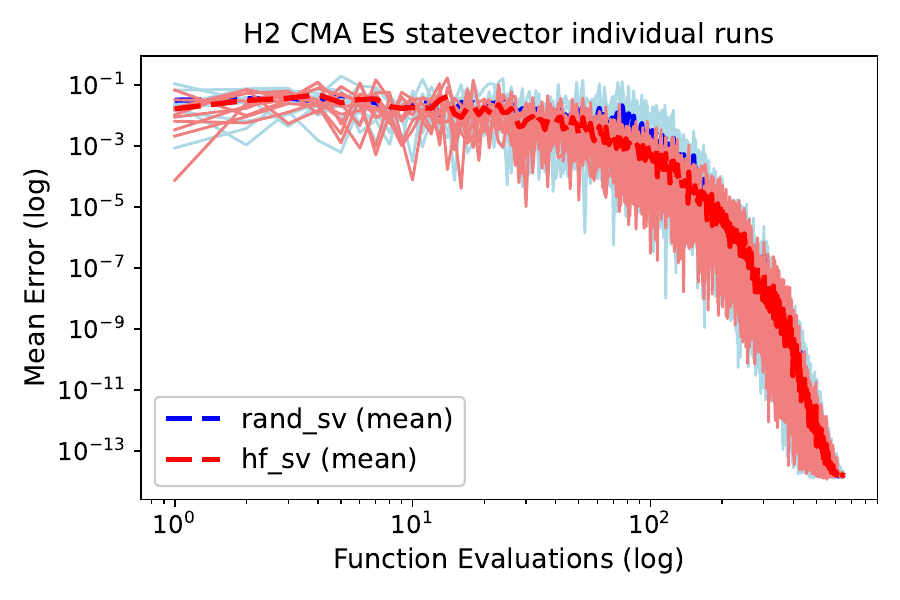}
\label{fig:cma_es_sv}
\end{subfigure}%
}%
\\[-9.5mm] 

\mbox{%
\begin{subfigure}[b]{0.49\linewidth}
\includegraphics[width=\linewidth]{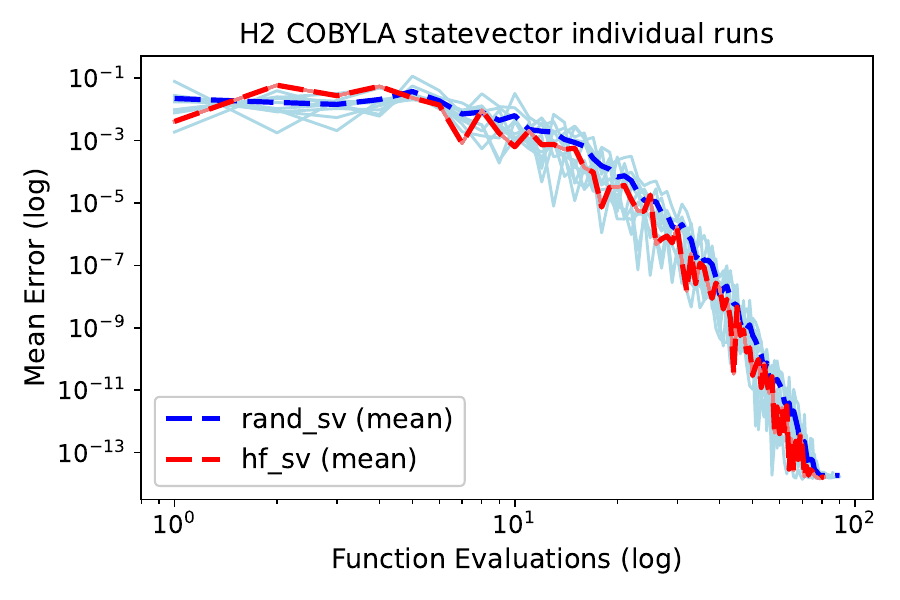}
\label{fig:cobyla_sv}
\end{subfigure}%
\begin{subfigure}[b]{0.49\linewidth}
\includegraphics[width=\linewidth]{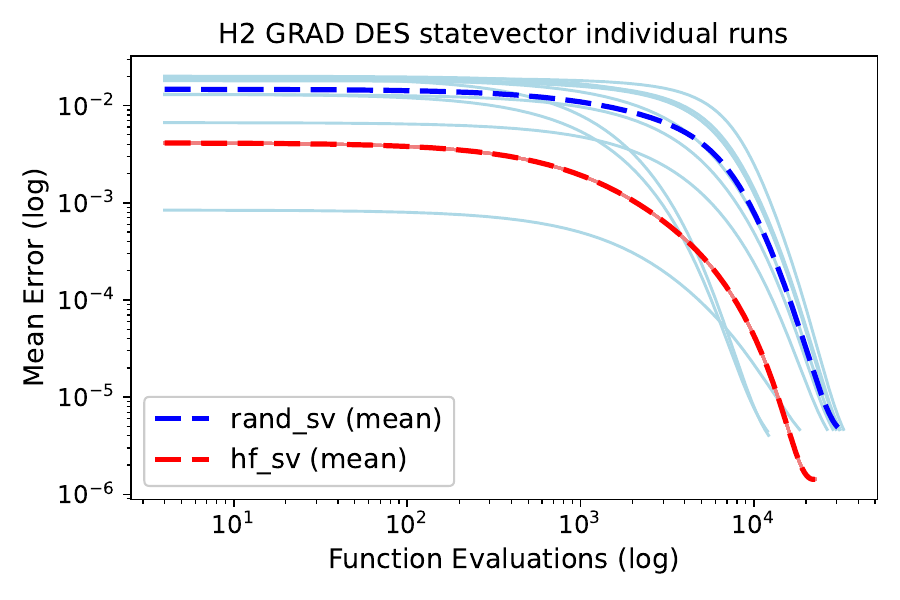}
\label{fig:grad_des_sv}
\end{subfigure}%
}%
\\[-9.5mm] 

\mbox{%
\begin{subfigure}[b]{0.49\linewidth}
\includegraphics[width=\linewidth]{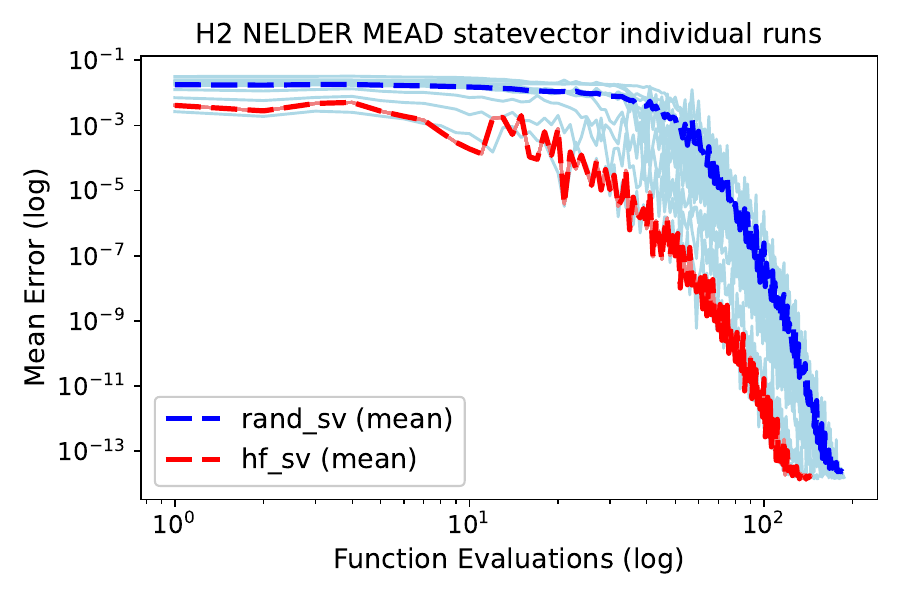}
\label{fig:nelder_mead_sv}
\end{subfigure}%
\begin{subfigure}[b]{0.49\linewidth}
\includegraphics[width=\linewidth]{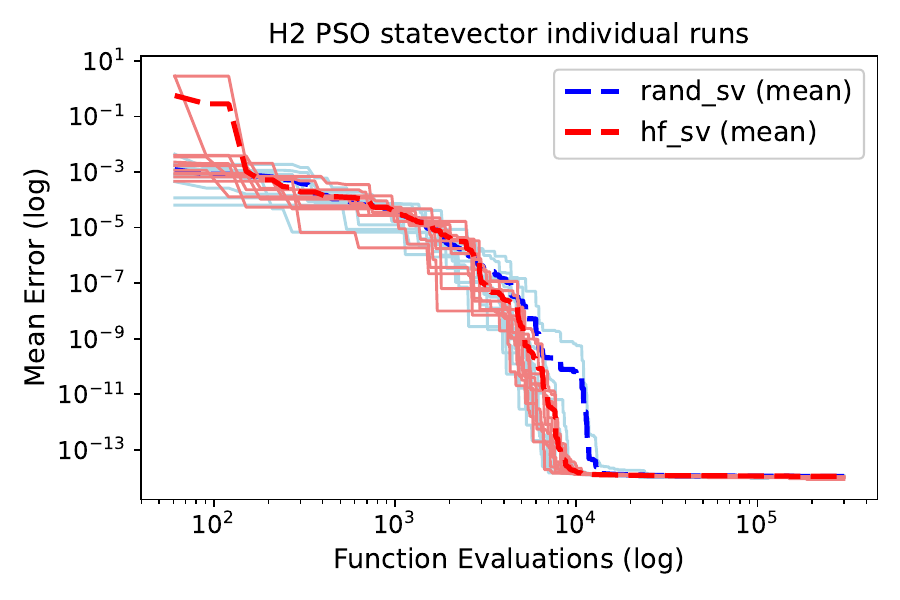}
\label{fig:pso_sv}
\end{subfigure}%
}%
\\[-9.5mm]

\mbox{%
\begin{subfigure}[b]{0.49\linewidth}
\includegraphics[width=\linewidth]{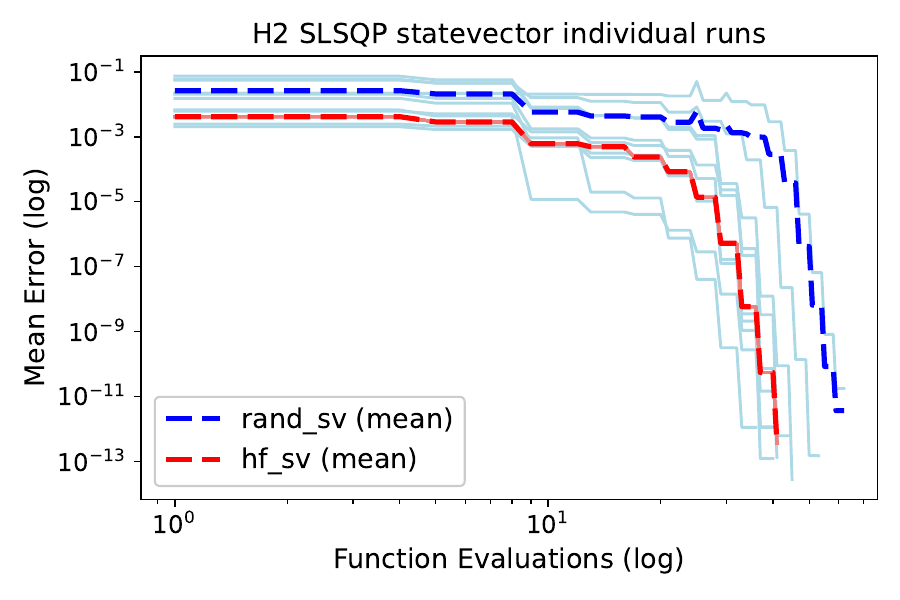}
\label{fig:slsqp_sv}
\end{subfigure}%
\begin{subfigure}[b]{0.49\linewidth}
\includegraphics[width=\linewidth]{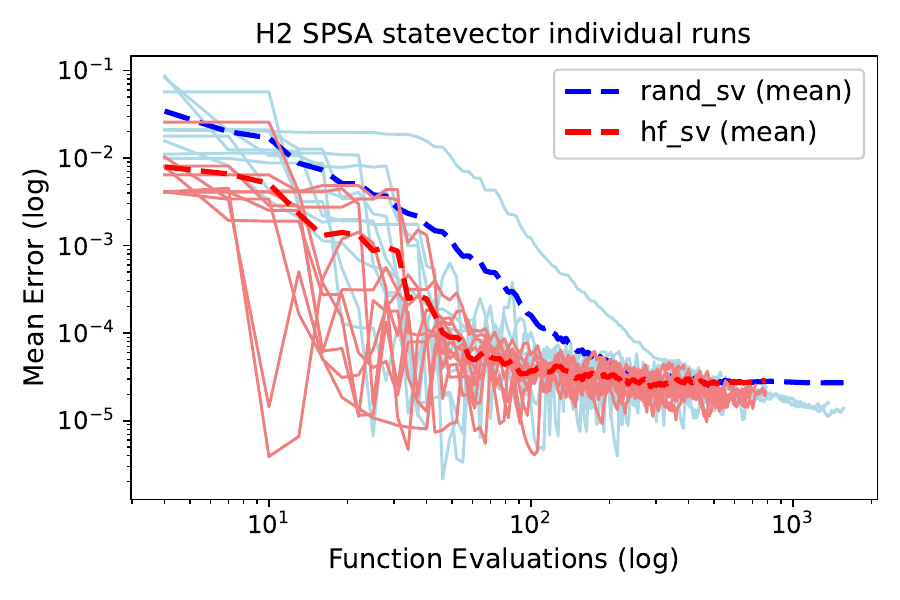}
\label{fig:spsa_sv}
\end{subfigure}%
}%
\\[-10.5mm]
\caption{H2 individual run convergence plots for statevector.}
\label{fig:h2_statevector_convergence}
\end{figure}

\begin{figure}[htbp]
\centering
\vspace*{-9.5mm} 
\mbox{%
\begin{subfigure}[b]{0.49\linewidth}
\includegraphics[width=\linewidth]{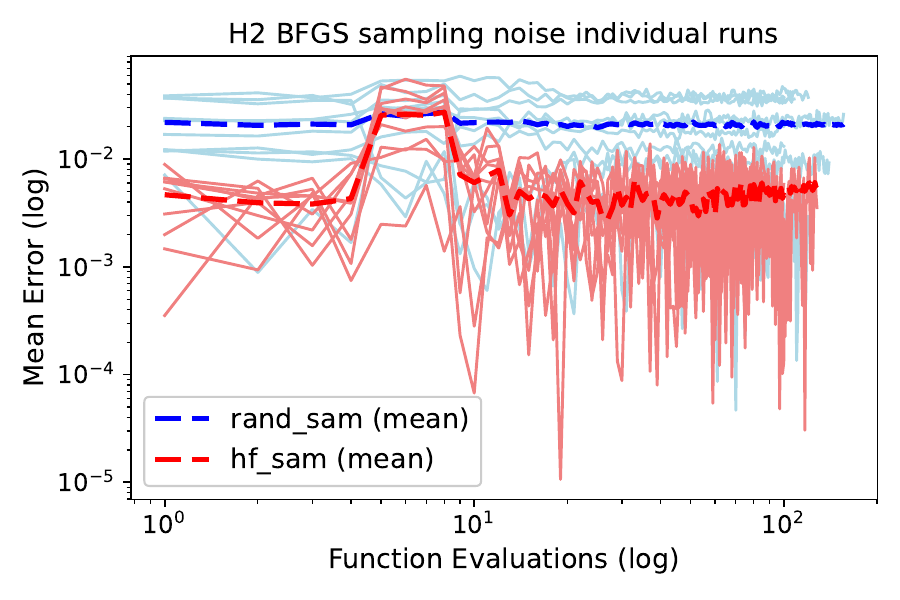}
\label{fig:bfgs_sam}
\end{subfigure}%
\begin{subfigure}[b]{0.49\linewidth}
\includegraphics[width=\linewidth]{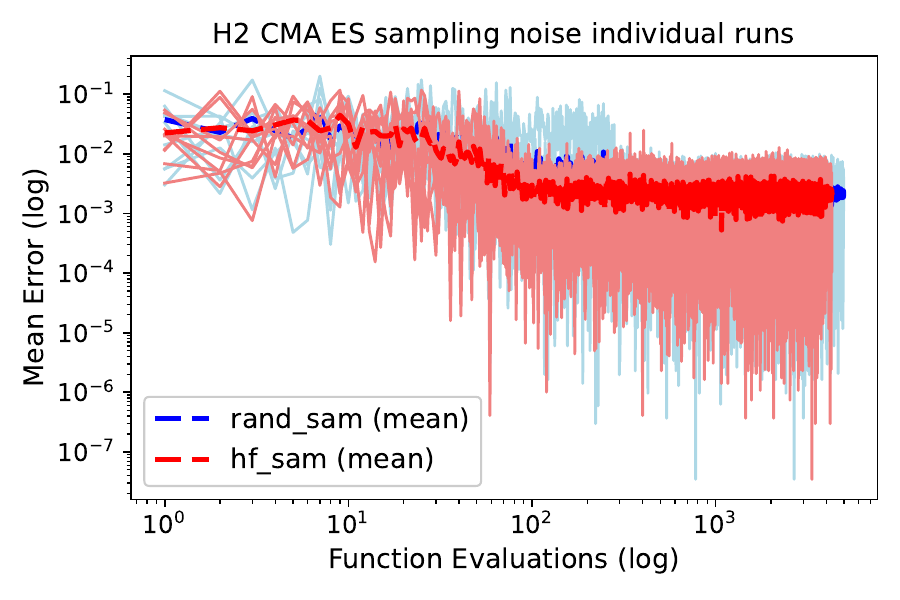}
\label{fig:cma_es_sam}
\end{subfigure}%
}%
\\[-9.5mm] 
\mbox{%
\begin{subfigure}[b]{0.49\linewidth}
\includegraphics[width=\linewidth]{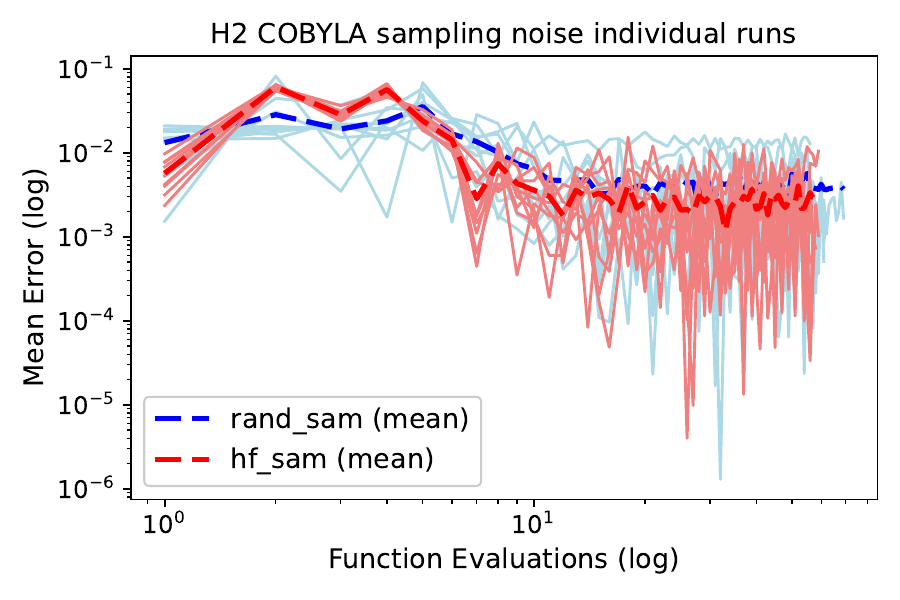}
\label{fig:cobyla_sam}
\end{subfigure}%
\begin{subfigure}[b]{0.49\linewidth}
\includegraphics[width=\linewidth]{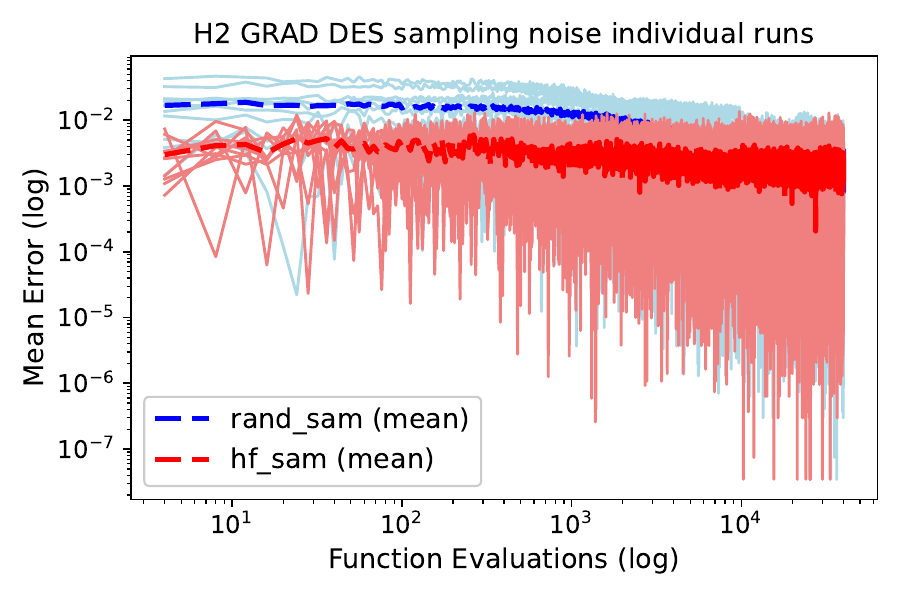}
\label{fig:grad_des_sam}
\end{subfigure}%
}%
\\[-9.5mm] 
\mbox{%
\begin{subfigure}[b]{0.49\linewidth}
\includegraphics[width=\linewidth]{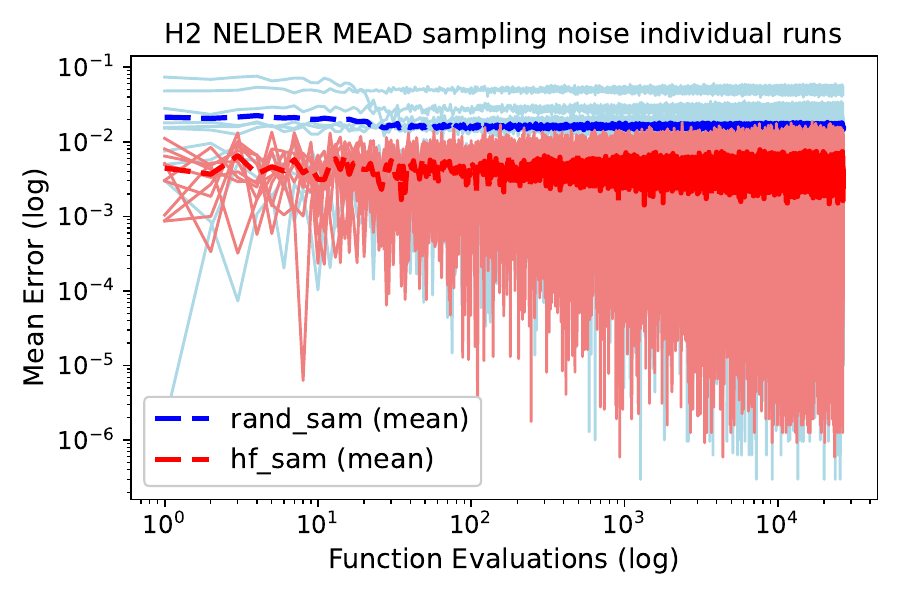}
\label{fig:nelder_mead_sam}
\end{subfigure}%
\begin{subfigure}[b]{0.49\linewidth}
\includegraphics[width=\linewidth]{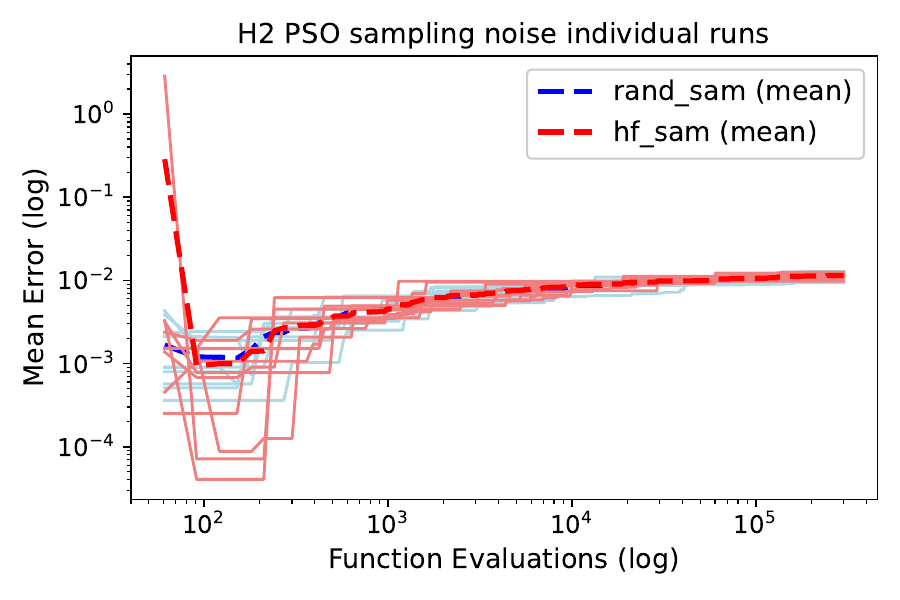}
\label{fig:pso_sam}
\end{subfigure}%
}%
\\[-9.5mm] 
\mbox{%
\begin{subfigure}[b]{0.49\linewidth}
\includegraphics[width=\linewidth]{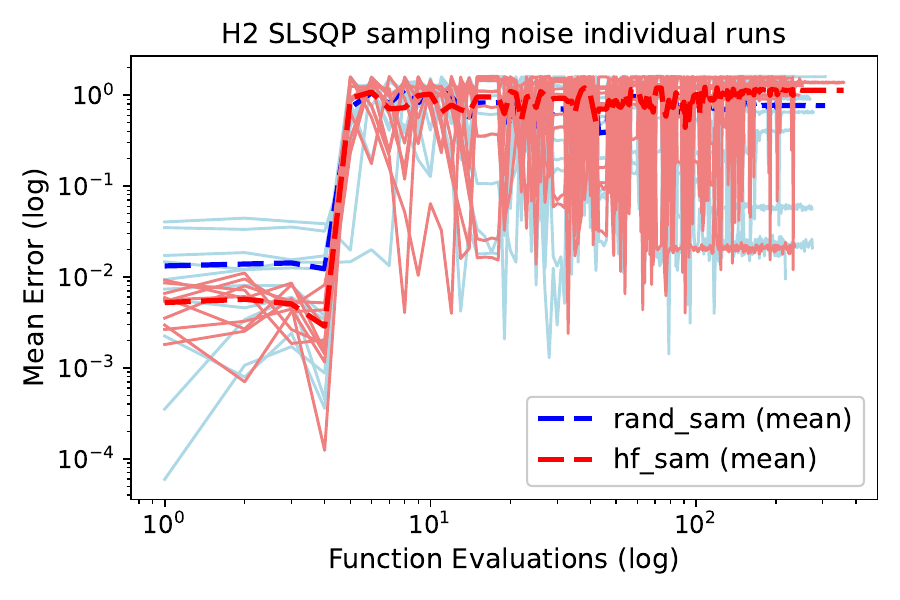}
\label{fig:slsqp_sam}
\end{subfigure}%
\begin{subfigure}[b]{0.49\linewidth}
\includegraphics[width=\linewidth]{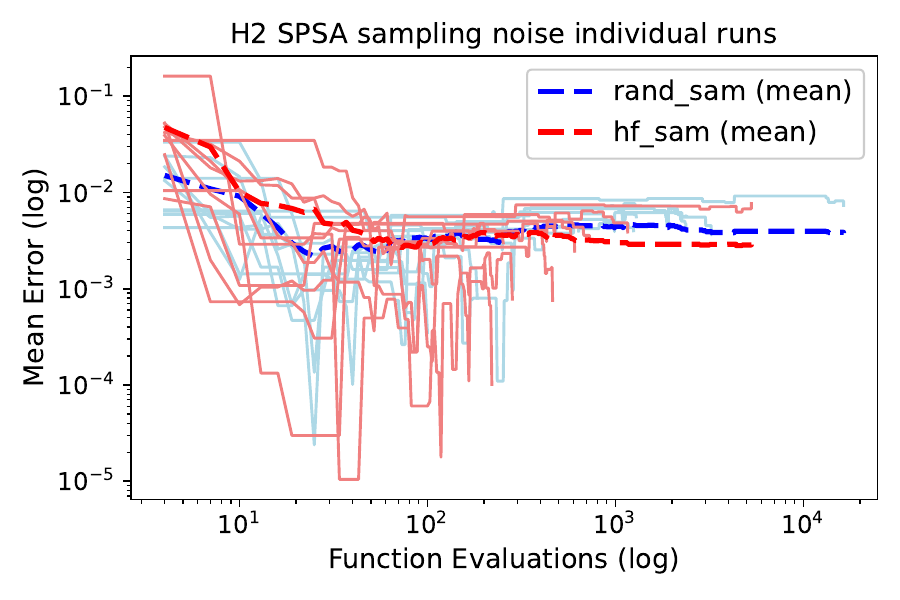}
\label{fig:spsa_sam}
\end{subfigure}%
}%
\\[-10.5mm]
\caption{H2 individual run convergence plots for sampling noise.}
\label{fig:sampling_noise_convergence}
\end{figure}

\newpage

\begin{figure}[htbp]
\centering
\vspace*{-3mm} 
\mbox{%
\begin{subfigure}[b]{0.49\linewidth}
\includegraphics[width=\linewidth]{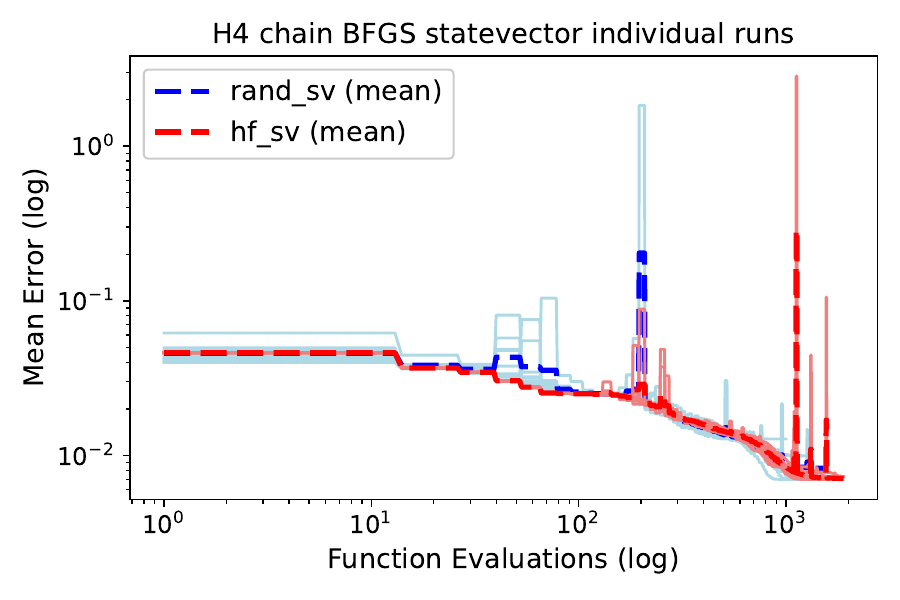}
\end{subfigure}%
\begin{subfigure}[b]{0.49\linewidth}
\includegraphics[width=\linewidth]{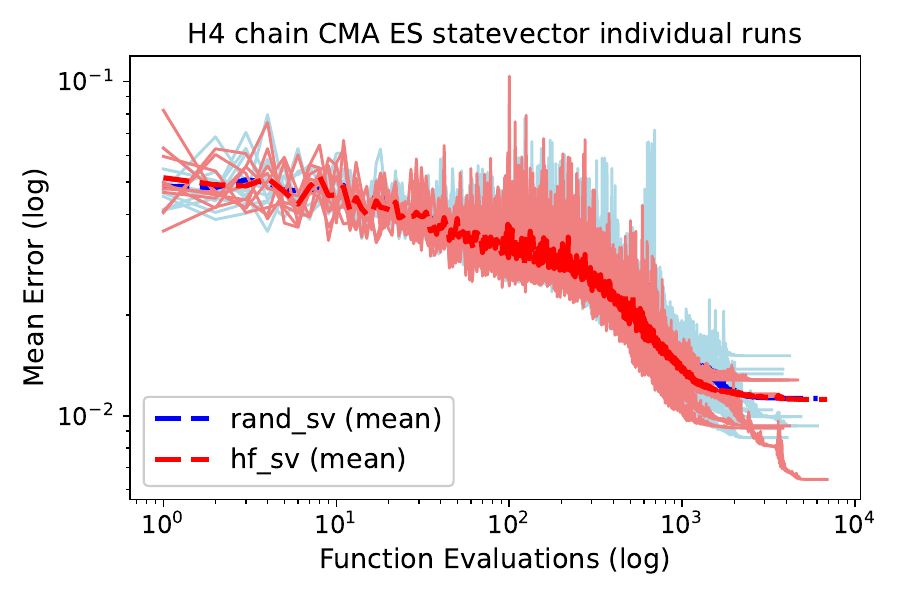}
\end{subfigure}%
}%
\\[-2mm]
\mbox{%
\begin{subfigure}[b]{0.49\linewidth}
\includegraphics[width=\linewidth]{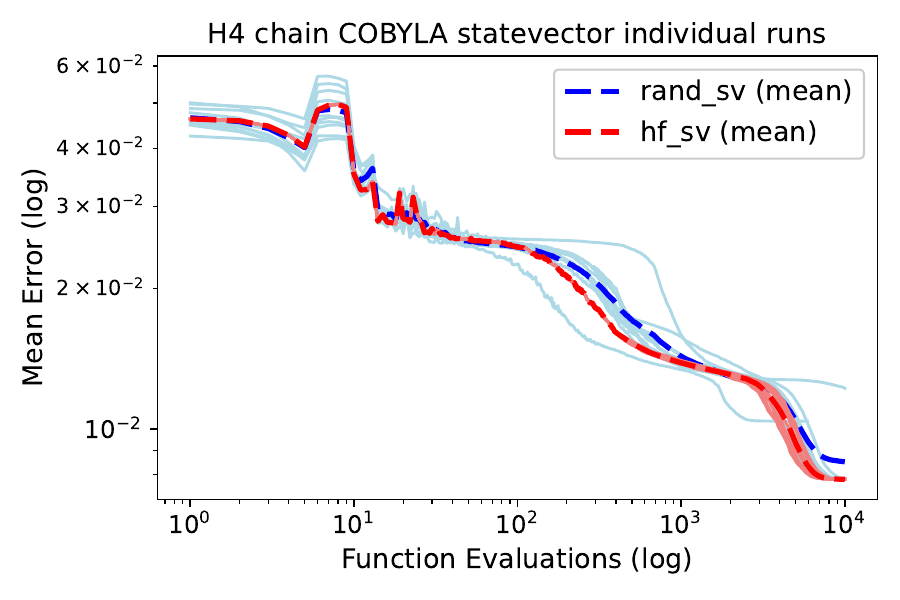}
\end{subfigure}%
\begin{subfigure}[b]{0.49\linewidth}
\includegraphics[width=\linewidth]{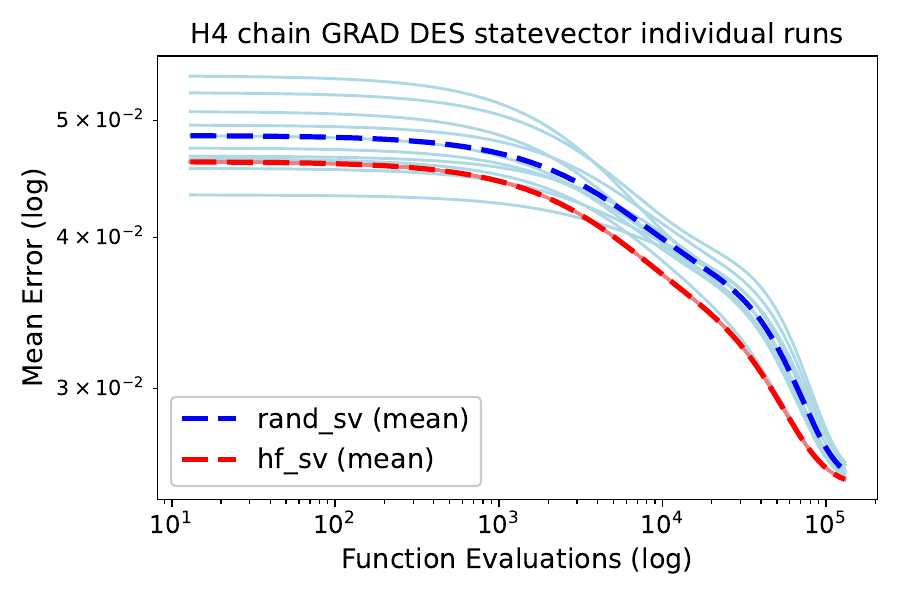}
\end{subfigure}%
}%
\\[-2mm]
\mbox{%
\begin{subfigure}[b]{0.49\linewidth}
\includegraphics[width=\linewidth]{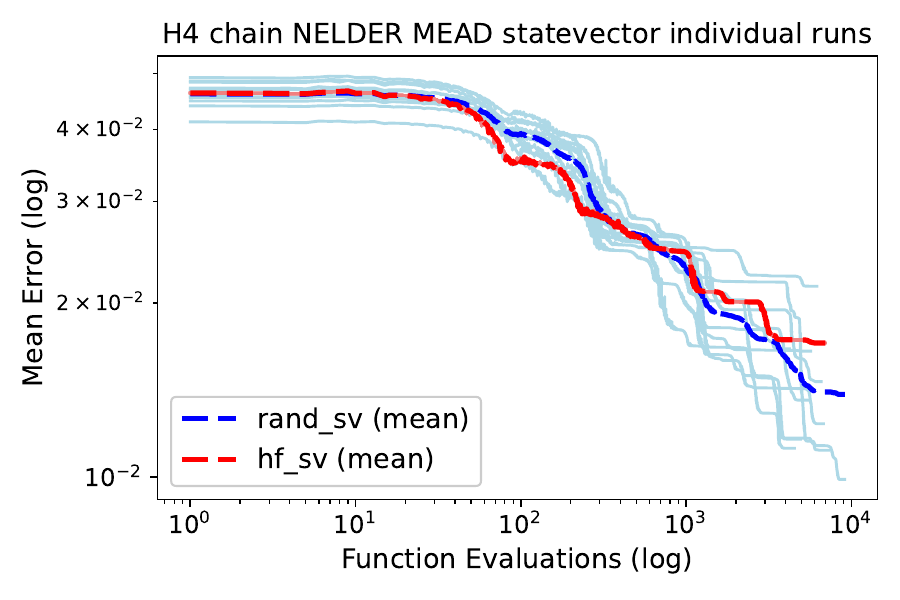}
\end{subfigure}%
\begin{subfigure}[b]{0.49\linewidth}
\includegraphics[width=\linewidth]{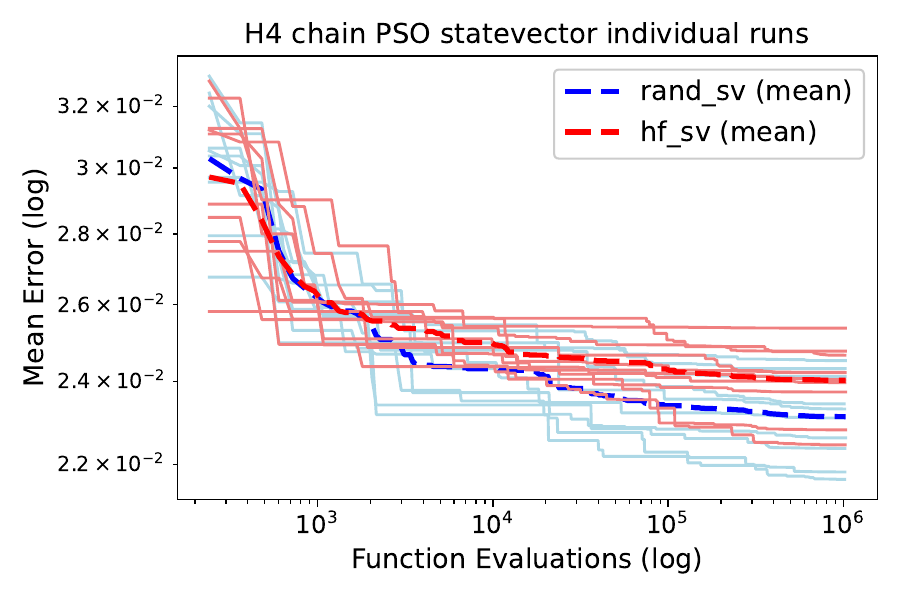}
\end{subfigure}%
}%
\\[-2mm]
\mbox{%
\begin{subfigure}[b]{0.49\linewidth}
\includegraphics[width=\linewidth]{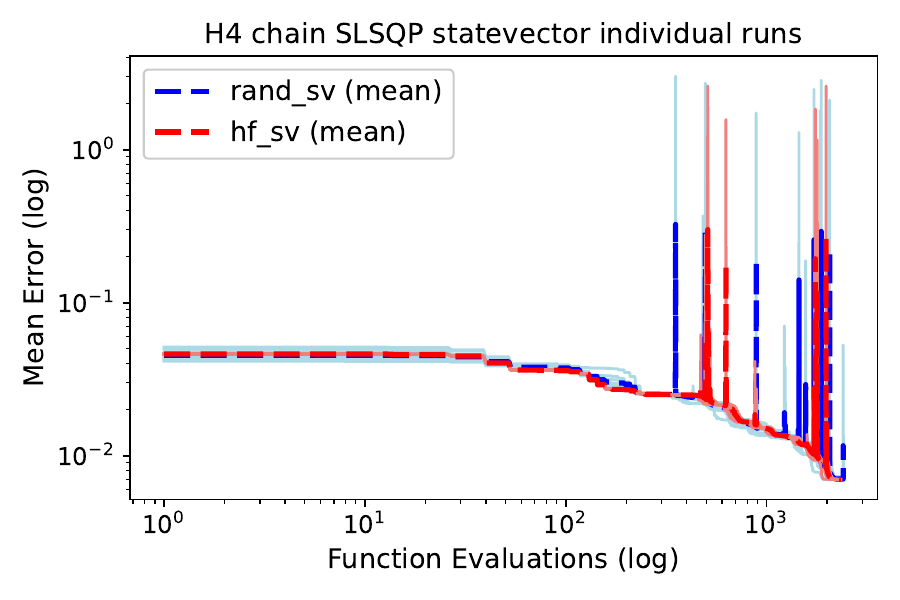}
\end{subfigure}%
\begin{subfigure}[b]{0.49\linewidth}
\includegraphics[width=\linewidth]{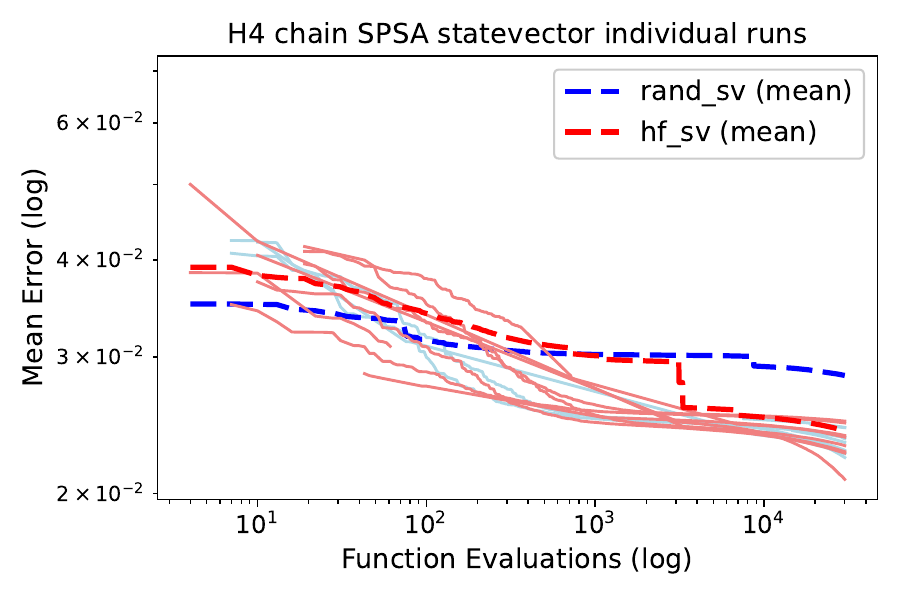}
\end{subfigure}%
}%
\\[-3.5mm]
\caption{H4 chain individual run convergence plots for statevector.}
\label{fig:h4_statevector_convergence}
\end{figure}

\begin{figure}[htbp]
\centering
\vspace*{-3mm} 
\mbox{%
\begin{subfigure}[b]{0.49\linewidth}
\includegraphics[width=\linewidth]{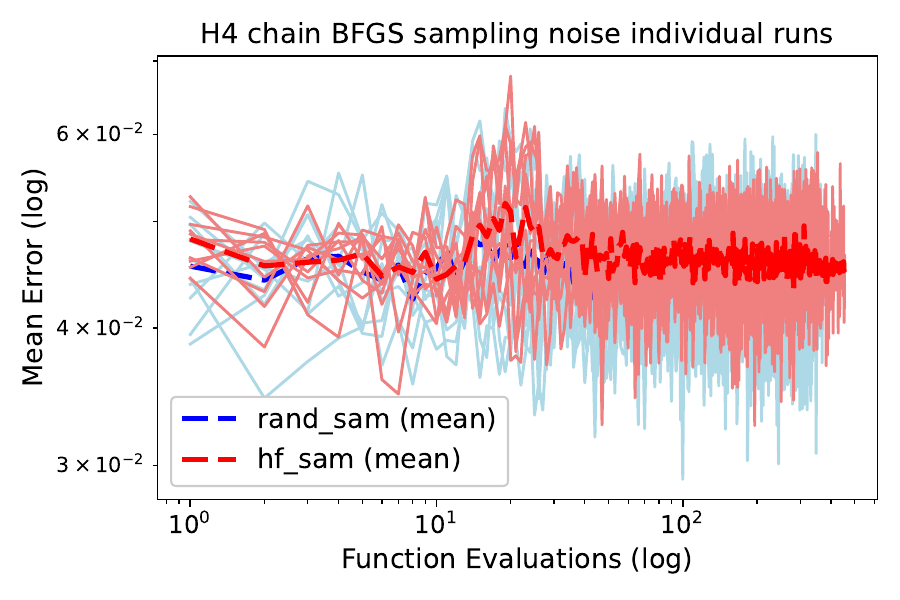}
\end{subfigure}%
\begin{subfigure}[b]{0.49\linewidth}
\includegraphics[width=\linewidth]{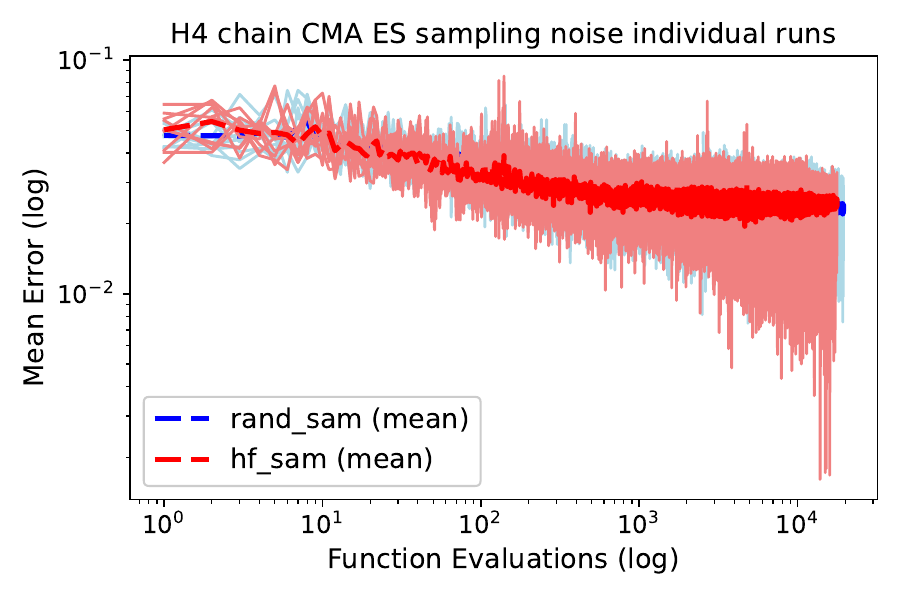}
\end{subfigure}%
}%
\\[-2mm]
\mbox{%
\begin{subfigure}[b]{0.49\linewidth}
\includegraphics[width=\linewidth]{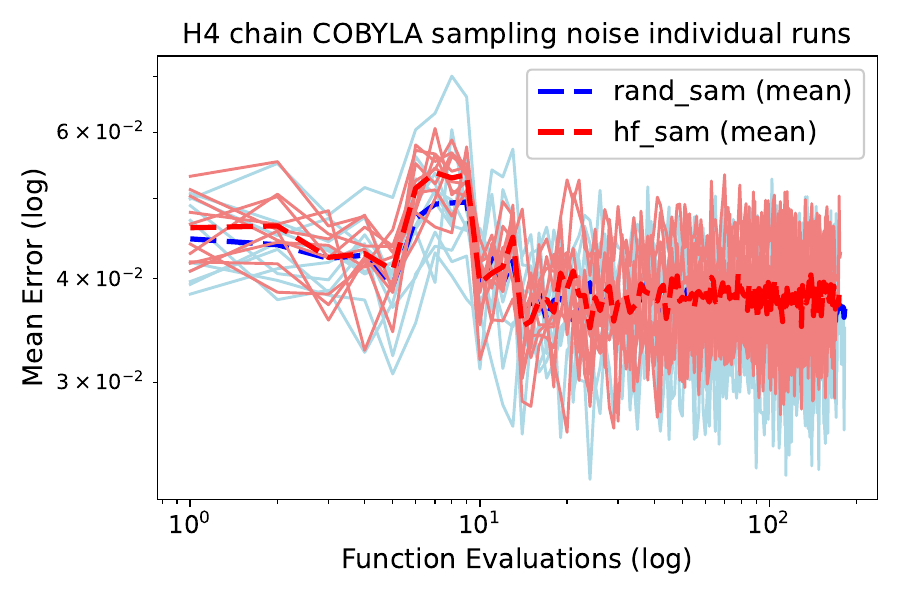}
\end{subfigure}%
\begin{subfigure}[b]{0.49\linewidth}
\includegraphics[width=\linewidth]{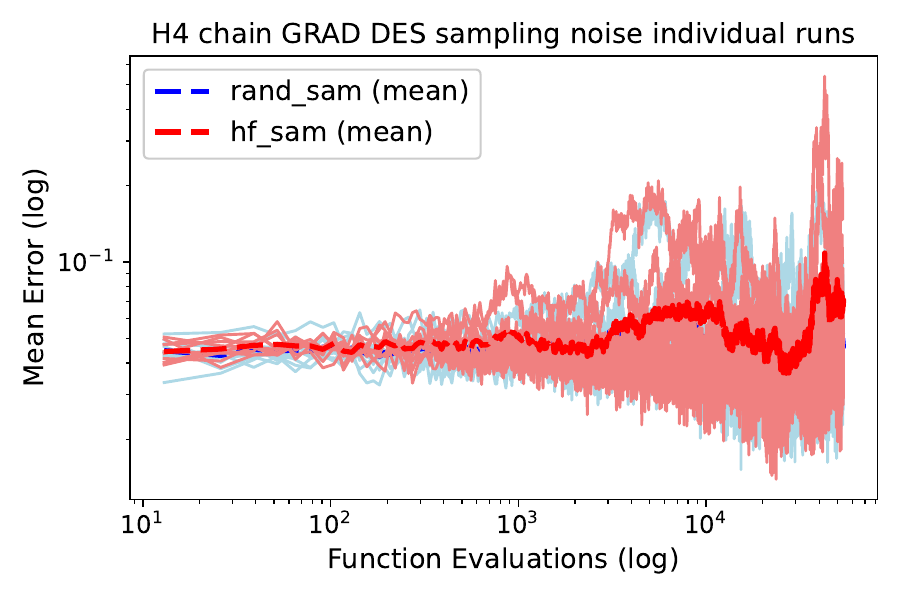}
\end{subfigure}%
}%
\\[-2mm]
\mbox{%
\begin{subfigure}[b]{0.49\linewidth}
\includegraphics[width=\linewidth]{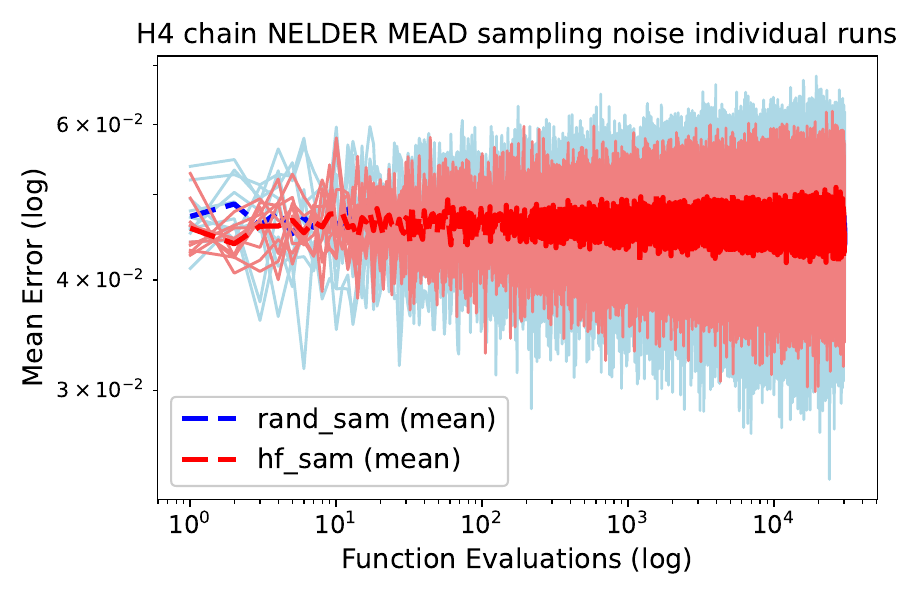}
\end{subfigure}%
\begin{subfigure}[b]{0.49\linewidth}
\includegraphics[width=\linewidth]{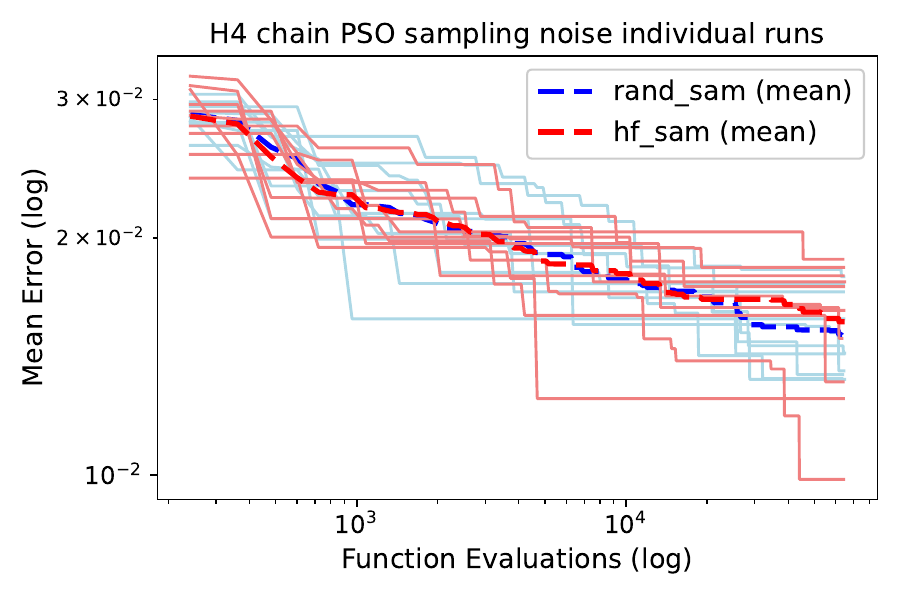}
\end{subfigure}%
}%
\\[-2mm]
\mbox{%
\begin{subfigure}[b]{0.49\linewidth}
\includegraphics[width=\linewidth]{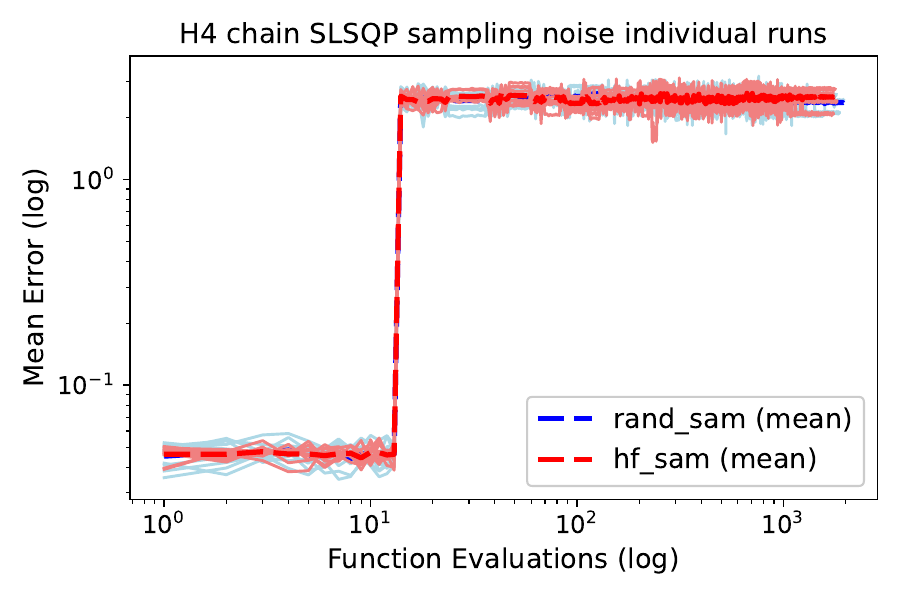}
\end{subfigure}%
\begin{subfigure}[b]{0.49\linewidth}
\includegraphics[width=\linewidth]{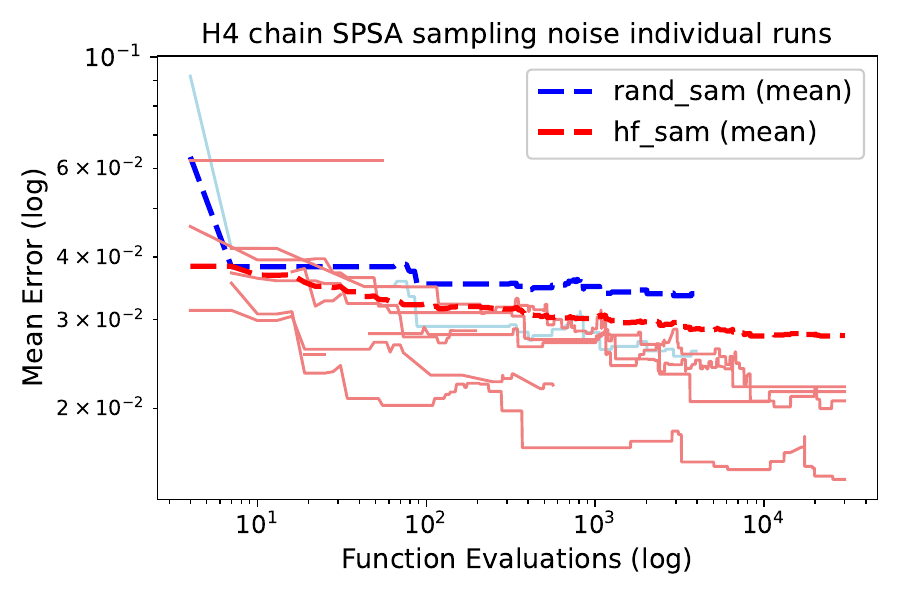}
\end{subfigure}%
}%
\\[-3.5mm]
\caption{H4 chain individual run convergence plots for sampling noise.}
\label{fig:h4_sampling_noise_convergence}
\end{figure}

\begin{figure}[htbp]
\centering
\vspace*{-3mm} 
\mbox{%
\begin{subfigure}[b]{0.49\linewidth}
\includegraphics[width=\linewidth]{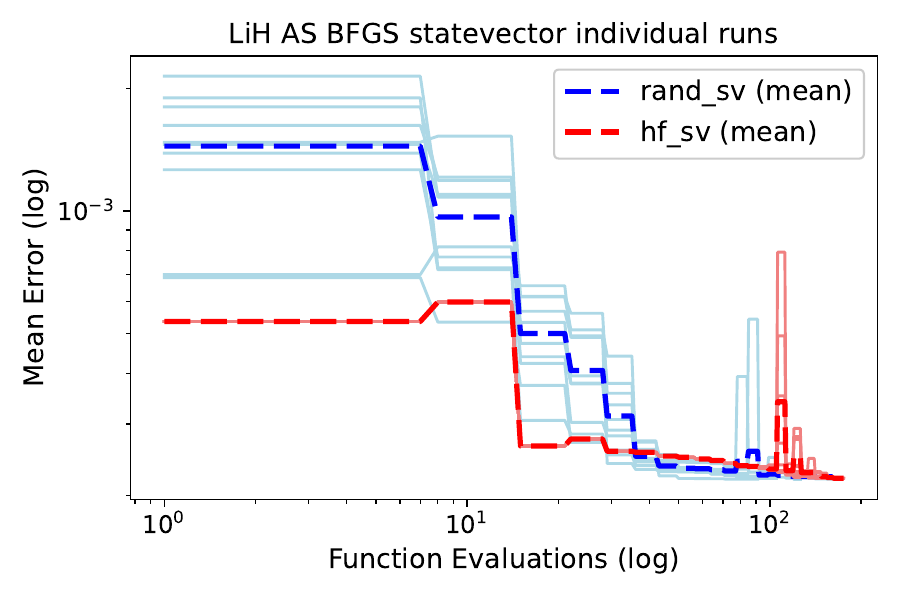}
\end{subfigure}%
\begin{subfigure}[b]{0.49\linewidth}
\includegraphics[width=\linewidth]{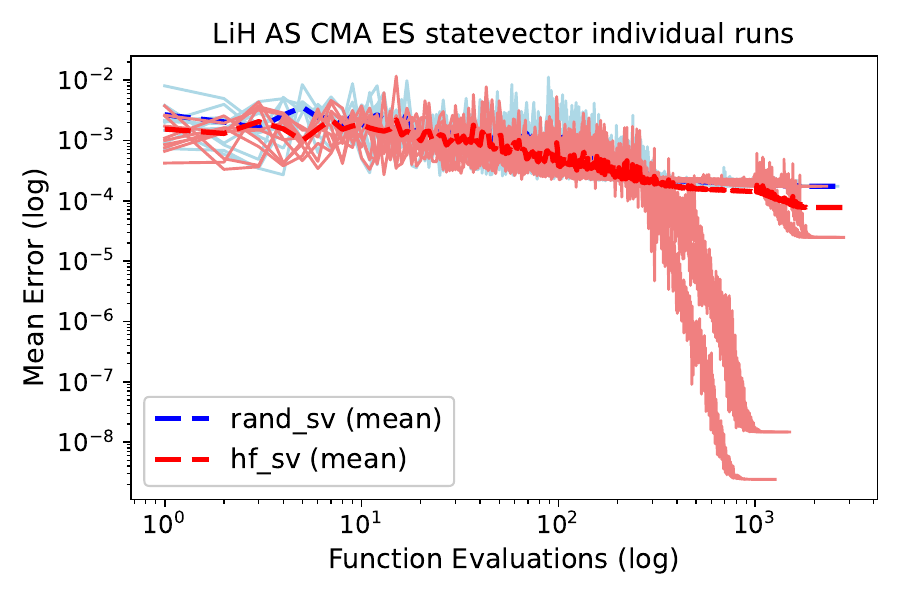}
\end{subfigure}%
}%
\\[-2mm]
\mbox{%
\begin{subfigure}[b]{0.49\linewidth}
\includegraphics[width=\linewidth]{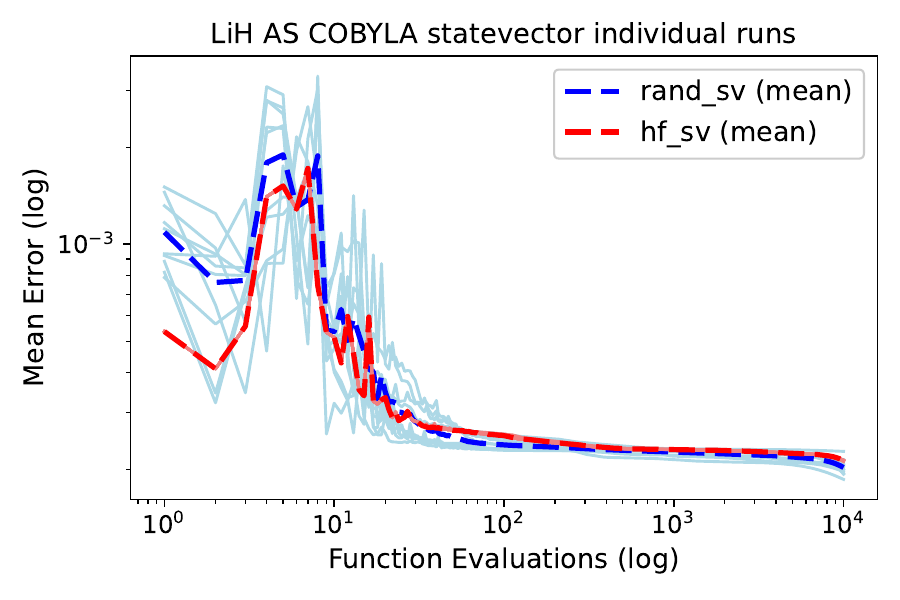}
\end{subfigure}%
\begin{subfigure}[b]{0.49\linewidth}
\includegraphics[width=\linewidth]{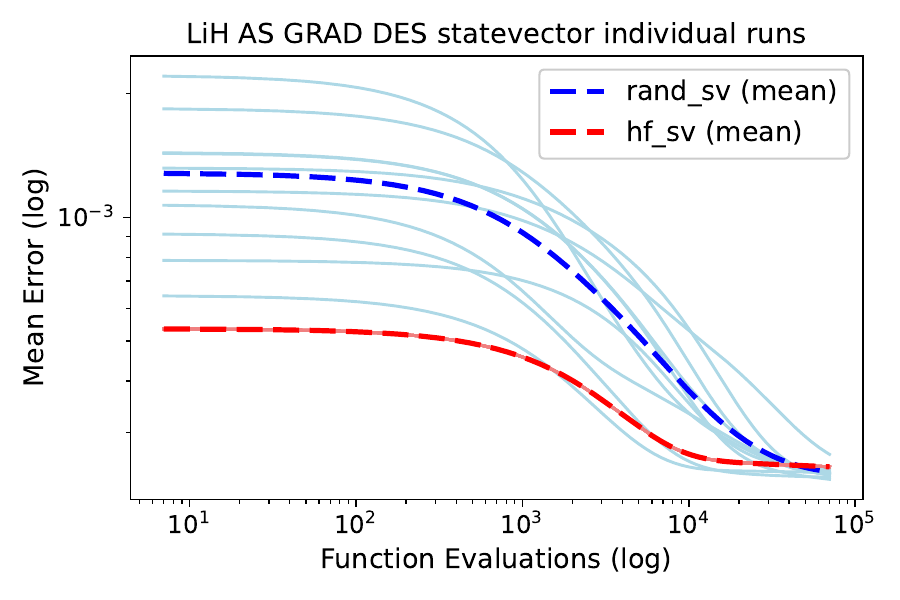}
\end{subfigure}%
}%
\\[-2mm]
\mbox{%
\begin{subfigure}[b]{0.49\linewidth}
\includegraphics[width=\linewidth]{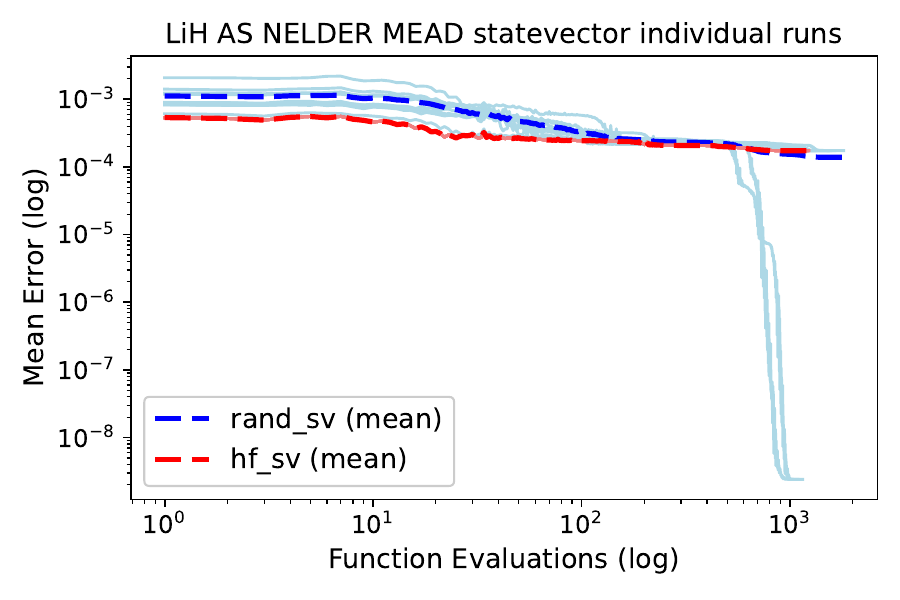}
\end{subfigure}%
\begin{subfigure}[b]{0.49\linewidth}
\includegraphics[width=\linewidth]{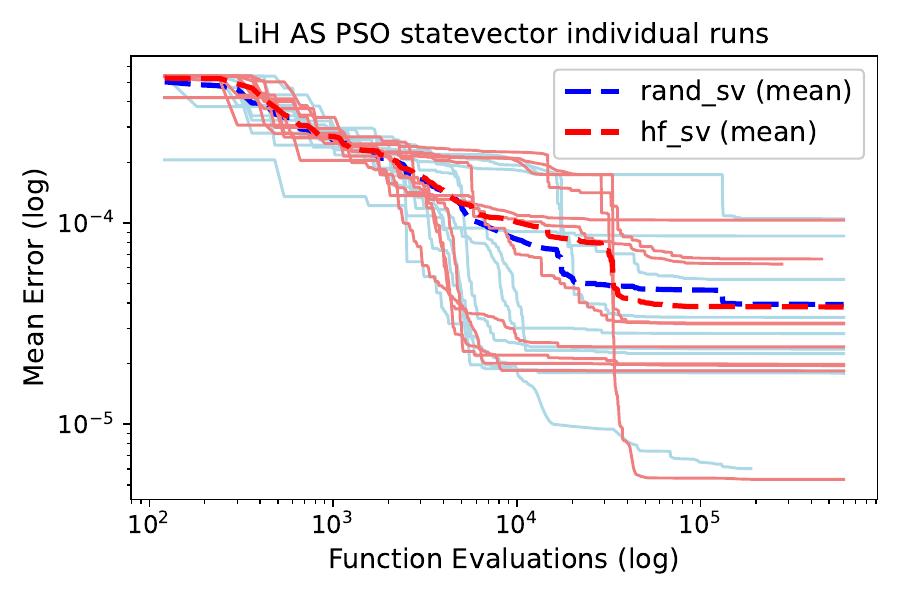}
\end{subfigure}%
}%
\\[-2mm]
\mbox{%
\begin{subfigure}[b]{0.49\linewidth}
\includegraphics[width=\linewidth]{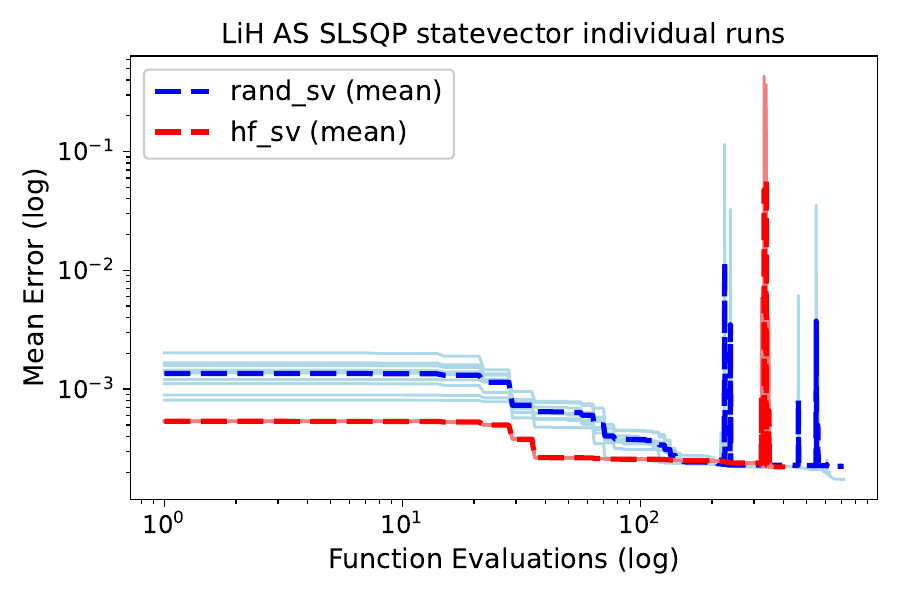}
\end{subfigure}%
\begin{subfigure}[b]{0.49\linewidth}
\includegraphics[width=\linewidth]{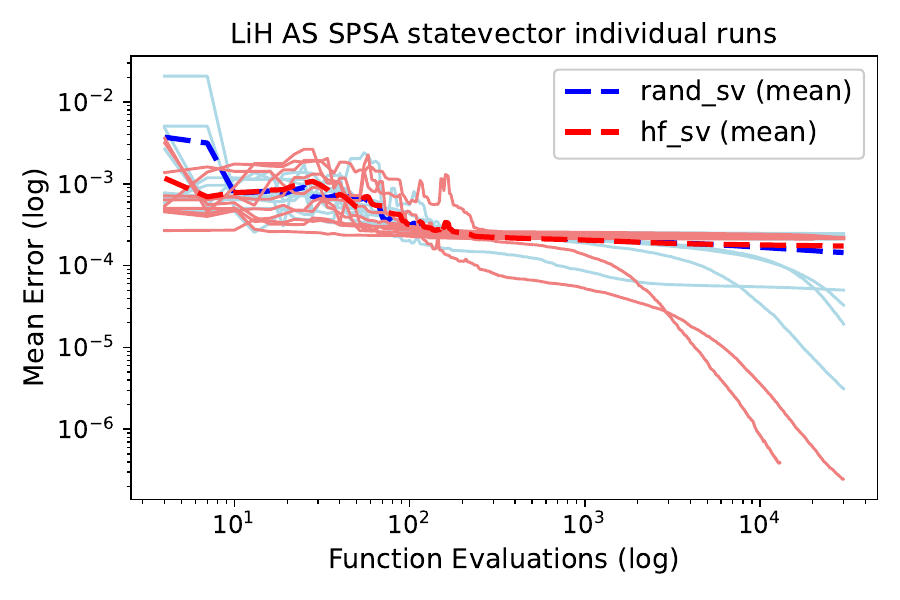}
\end{subfigure}%
}%
\\[-3.5mm]
\caption{LiH active space individual run convergence plots for statevector.}
\label{lih_as_fig:statevector_convergence}
\end{figure}

\begin{figure}[htbp]
\centering
\vspace*{-3mm} 
\mbox{%
\begin{subfigure}[b]{0.49\linewidth}
\includegraphics[width=\linewidth]{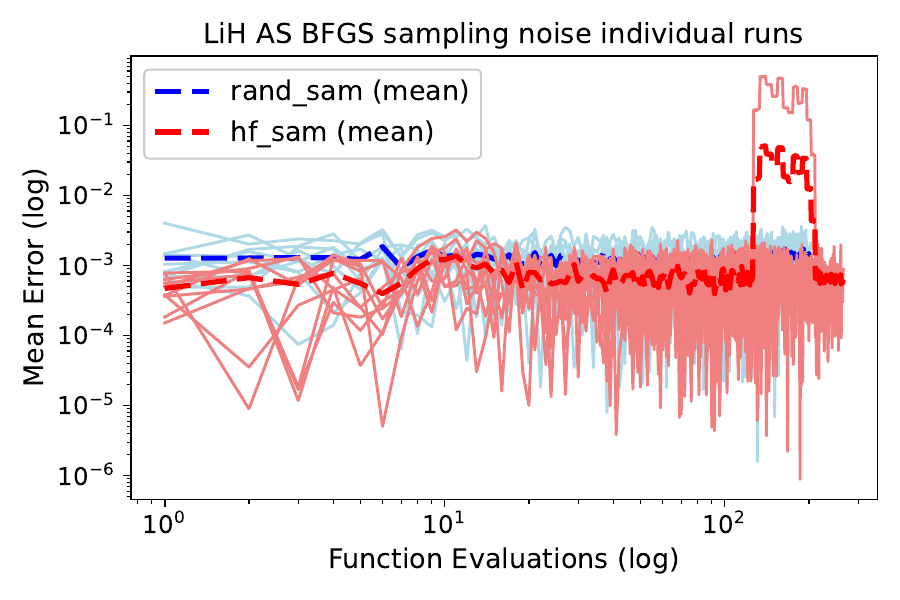}
\end{subfigure}%
\begin{subfigure}[b]{0.49\linewidth}
\includegraphics[width=\linewidth]{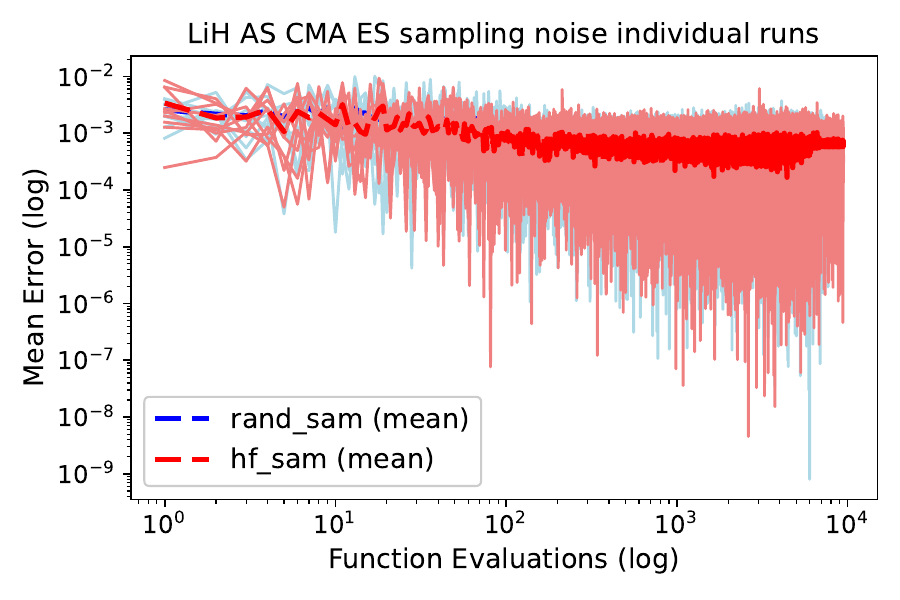}
\end{subfigure}%
}%
\\[-2mm]
\mbox{%
\begin{subfigure}[b]{0.49\linewidth}
\includegraphics[width=\linewidth]{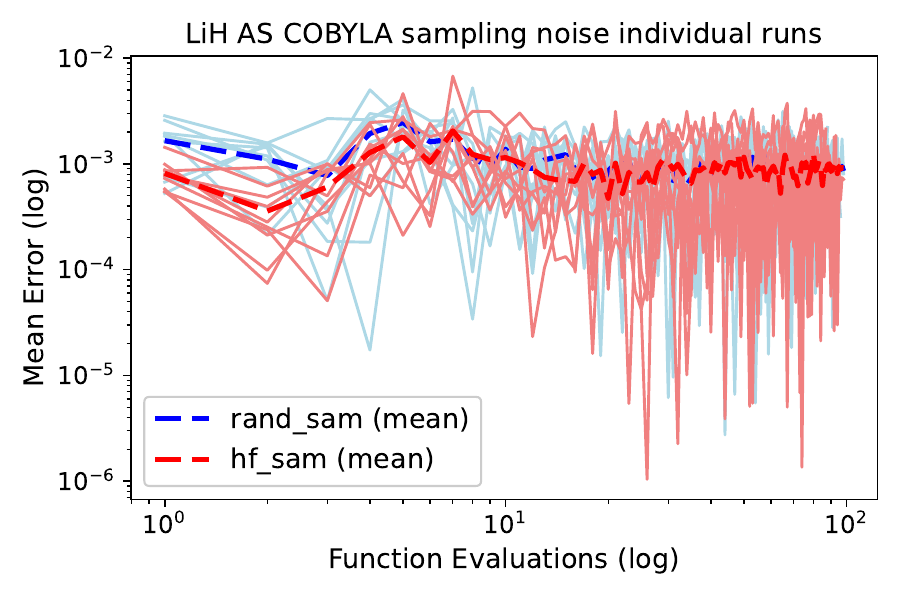}
\end{subfigure}%
\begin{subfigure}[b]{0.49\linewidth}
\includegraphics[width=\linewidth]{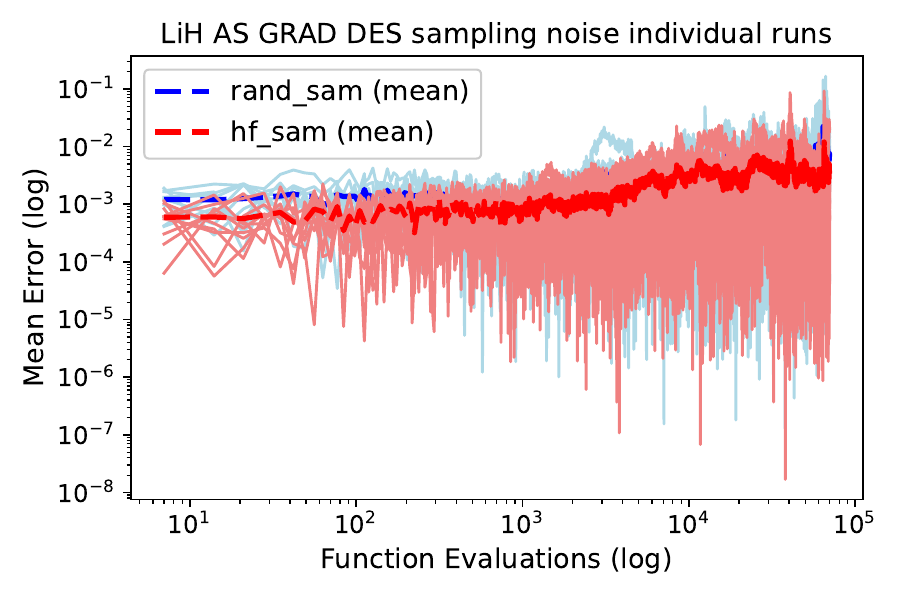}
\end{subfigure}%
}%
\\[-2mm]
\mbox{%
\begin{subfigure}[b]{0.49\linewidth}
\includegraphics[width=\linewidth]{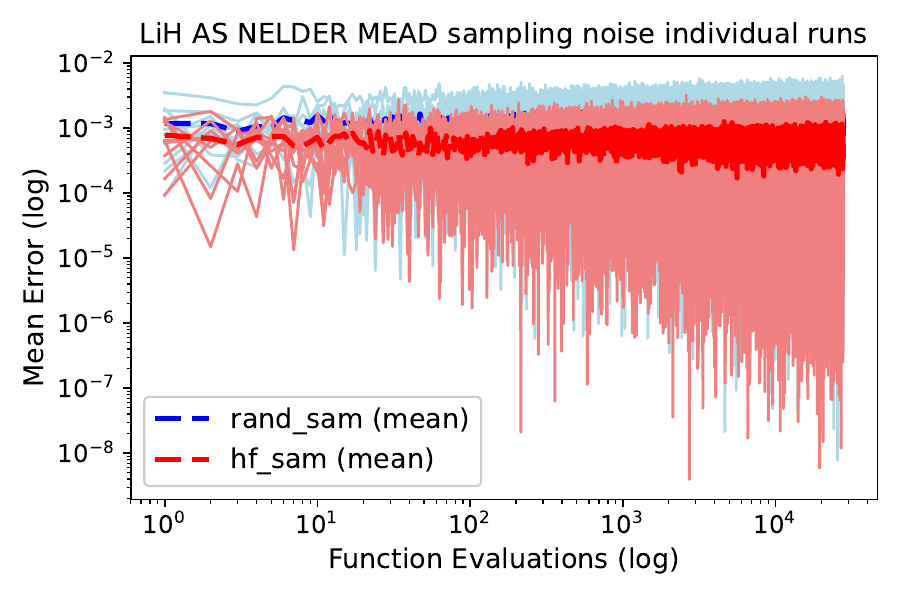}
\end{subfigure}%
\begin{subfigure}[b]{0.49\linewidth}
\includegraphics[width=\linewidth]{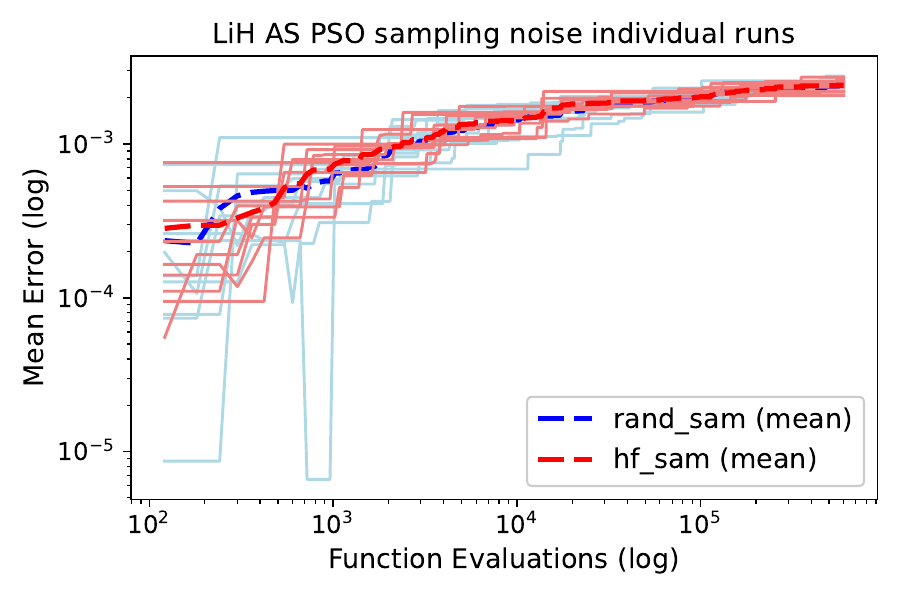}
\end{subfigure}%
}%
\\[-2mm]
\mbox{%
\begin{subfigure}[b]{0.49\linewidth}
\includegraphics[width=\linewidth]{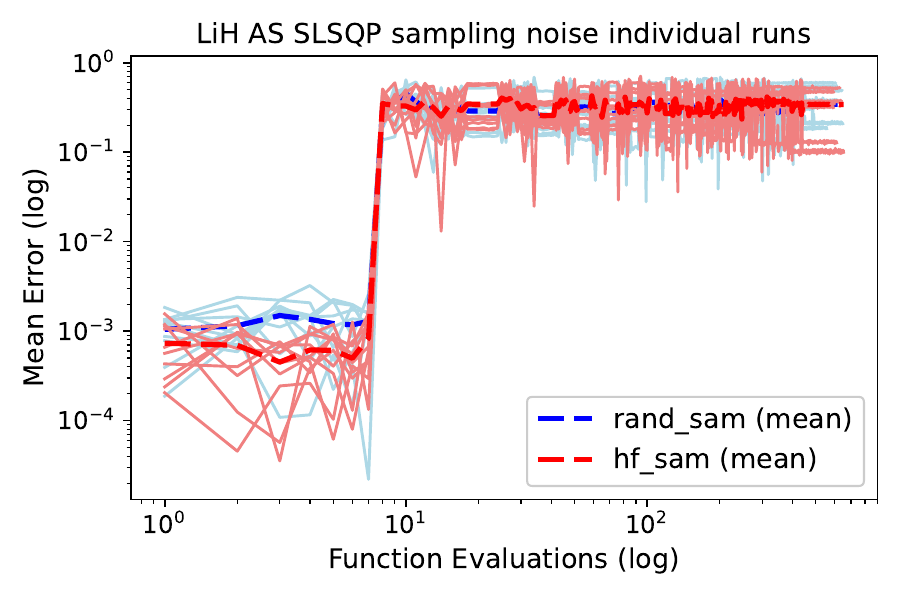}
\end{subfigure}%
\begin{subfigure}[b]{0.49\linewidth}
\includegraphics[width=\linewidth]{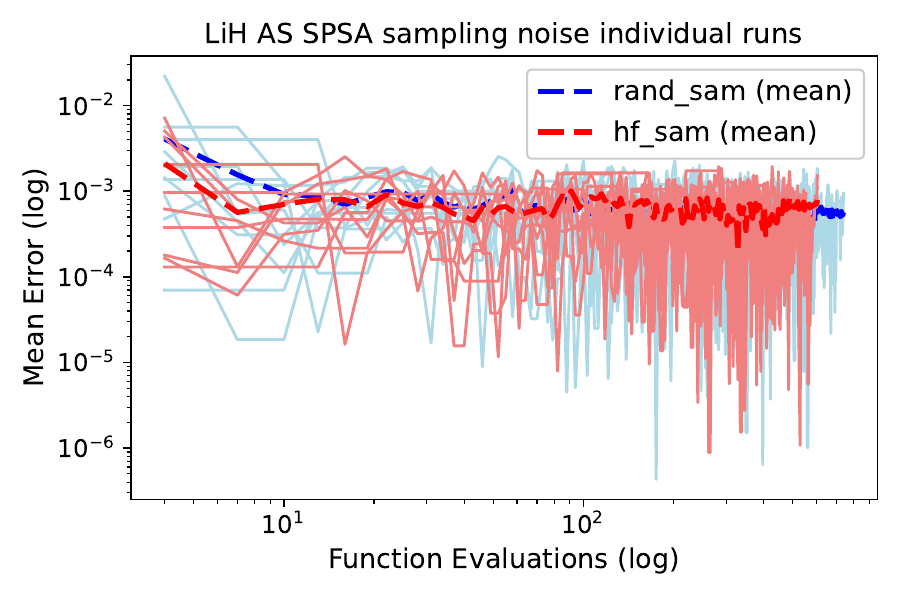}
\end{subfigure}%
}%
\\[-3.5mm]
\caption{LiH active space individual run convergence plots for sampling noise.}
\label{fig:lih_as_sampling_noise_convergence}
\end{figure}

\begin{figure}[htbp]
\centering
\vspace*{-3mm} 
\mbox{%
\begin{subfigure}[b]{0.49\linewidth}
\includegraphics[width=\linewidth]{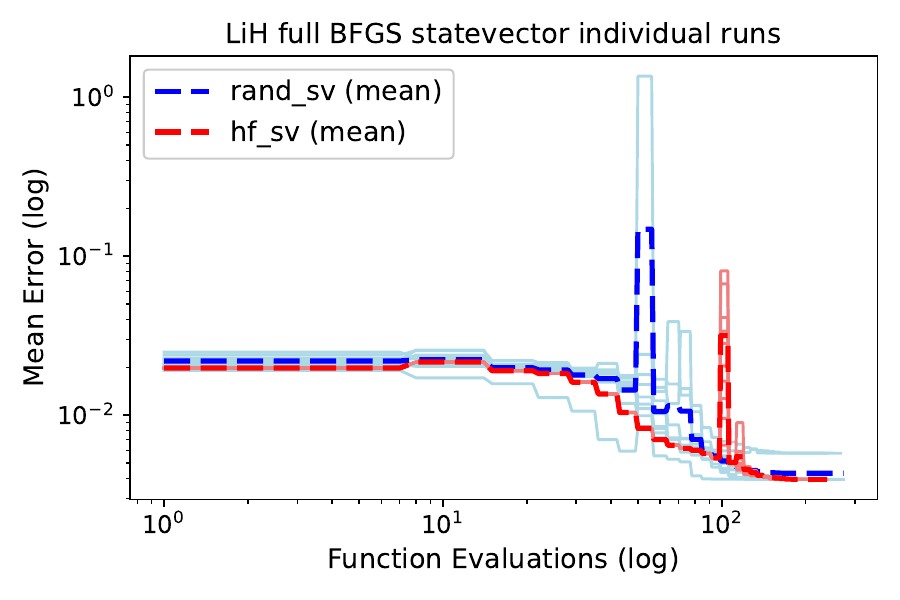}
\end{subfigure}%
\begin{subfigure}[b]{0.49\linewidth}
\includegraphics[width=\linewidth]{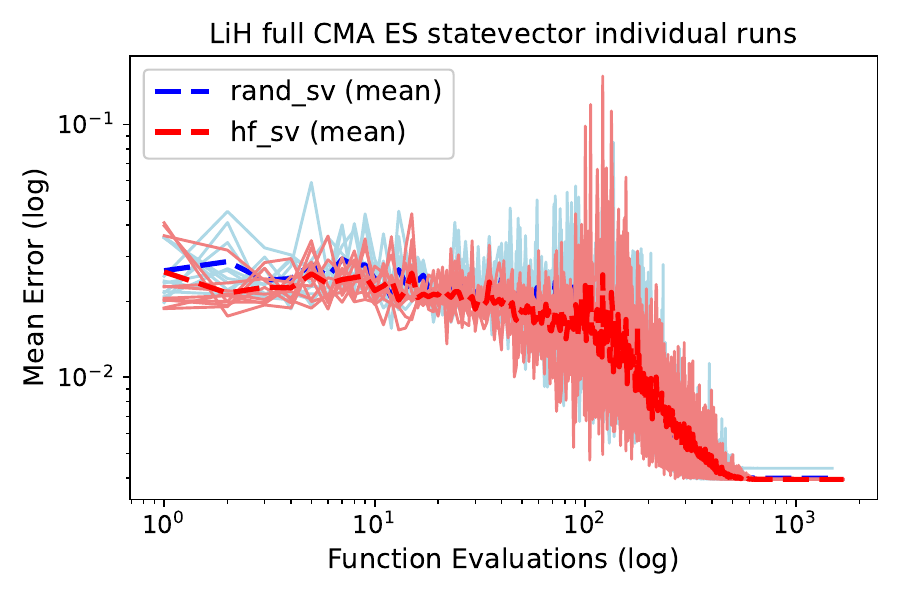}
\end{subfigure}%
}%
\\[-2mm]
\mbox{%
\begin{subfigure}[b]{0.49\linewidth}
\includegraphics[width=\linewidth]{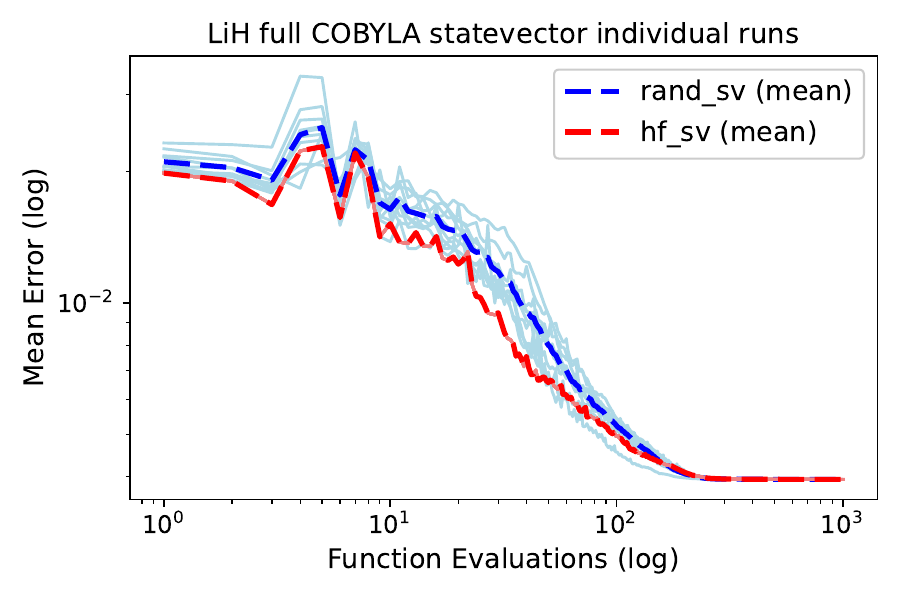}
\end{subfigure}%
\begin{subfigure}[b]{0.49\linewidth}
\includegraphics[width=\linewidth]{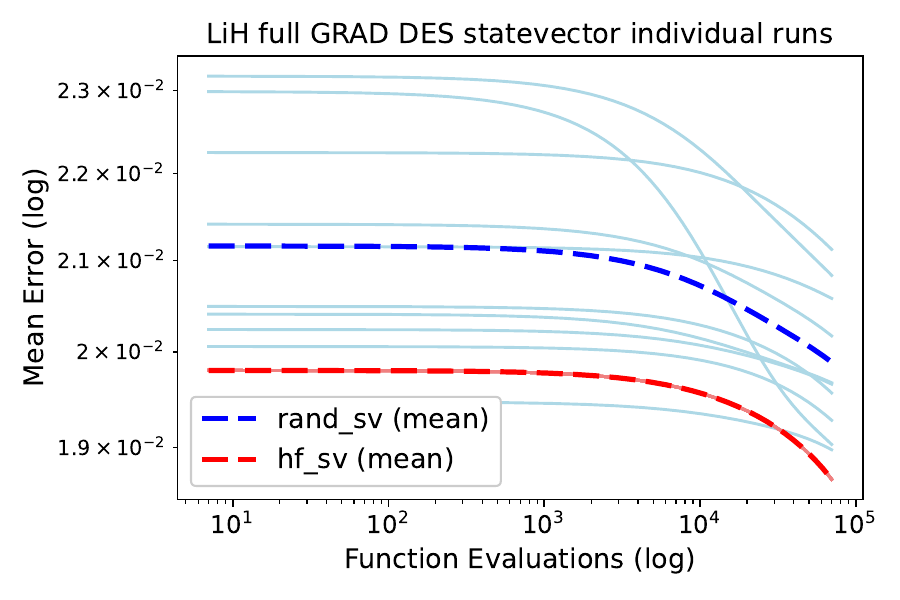}
\end{subfigure}%
}%
\\[-2mm]
\mbox{%
\begin{subfigure}[b]{0.49\linewidth}
\includegraphics[width=\linewidth]{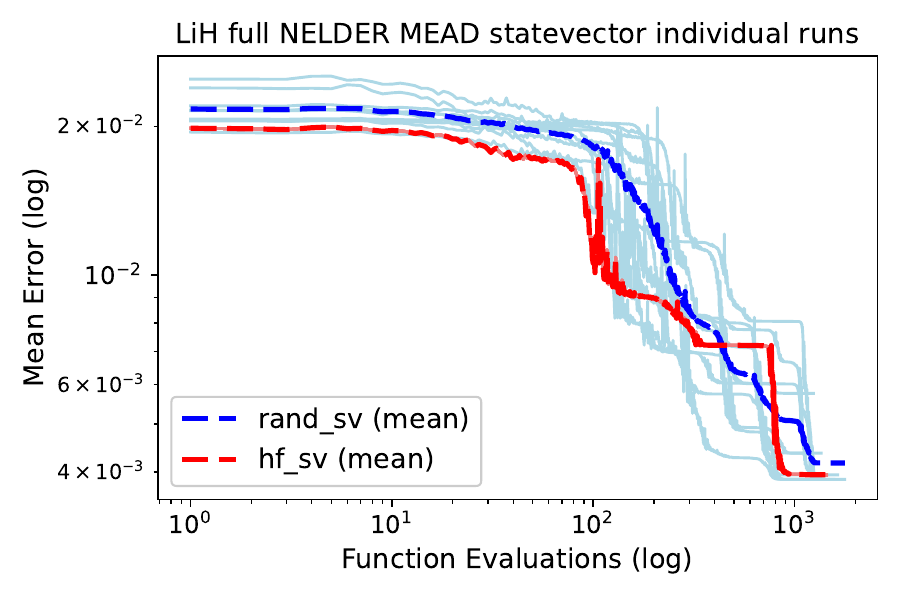}
\end{subfigure}%
\begin{subfigure}[b]{0.49\linewidth}
\includegraphics[width=\linewidth]{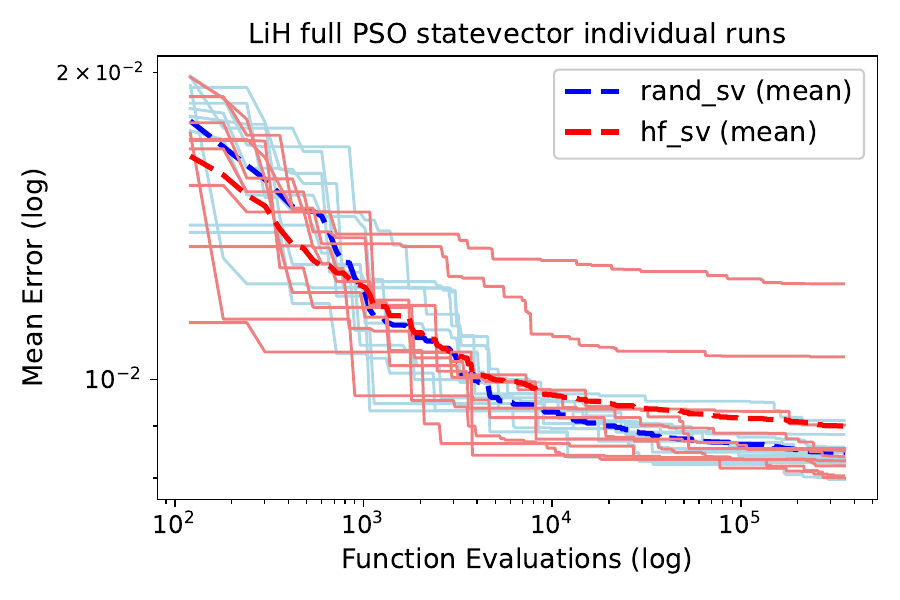}
\end{subfigure}%
}%
\\[-2mm]
\mbox{%
\begin{subfigure}[b]{0.49\linewidth}
\includegraphics[width=\linewidth]{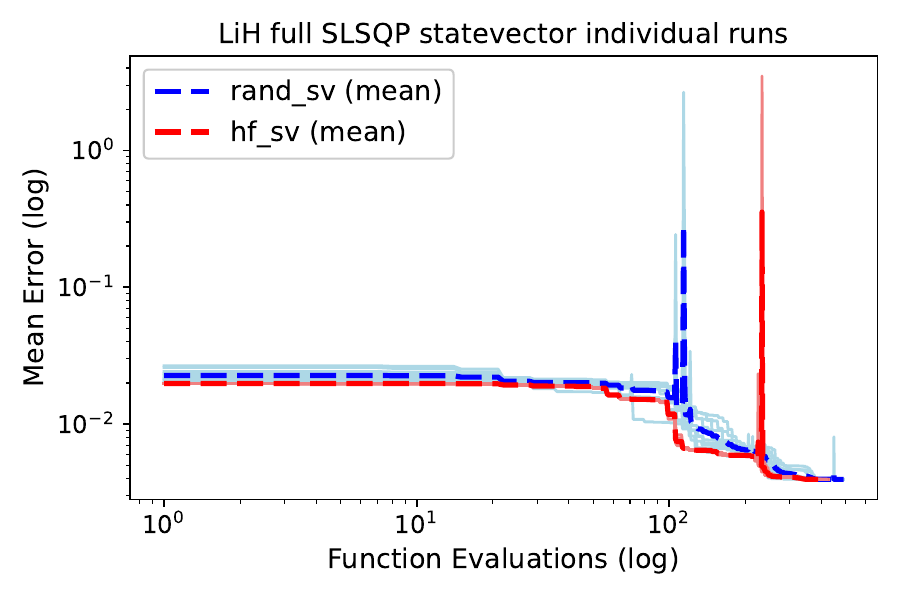}
\end{subfigure}%
\begin{subfigure}[b]{0.49\linewidth}
\includegraphics[width=\linewidth]{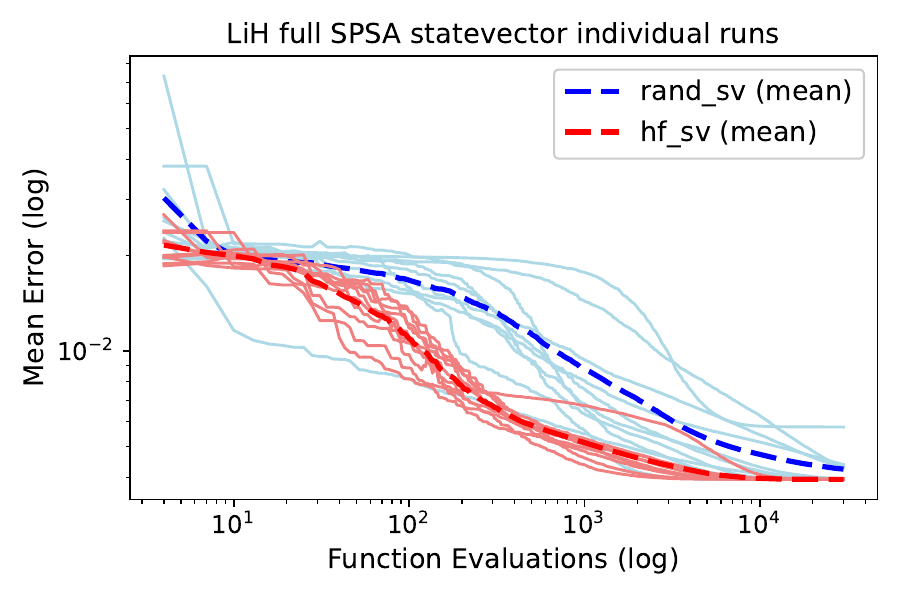}
\end{subfigure}%
}%
\\[-3.5mm]
\caption{LiH full space individual run convergence plots for statevector.}
\label{fig:lih_statevector_convergence}
\end{figure}

\begin{figure}[htbp]
\centering
\vspace*{-3mm} 
\mbox{%
\begin{subfigure}[b]{0.49\linewidth}
\includegraphics[width=\linewidth]{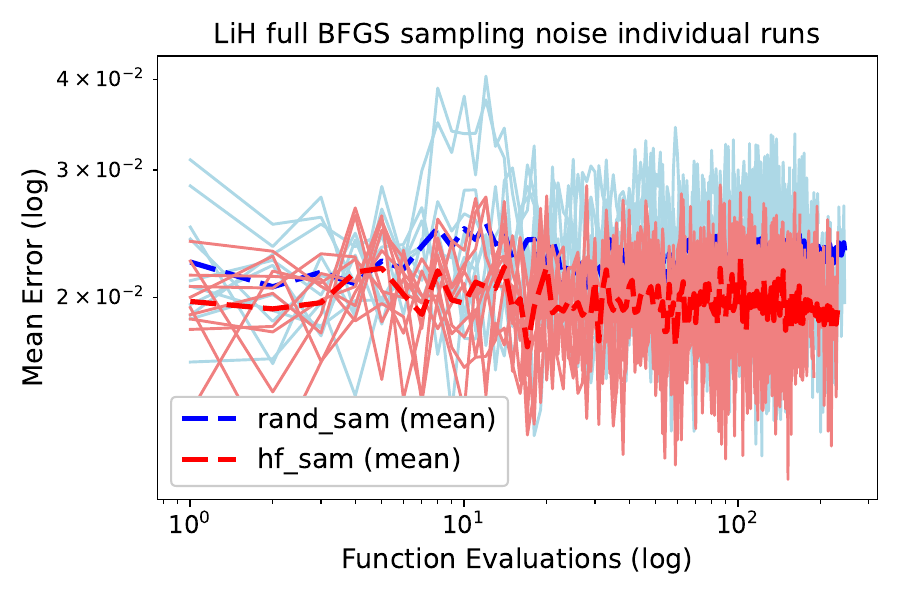}
\end{subfigure}%
\begin{subfigure}[b]{0.49\linewidth}
\includegraphics[width=\linewidth]{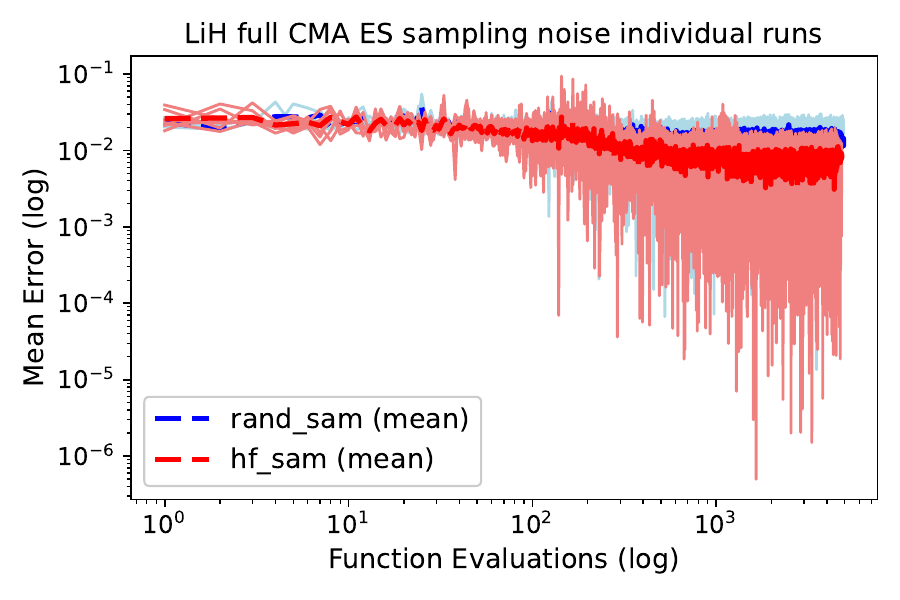}
\end{subfigure}%
}%
\\[-2mm]
\mbox{%
\begin{subfigure}[b]{0.49\linewidth}
\includegraphics[width=\linewidth]{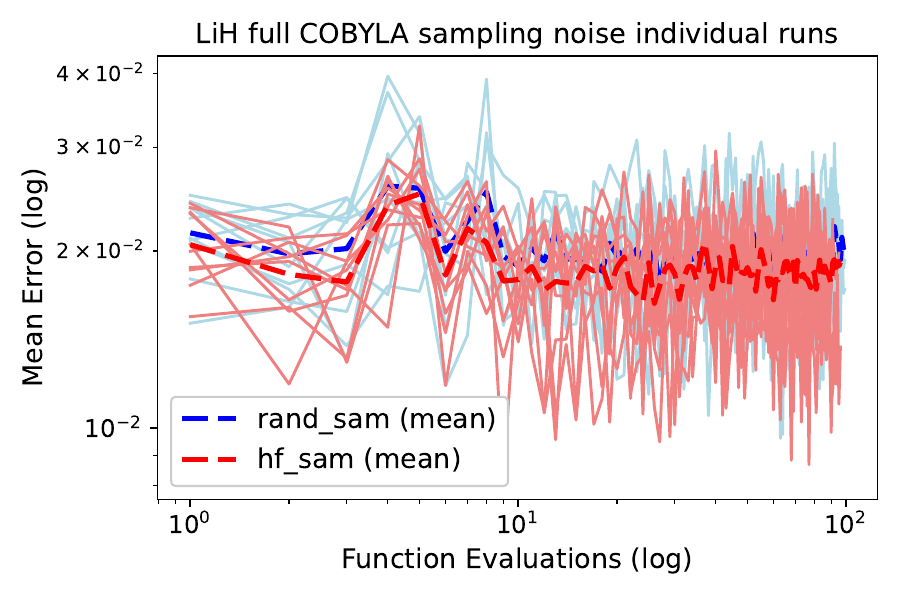}
\end{subfigure}%
\begin{subfigure}[b]{0.49\linewidth}
\includegraphics[width=\linewidth]{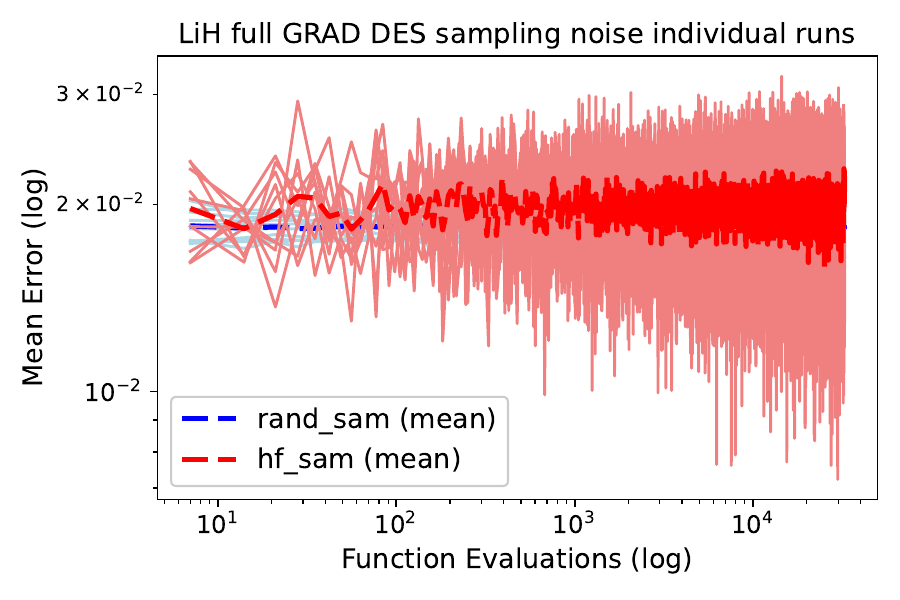}
\end{subfigure}%
}%
\\[-2mm]
\mbox{%
\begin{subfigure}[b]{0.49\linewidth}
\includegraphics[width=\linewidth]{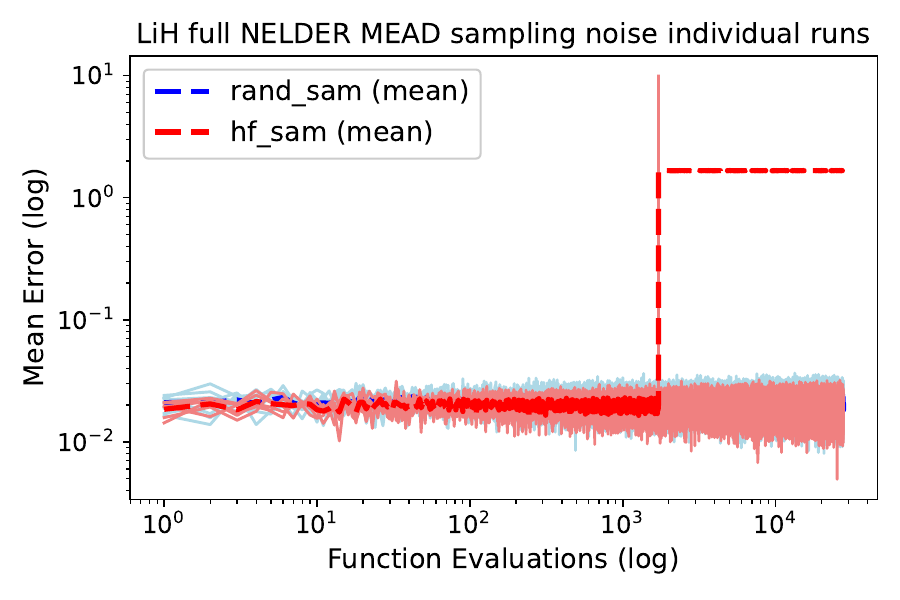}
\end{subfigure}%
\begin{subfigure}[b]{0.49\linewidth}
\includegraphics[width=\linewidth]{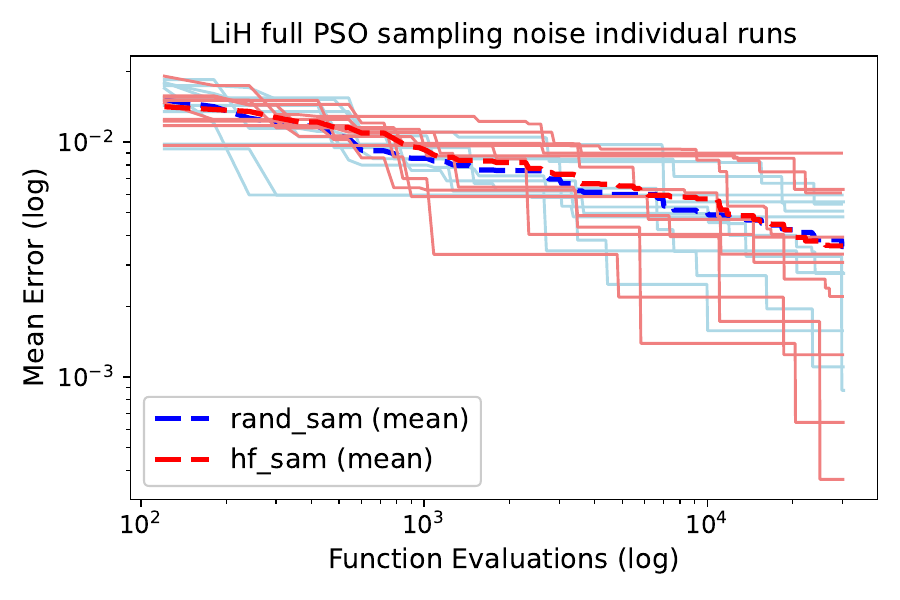}
\end{subfigure}%
}%
\\[-2mm]
\mbox{%
\begin{subfigure}[b]{0.49\linewidth}
\includegraphics[width=\linewidth]{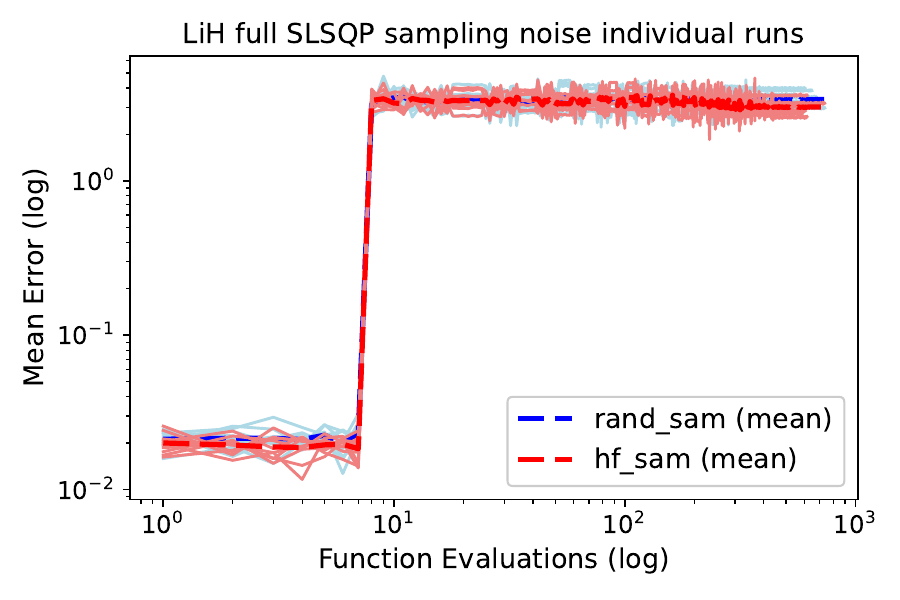}
\end{subfigure}%
\begin{subfigure}[b]{0.49\linewidth}
\includegraphics[width=\linewidth]{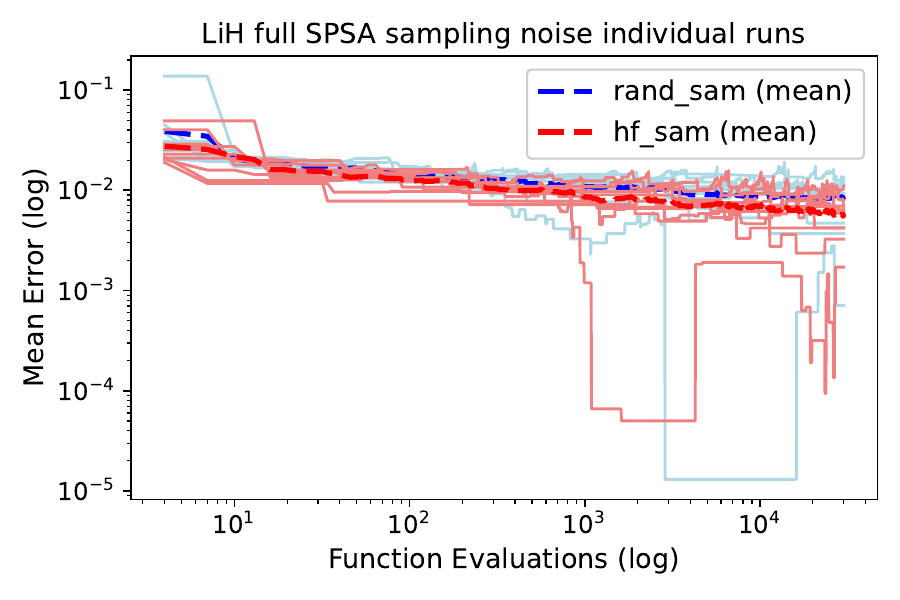}
\end{subfigure}%
}%
\\[-3.5mm]
\caption{LiH full space individual run convergence plots for sampling noise.}
\label{fig:lih_sampling_noise_convergence}
\end{figure}

\section{Population means}
\label{sec:pop_mean}

The supplementary population size analysis in \cref{fig:error_analysis_7,fig:error_analysis_50,fig:error_analysis_100} reveals systematic noise suppression through population averaging across different optimization scales. For the smallest population size of 7 (\cref{fig:error_analysis_7}), the mean-based approach already demonstrates superior performance compared to the best-value selection, particularly evident in the bottom panel where the black line (iteration means) maintains consistently lower errors than the red line (best values) across all shot budgets. The blue line representing averages from the best iteration shows significant fluctuations around the noise floor, indicating instability in conventional optimization approaches.

This behavior becomes more pronounced with larger populations. At size 25 (main text results) and particularly for the 50-individual case (\cref{fig:error_analysis_50}), the mean-based strategy achieves near-perfect alignment with the theoretical noise floor. The high-shot-count regime (30,000 shots) reveals particularly instructive behavior: populations of 25 and above show transient convergence to zero error, with the 100-individual case (\cref{fig:error_analysis_100}) demonstrating both the benefits and limitations of large populations. Here, rapid convergence occurs by iteration 5, maintaining near-zero error until approximately iteration 20, after which noise-induced divergence becomes apparent. This early convergence followed by overfitting suggests an optimal population size window exists that balances convergence speed with noise resilience.

\FloatBarrier

\begin{figure}[htpb]
\centering
\includegraphics[width=0.75\textwidth]{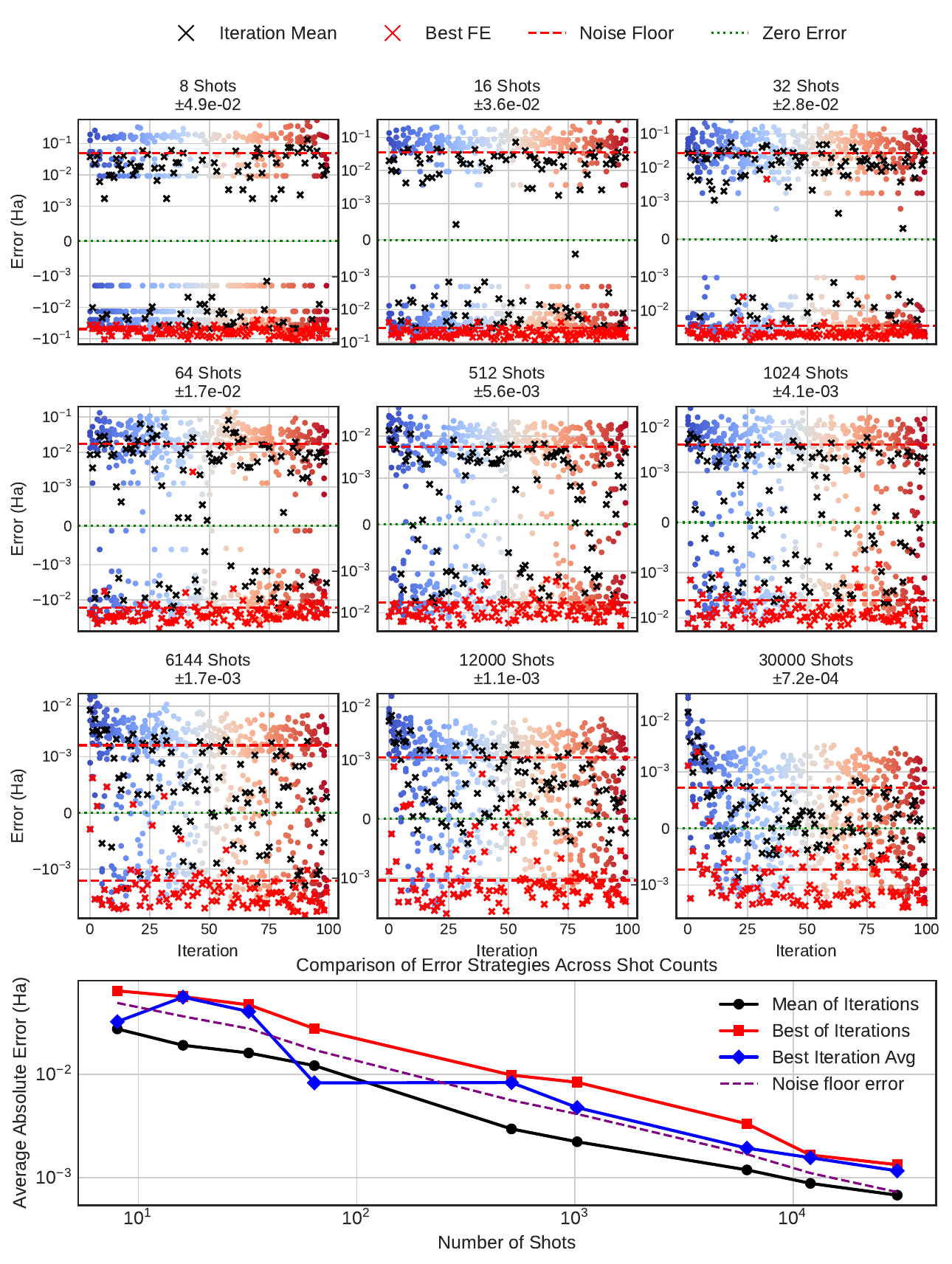}
\caption{Energy error progression for \ce{H2} using \ac{tvha} with \ac{cmaes} population size 7. Top: FEs of all individuals (colored points), average of FEs in iteration (black crosses), best (lowest) \ac{fe} in iteration (red crosses) and noise floor (red dashed lines). Bottom: Aggregated absolute errors for the mean of FEs in iteration (black line), aggregated absolute errors for the best \ac{fe} in iteration (red line), and the best iteration average (blue) approaches compared to the computed noise floor error (purple).}
\label{fig:error_analysis_7}
\end{figure}

\begin{figure}[htpb]
\centering
\includegraphics[width=0.75\textwidth]{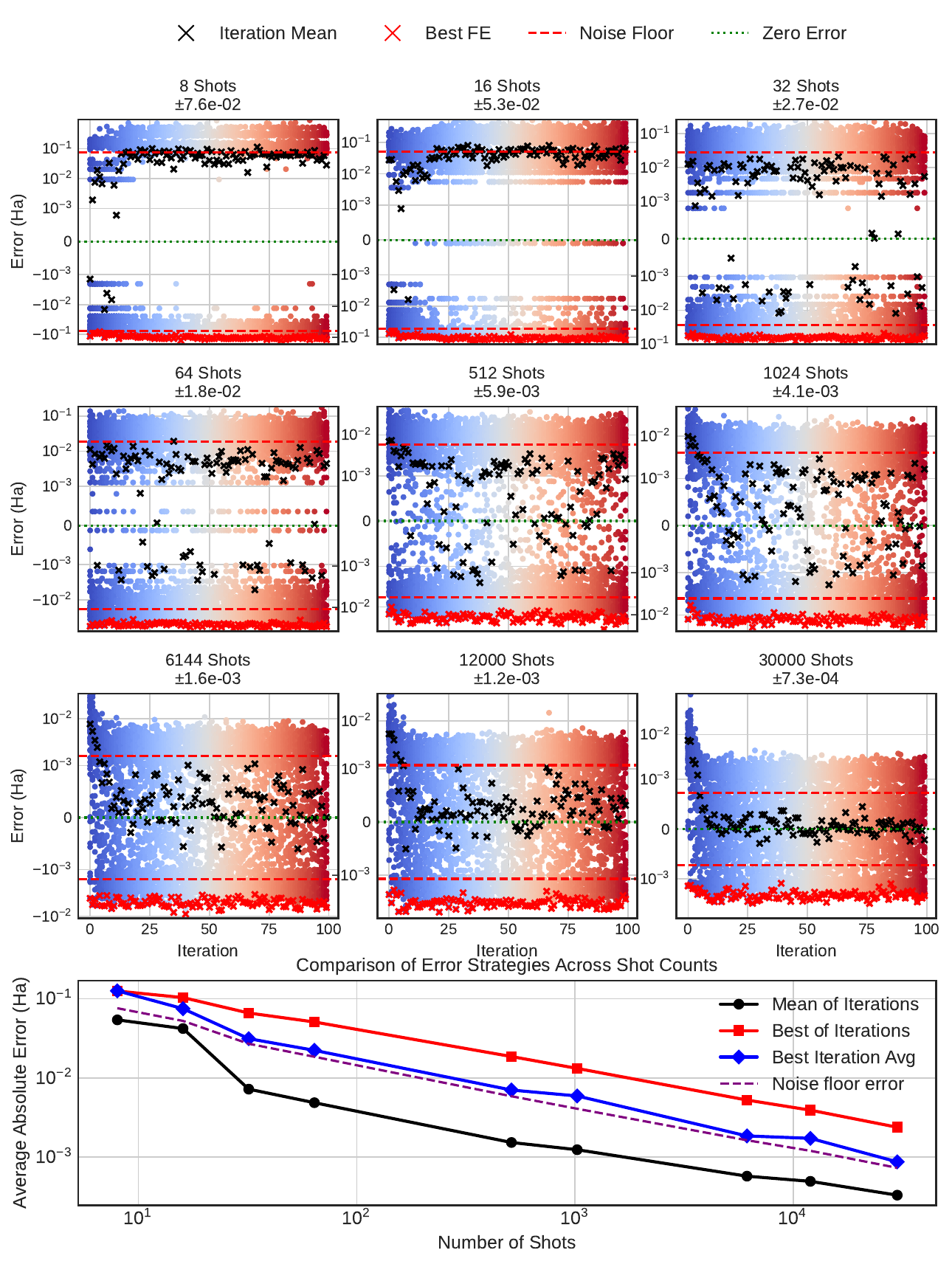}
\caption{Energy error progression for \ce{H2} using \ac{tvha} with \ac{cmaes} population size 50. Top: FEs of all individuals (colored points), average of FEs in iteration (black crosses), best (lowest) \ac{fe} in iteration (red crosses) and noise floor (red dashed lines). Bottom: Aggregated absolute errors for the mean of FEs in iteration (black line), aggregated absolute errors for the best \ac{fe} in iteration (red line), and the best iteration average (blue) approaches compared to the computed noise floor error (purple).}
\label{fig:error_analysis_50}
\end{figure}

\begin{figure}[htpb]
\centering
\includegraphics[width=0.75\textwidth]{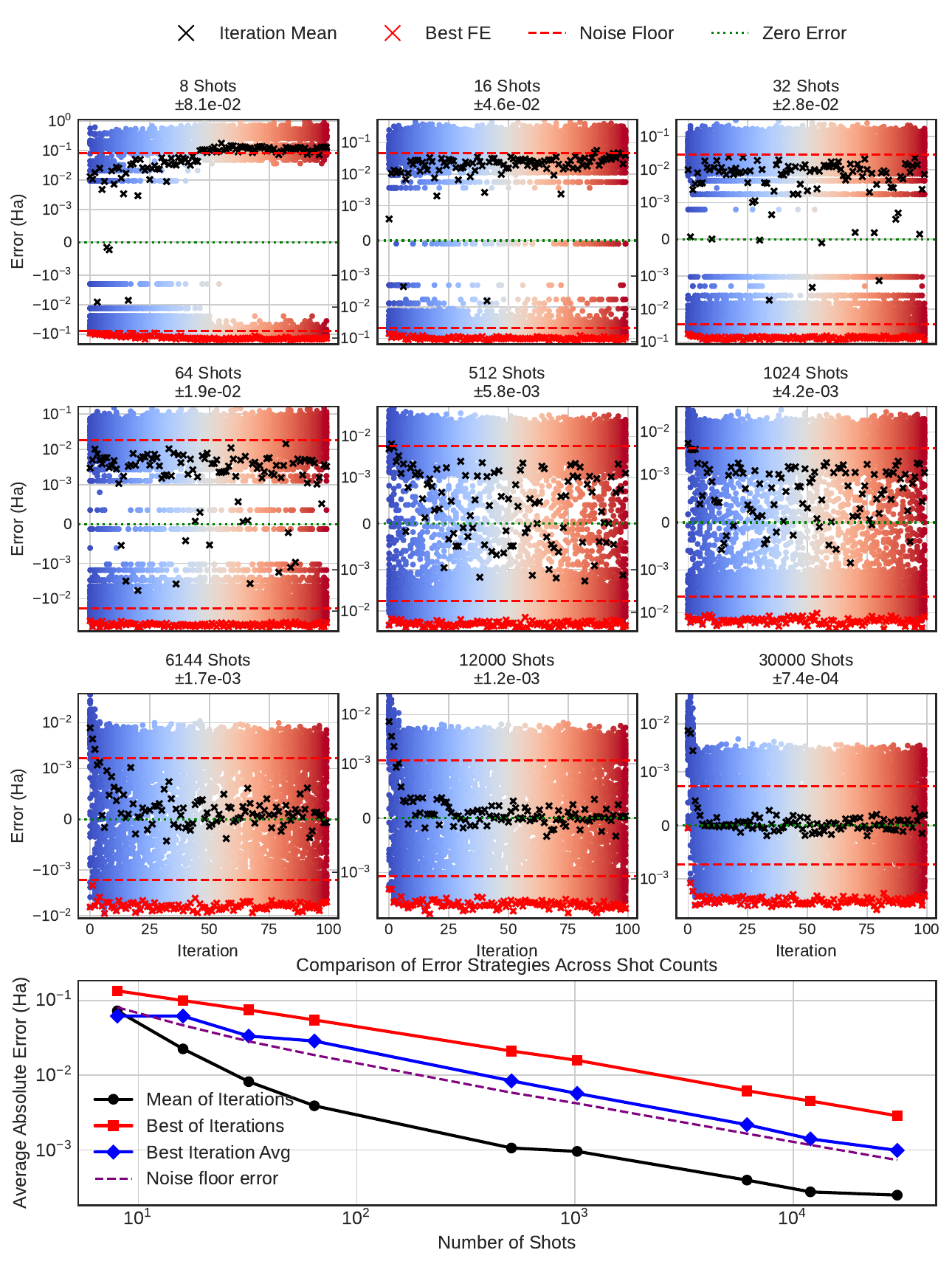}
\caption{Energy error progression for \ce{H2} using \ac{tvha} with \ac{cmaes} population size 100. Top: FEs of all individuals (colored points), average of FEs in iteration (black crosses), best (lowest) \ac{fe} in iteration (red crosses) and noise floor (red dashed lines). Bottom: Aggregated absolute errors for the mean of FEs in iteration (black line), aggregated absolute errors for the best \ac{fe} in iteration (red line), and the best iteration average (blue) approaches compared to the computed noise floor error (purple).}
\label{fig:error_analysis_100}
\end{figure}

\FloatBarrier

\section{Molecular configurations}
\label{app:molecular_config}

\begin{table}[htbp]
\centering
\caption{Molecular configurations and computational parameters used in quantum simulations (transposed).}
\label{tab:molecular_configurations_transposed}
\begin{tabular}{lcccc}
\hline
\textbf{Property} & \textbf{\ce{H2}} & \textbf{\ce{LiH} (Active Space)} & \textbf{\ce{LiH}} & \textbf{\ce{H4}} \\
\hline
Bond Length(s) (\AA) & 0.74279 & 1.596 & 1.596 & 0.7, 0.7, 0.768$^b$ \\
Charge & 0 & 0 & 0 & 0 \\
Multiplicity & 1 & 1 & 1 & 1 \\
Basis Set & STO-3G & STO-3G & STO-3G & STO-3G \\
Active Space & Full & (2,2) in 3 orbitals$^a$ & Full & Full \\
\hline
\end{tabular}
\par
\footnotesize
$^a$ Active space includes orbitals [1,2,3] with 2 alpha and 2 beta electrons.\\
$^b$ Consecutive H–H bond distances in the linear chain configuration.
\end{table}

\section{Convergence criteria}
\label{app:convergence}
Aside from the maximum number of function evaluation, which was the same across all optimizers, every optimizer had its own convergence criterion. The values were taken from a priori testing of the optimizer's settings.

For \ac{bfgs} the convergence criterion was based on parameter \texttt{ftol}=$10^{-10}$ and defined as
\begin{equation}
    \frac{f_k - f_{k+1}}{\max\left( |f_k|,\, |f_{k+1}|,\, 1 \right)} \leq \text{ftol},
\end{equation}
where $f_k$ is the function evaluation value in $k$-th iteration.

For \ac{cobyla}, the convergence criterion was based on the optimization tolerance 
\texttt{tol} = $10^{-7}$ and defined as
\begin{equation}
    \Delta f_k \leq \text{tol},
\end{equation}
where $\Delta f_k = |f_{k+1} - f_k|$ is the change in the objective function value between successive iterations. 

For \ac{gd}, the convergence criterion was based on the parameter tolerance \texttt{tol} = $5 \times 10^{-4}$ and defined as
\begin{equation}
    \lVert \boldsymbol{\theta}_{k+1} - \boldsymbol{\theta}_k \rVert \leq \text{tol},
\end{equation}
where $\boldsymbol{\theta}_k$ denotes the parameter vector at the $k$-th iteration.

For \ac{cmaes}, the convergence criteria were based on the tolerances  
\(\texttt{tolx} = 10^{-12}\) and \(\texttt{tolfun} = 10^{-12}\), and defined as  
\begin{equation}
    \begin{aligned}
      \max\limits_{i}\sigma_i &\;\le\; \texttt{tolx}, \\
      \mathrm{range}\!\Bigl(f_{k}^{(1:\lambda)}\Bigr) &\;\le\; \texttt{tolfun},
    \end{aligned}
\end{equation}
where \(\sigma_i\) are the standard deviations of the search distribution in each coordinate and  \(\mathrm{range}(f_{k}^{(1:\lambda)})\) denotes the range of objective-function values among the current population of size \(\lambda\).

For \ac{nelder}, the convergence criterion was based on the parameter tolerance 
\texttt{xatol} = $10^{-6}$ and defined as
\begin{equation}
    \lVert \boldsymbol{x}_{k+1} - \boldsymbol{x}_k \rVert \leq \text{xatol},
\end{equation}
where $\boldsymbol{x}_k$ denotes the parameter vector at the $k$-th iteration.

For \ac{pso}, the convergence criterion was based on the number of non-clustered particles and a velocity convergence threshold. The algorithm terminates when
\begin{equation}
    \sum_{i=1}^{N} c_i < \text{cluster\_threshold}
    \quad \text{and} \quad
    \max(\mathbf{v}) < \text{velocity\_convergence},
\end{equation}
where $c_i$ indicates whether the $i$-th particle is unclustered ($c_i = 1$ if unclustered, $c_i = 0$ otherwise), $\mathbf{v}$ is the particle velocity vector. 
The following values were set,\texttt{velocity\_convergence} = $10^{-5}$ and \texttt{cluster\_treshold} =$1$.

For \ac{slsqp}, the convergence criterion was based on the function tolerance 
\texttt{ftol} = $10^{-10}$ and defined as
\begin{equation}
    \frac{f_k - f_{k+1}}{\max\left( |f_k|,\, |f_{k+1}|,\, 1 \right)} \leq \text{ftol},
\end{equation}
where $f_k$ denotes the objective function value at the $k$-th iteration.

For \ac{spsa}, the convergence criterion was based on the \texttt{termination\_checker} 
and the \texttt{allowed\_increase} parameter. The optimization process terminates when 
the improvement in the objective function stagnates over the last $N = 100$ iterations 
and the fitted trend of objective values indicates an increasing slope:
\begin{equation}
    \text{slope} = \frac{1}{N} \, \frac{d f_k}{d k} > 0,
\end{equation}
where the slope is obtained from a linear regression of the last $N$ objective values 
$\{ f_{k-N+1}, \ldots, f_k \}$. The parameter \texttt{allowed\_increase} = $10^{-3}$ 
controls the acceptable increase in the objective value during optimization.

\bibliographystyle{iopart-num} 
\bibliography{optimization_methods,main_references,vha,newbib}

@article{tilly2022variational,
  title={The variational quantum eigensolver: a review of methods and best practices},
  author={Tilly, Jules and Chen, Hongxiang and Cao, Shuxiang and Picozzi, Dario and Setia, Kanav and Li, Ying and Grant, Edward and Wossnig, Leonard and Rungger, Ivan and Booth, George H and others},
  journal={Physics Reports},
  volume={986},
  pages={1--128},
  year={2022},
  publisher={Elsevier}
}

@article{beyer2001theory,
  title={Theory of evolution strategies—A tutorial},
  author={Beyer, H-G and Arnold, Dirk V},
  journal={Theoretical aspects of evolutionary computing},
  pages={109--133},
  year={2001},
  publisher={Springer}
}

@article{peruzzo2014variational,
  title={A variational eigenvalue solver on a photonic quantum processor},
  author={Peruzzo, Alberto and McClean, Jarrod and Shadbolt, Peter and Yung, Man-Hong and Zhou, Xiao-Qi and Love, Peter J and Aspuru-Guzik, Al\'{a}n and O'Brien, Jeremy L},
  journal={Nature Communications},
  volume={5},
  pages={4213},
  year={2014},
  publisher={Nature Publishing Group}
}

@article{temme2017error,
  title={Error mitigation for short-depth quantum circuits},
  author={Temme, Kristan and Bravyi, Sergey and Gambetta, Jay M},
  journal={Physical Review Letters},
  volume={119},
  number={18},
  pages={180509},
  year={2017},
  publisher={APS}
}

@article{li2017efficient,
  title={Efficient variational quantum simulator incorporating active error minimization},
  author={Li, Ying and Benjamin, Simon C},
  journal={Physical Review X},
  volume={7},
  number={2},
  pages={021050},
  year={2017},
  publisher={APS}
}

@inproceedings{saib2021effect,
  title={The effect of noise on the performance of variational algorithms for quantum chemistry},
  author={Saib, Waheeda and Wallden, Petros and Akhalwaya, Ismail},
  booktitle={2021 IEEE International Conference on Quantum Computing and Engineering (QCE)},
  pages={42--53},
  year={2021},
  organization={IEEE}
}

@article{chi2026variational,
  title={Variational quantum algorithms with invariant probabilistic error cancellation on noisy quantum processors},
  author={Chi, Yulin and Shi, Hongyi and Zheng, Wen and Cai, Haoyang and Zhang, Yu and Tan, Xinsheng and Li, Shaoxiong and Wang, Jianwei and Cui, Jiangyu and Yung, Man-Hong and others},
  journal={Science China Physics, Mechanics \& Astronomy},
  volume={69},
  number={1},
  pages={210312},
  year={2026},
  publisher={Springer}
}

@article{sagastizabal2019experimental,
  title={Experimental error mitigation via symmetry verification in a variational quantum eigensolver},
  author={Sagastizabal, Ramiro and Bonet-Monroig, Xavier and Singh, Malay and Rol, M Adriaan and Bultink, CC and Fu, Xiang and Price, CH and Ostroukh, VP and Muthusubramanian, N and Bruno, A and others},
  journal={Physical Review A},
  volume={100},
  number={1},
  pages={010302},
  year={2019},
  publisher={APS}
}

@article{oliv2022evaluating,
  title={Evaluating the impact of noise on the performance of the variational quantum eigensolver},
  author={Oliv, Marita and Matic, Andrea and Messerer, Thomas and Lorenz, Jeanette Miriam},
  journal={arXiv preprint arXiv:2209.12803},
  year={2022}
}

@article{cai2023quantum,
  title={Quantum error mitigation},
  author={Cai, Zhenyu and Babbush, Ryan and Benjamin, Simon C and Endo, Suguru and Huggins, William J and Li, Ying and McClean, Jarrod R and O’Brien, Thomas E},
  journal={Reviews of Modern Physics},
  volume={95},
  number={4},
  pages={045005},
  year={2023},
  publisher={APS}
}

@article{hansen2008method,
  title={A method for handling uncertainty in evolutionary optimization with an application to feedback control of combustion},
  author={Hansen, Nikolaus and Niederberger, Andr{\'e} SP and Guzzella, Lino and Koumoutsakos, Petros},
  journal={IEEE Transactions on Evolutionary Computation},
  volume={13},
  number={1},
  pages={180--197},
  year={2008},
  publisher={IEEE}
}

@inproceedings{hellwig2016evolution,
  title={Evolution under strong noise: A self-adaptive evolution strategy can reach the lower performance bound-the pccmsa-es},
  author={Hellwig, Michael and Beyer, Hans-Georg},
  booktitle={International Conference on Parallel Problem Solving from Nature},
  pages={26--36},
  year={2016},
  organization={Springer}
}

@article{bonet2023performance,
  title={Performance comparison of optimization methods on variational quantum algorithms},
  author={Bonet-Monroig, Xavier and Wang, Hao and Vermetten, Diederick and Senjean, Bruno and Moussa, Charles and B{\"a}ck, Thomas and Dunjko, Vedran and O'Brien, Thomas E},
  journal={Physical Review A},
  volume={107},
  number={3},
  pages={032407},
  year={2023},
  publisher={APS}
}

@article{reach,
  title = {Reachability Deficits in Quantum Approximate Optimization},
  author = {Akshay, V. and Philathong, H. and Morales, M. E. S. and Biamonte, J. D.},
  journal = {Phys. Rev. Lett.},
  volume = {124},
  issue = {9},
  pages = {090504},
  numpages = {5},
  year = {2020},
  month = {Mar},
  publisher = {American Physical Society},
  doi = {10.1103/PhysRevLett.124.090504},
  url = {https://link.aps.org/doi/10.1103/PhysRevLett.124.090504}
}

@article{Cerezo2021,
  author = {Cerezo, M. and Sone, Akira and Volkoff, Tyler and Cincio, Lukasz and Coles, Patrick J.},
  title = {Cost function dependent barren plateaus in shallow parametrized quantum circuits},
  journal = {Nature Communications},
  volume = {12},
  number = {1},
  pages = {1791},
  year = {2021},
  doi = {10.1038/s41467-021-21728-w},
  url = {https://doi.org/10.1038/s41467-021-21728-w},
  issn = {2041-1723}
}

@article{huggins2021efficient,
author = {Huggins, William J. and McClean, Jarrod R. and Rubin, Nicholas C. and Jiang, Zhang and Wiebe, Nathan and Whaley, K. Birgitta and Babbush, Ryan},
title = {Efficient and noise resilient measurements for quantum chemistry on near-term quantum computers},
journal = {npj Quantum Information},
volume = {7},
number = {1},
pages = {23},
year = {2021},
doi = {10.1038/s41534-020-00341-7},
url = {https://doi.org/10.1038/s41534-020-00341-7},
issn = {2056-6387}
}

@article{nature_comp_sci_2024,
  author = {Barron, Samantha V. and Egger, Daniel J. and Pelofske, Elijah and B{\"a}rtschi, Andreas and Eidenbenz, Stephan and Lehmkuehler, Matthis and Woerner, Stefan},
  title = {Provable bounds for noise-free expectation values computed from noisy samples},
  journal = {Nature Computational Science},
  volume = {4},
  number = {11},
  pages = {865--875},
  year = {2024},
  doi = {10.1038/s43588-024-00709-1},
  url = {https://doi.org/10.1038/s43588-024-00709-1},
  issn = {2662-8457}
}

@misc{hopso_vqe_2025,
      title={HOPSO: A Robust Classical Optimizer for VQE}, 
      author={Ijaz Ahamed Mohammad and Yury Chernyak and Martin Plesch},
      year={2025},
      eprint={2508.13651},
      archivePrefix={arXiv},
      primaryClass={quant-ph},
      url={https://arxiv.org/abs/2508.13651}, 
}

@misc{noise_impact_vqe_2022,
      title={Evaluating the impact of noise on the performance of the Variational Quantum Eigensolver}, 
      author={Marita Oliv and Andrea Matic and Thomas Messerer and Jeanette Miriam Lorenz},
      year={2022},
      eprint={2209.12803},
      archivePrefix={arXiv},
      primaryClass={quant-ph},
      url={https://arxiv.org/abs/2209.12803}, 
}

@article{zeng2021simulating,
author = {Zeng, Jinfeng and Wu, Zipeng and Cao, Chenfeng and Zhang, Chao and Hou, Shi-Yao and Xu, Pengxiang and Zeng, Bei},
title = {Simulating noisy variational quantum eigensolver with local noise models},
journal = {Quantum Engineering},
volume = {3},
number = {4},
pages = {e77},
doi = {https://doi.org/10.1002/que2.77},
url = {https://onlinelibrary.wiley.com/doi/abs/10.1002/que2.77},
eprint = {https://onlinelibrary.wiley.com/doi/pdf/10.1002/que2.77},
year = {2021}
}

@article{Park2024hamiltonian,
  doi = {10.22331/q-2024-02-01-1239},
  url = {https://doi.org/10.22331/q-2024-02-01-1239},
  title = {Hamiltonian variational ansatz without barren plateaus},
  author = {Park, Chae-Yeun and Killoran, Nathan},
  journal = {{Quantum}},
  issn = {2521-327X},
  publisher = {{Verein zur F{\"{o}}rderung des Open Access Publizierens in den Quantenwissenschaften}},
  volume = {8},
  pages = {1239},
  month = feb,
  year = {2024}
}

@article{boy2025energy,
  title={Energy Landscapes for the Unitary Coupled Cluster Ansatz},
  author={Boy, Choy and Filip, Maria-Andreea and Wales, David J},
  journal={Journal of Chemical Theory and Computation},
  volume={21},
  number={4},
  pages={1739--1751},
  year={2025},
  publisher={ACS Publications}
}

@misc{qiskit,
      title={Quantum computing with {Q}iskit},
      author={Javadi-Abhari, Ali and Treinish, Matthew and Krsulich, Kevin and Wood, Christopher J. and Lishman, Jake and Gacon, Julien and Martiel, Simon and Nation, Paul D. and Bishop, Lev S. and Cross, Andrew W. and Johnson, Blake R. and Gambetta, Jay M.},
      year={2024},
      doi={10.48550/arXiv.2405.08810},
      eprint={2405.08810},
      archivePrefix={arXiv},
      primaryClass={quant-ph}
}

@article{sun2018pyscf,
  title={PySCF: the Python-based simulations of chemistry framework},
  author={Sun, Qiming and Berkelbach, Timothy C and Blunt, Nick S and Booth, George H and Guo, Sheng and Li, Zhendong and Liu, Junzi and McClain, James D and Sayfutyarova, Elvira R and Sharma, Sandeep and others},
  journal={Wiley Interdisciplinary Reviews: Computational Molecular Science},
  volume={8},
  number={1},
  pages={e1340},
  year={2018},
  publisher={Wiley Online Library}
}

@ARTICLE{2020SciPy-NMeth,
  author  = {Virtanen, Pauli and Gommers, Ralf and Oliphant, Travis E. and
            Haberland, Matt and Reddy, Tyler and Cournapeau, David and
            Burovski, Evgeni and Peterson, Pearu and Weckesser, Warren and
            Bright, Jonathan and {van der Walt}, St{\'e}fan J. and
            Brett, Matthew and Wilson, Joshua and Millman, K. Jarrod and
            Mayorov, Nikolay and Nelson, Andrew R. J. and Jones, Eric and
            Kern, Robert and Larson, Eric and Carey, C J and
            Polat, {\.I}lhan and Feng, Yu and Moore, Eric W. and
            {VanderPlas}, Jake and Laxalde, Denis and Perktold, Josef and
            Cimrman, Robert and Henriksen, Ian and Quintero, E. A. and
            Harris, Charles R. and Archibald, Anne M. and
            Ribeiro, Ant{\^o}nio H. and Pedregosa, Fabian and
            {van Mulbregt}, Paul and {SciPy 1.0 Contributors}},
  title   = {{{SciPy} 1.0: Fundamental Algorithms for Scientific
            Computing in Python}},
  journal = {Nature Methods},
  year    = {2020},
  volume  = {17},
  pages   = {261--272},
  adsurl  = {https://rdcu.be/b08Wh},
  doi     = {10.1038/s41592-019-0686-2},
}

@misc{hansen2019pycma,
  author       = {Nikolaus Hansen and Youhei Akimoto and Petr Baudis},
  title        = {{CMA-ES/pycma} on {G}ithub},
  howpublished = {Zenodo, DOI:10.5281/zenodo.2559634},
  month        = feb,
  year         = 2019,
  doi          = {10.5281/zenodo.2559634},
  url          = {https://doi.org/10.5281/zenodo.2559634},
}

@article{lavrijsen2020classopt,
  title={Classical Optimizers for Noisy Intermediate-Scale Quantum Devices},
  author={Lavrijsen, Wim and Tudor, Ana and Müller, Juliane and Iancu, Costin and de Jong, Wibe},
  journal={Proceedings of the IEEE International Conference on Quantum Computing and Engineering},
  pages={267--277},
  year={2020},
  publisher={IEEE}
}

@article{nannicini2019performance,
  title={Performance of hybrid quantum-classical variational heuristics for combinatorial optimization},
  author={Nannicini, Giacomo},
  journal={Physical Review E},
  volume={99},
  number={1},
  pages={013304},
  year={2019},
  publisher={APS}
}

@article{kandala2017hardware,
  title={Hardware-efficient variational quantum eigensolver for small molecules and quantum magnets},
  author={Kandala, Abhinav and Mezzacapo, Antonio and Temme, Kristan and Takita, Maika and Brink, Markus and Chow, Jerry M and Gambetta, Jay M},
  journal={Nature},
  volume={549},
  number={7671},
  pages={242--246},
  year={2017},
  publisher={Nature Publishing Group}
}

@article{ahamed2025hopso,
  title={HOPSO: A Robust Classical Optimizer for VQE},
  author={Ahamed Mohammad, Ijaz and Chernyak, Yury and Plesch, Martin},
  journal={arXiv e-prints},
  pages={arXiv--2508},
  year={2025}
}

@article{BenchmarkKitaev,
  title={Benchmarking variational quantum eigensolvers for the square-octagon-lattice Kitaev model},
  author={Li, Andy~C.~Y. and Alam, M.~Sohaib and Iadecola, Thomas and Jahin, Ammar and Job, Joshua and Kurkcuoglu, Doga~Murat and Li, Richard and Orth, Peter~P. and {\"Ozg{\"u}ler}, A.~Bar{\i}{\c{s}} and Perdue, Gabriel~N. and Tubman, Norm~M.},
  journal={Phys. Rev. Res.},
  volume={5},
  issue={3},
  pages={033071},
  numpages={17},
  year={2023},
  month={Aug},
  publisher={American Physical Society},
  doi={10.1103/PhysRevResearch.5.033071},
  url={https://link.aps.org/doi/10.1103/PhysRevResearch.5.033071}
}

@article{tang2021qubit,
  title={qubit-ADAPT-VQE: An adaptive algorithm for constructing hardware-efficient ansätze on a quantum processor},
  author={Tang, Harper L and Shkolnikov, V.O. and Barron, George S and Harper, Herschel R and Garrison, Nicholas J and Babbush, Ryan and McClean, Jarrod R},
  journal={PRX Quantum},
  volume={2},
  number={2},
  pages={020310},
  year={2021},
  publisher={APS}
}

@article{Preskill2018,
  author   = {John Preskill},
  title    = {Quantum Computing in the NISQ era and beyond},
  journal  = {Quantum},
  year     = {2018},
  volume   = {2},
  pages    = {79},
  doi      = {10.22331/q-2018-08-06-79}
}

@article{Peruzzo2014,
  author   = {Alberto Peruzzo and Jarrod R. McClean and Peter Shadbolt and Man-Hong Yung and Xiao-Qi Zhou and Peter J. Love and Al{\'a}n Aspuru-Guzik and Jeremy L. O'Brien},
  title    = {A variational eigenvalue solver on a photonic quantum processor},
  journal  = {Nature Communications},
  year     = {2014},
  volume   = {5},
  pages    = {4213},
  doi      = {10.1038/ncomms5213}
}

@article{Wecker2015,
  author   = {Dave Wecker and Matthew B. Hastings and Matthias Troyer},
  title    = {Progress towards practical quantum variational algorithms},
  journal  = {Physical Review A},
  year     = {2015},
  volume   = {92},
  pages    = {042303},
  doi      = {10.1103/PhysRevA.92.042303}
}

@article{McClean2016,
  author   = {Jarrod R. McClean and Jonathan Romero and Ryan Babbush and Al{\'a}n Aspuru-Guzik},
  title    = {The theory of variational hybrid quantum-classical algorithms},
  journal  = {New Journal of Physics},
  year     = {2016},
  volume   = {18},
  pages    = {023023},
  doi      = {10.1088/1367-2630/18/2/023023}
}

@article{Kandala2017,
  author   = {Abhinav Kandala and Antonio Mezzacapo and Kristan Temme and Maika Takita and Markus Brink and Jerry M. Chow and Jay M. Gambetta},
  title    = {Hardware-efficient variational quantum eigensolver for small molecules and quantum magnets},
  journal  = {Nature},
  year     = {2017},
  volume   = {549},
  number   = {7671},
  pages    = {242--246},
  doi      = {10.1038/nature23879}
}

@article{Temme2017,
  author   = {Kristan Temme and Sergey Bravyi and Jay M. Gambetta},
  title    = {Error Mitigation for Short-Depth Quantum Circuits},
  journal  = {Physical Review Letters},
  year     = {2017},
  volume   = {119},
  pages    = {180509},
  doi      = {10.1103/PhysRevLett.119.180509}
}

@article{Stokes2020,
  author   = {James Stokes and Josh Izaac and Nathan Killoran and Giuseppe Carleo},
  title    = {Quantum Natural Gradient},
  journal  = {Quantum},
  year     = {2020},
  volume   = {4},
  pages    = {269},
  doi      = {10.22331/q-2020-05-25-269}
}

@article{BPreview,
  author  = {Larocca, Martín and Thanasilp, Supanut and Wang, Samson and Sharma, Kunal and Biamonte, Jacob and Coles, Patrick J. and Cincio, Lukasz and McClean, Jarrod R. and Holmes, Zoë and Cerezo, M.},
  title   = {Barren plateaus in variational quantum computing},
  journal = {Nature Reviews Physics},
  volume  = {7},
  number  = {4},
  pages   = {174--189},
  year    = {2025},
  month   = apr,
  doi     = {10.1038/s42254-025-00813-9},
  url     = {https://doi.org/10.1038/s42254-025-00813-9},
  issn    = {2522-5820}
}

@article{arrasmith2020operator,
  title={Operator Sampling for Shot-frugal Optimization in Variational Algorithms},
  author={Arrasmith, Andrew and Cincio, Lukasz and Sornborger, Andrew T and Zurek, Wojciech H and Coles, Patrick J},
  journal={arXiv preprint arXiv:2004.06252},
  year={2020}
}

@article{mcclean2018barren,
  title={Barren plateaus in quantum neural network training landscapes},
  author={McClean, Jarrod R and Boixo, Sergio and Smelyanskiy, Vadim N and Babbush, Ryan and Neven, Hartmut},
  journal={Nature Communications},
  volume={9},
  number={1},
  pages={4812},
  year={2018},
  publisher={Nature Publishing Group}
}

@article{zhu2020training,
  title={Training of quantum circuits on a hybrid quantum computer},
  author={Zhu, Dongxiao and Linke, Norbert M and Benedetti, Marcello and Landsman, Kevin A and Nguyen, Nhung H and Alderete, C Huerta and Perdomo-Ortiz, Alejandro and Korda, Nicolas and Garfoot, Andrew and Brecque, Charles and others},
  journal={Science Advances},
  volume={5},
  number={10},
  pages={eaaw9918},
  year={2020},
  publisher={American Association for the Advancement of Science}
}

@article{smart2021efficient,
  title={Efficient two-electron approach for quantum computation of molecular electrons},
  author={Smart, Scott D and Mazziotti, David A},
  journal={Journal of Chemical Theory and Computation},
  volume={17},
  number={5},
  pages={3152--3159},
  year={2021},
  publisher={ACS Publications}
}

@article{yuan2024quantifying,
  title={Quantifying the advantages of applying quantum approximate algorithms to portfolio optimisation},
  author={Yuan, Haomu and Long, Christopher K and Lepage, Hugo V and Barnes, Crispin HW},
  journal={arXiv preprint arXiv:2410.16265},
  year={2024}
}

@article{novak2025reliable,
  title={Reliable Optimization Under Noise in Quantum Variational Algorithms},
  author={Nov{\'a}k, Vojt{\v{e}}ch and Ill{\'e}sov{\'a}, Silvie and Bezd{\v{e}}k, Tom{\'a}{\v{s}} and Zelinka, Ivan and Beseda, Martin},
  journal={arXiv preprint arXiv:2511.08289},
  year={2025},
  doi={10.48550/arXiv.2511.08289},
  url={https://doi.org/10.48550/arXiv.2511.08289}
}

@article{novak2025optimization,
  title={Optimization Strategies for Variational Quantum Algorithms in Noisy Landscapes},
  author={Nov{\'a}k, Vojt{\v{e}}ch and Zelinka, Ivan and Sn{\'a}{\v{s}}el, V{\'a}clav},
  journal={arXiv preprint arXiv:2506.01715},
  year={2025},
  doi={10.48550/arXiv.2506.01715},
  url={https://doi.org/10.48550/arXiv.2506.01715}
}

@article{novak2025predicting,
  title={Predicting Post-Surgical Complications with Quantum Neural Networks: A Clinical Study on Anastomotic Leak},
  author={Nov{\'a}k, Vojt{\v{e}}ch and Zelinka, Ivan and P{\v{r}}ibylov{\'a}, Lenka and Mart{\'i}nek, Lubom{\'i}r and Ben{\v{c}}urik, V{\'a}clav},
  journal={arXiv preprint arXiv:2506.01708},
  year={2025},
  doi={10.48550/arXiv.2506.01708},
  url={https://doi.org/10.48550/arXiv.2506.01708}
}

@article{illesova2025statistical,
  title={Statistical Benchmarking of Optimization Methods for Variational Quantum Eigensolver under Quantum Noise},
  author={Ill{\'e}sov{\'a}, Silvie and Bezd{\v{e}}k, Tom{\'a}{\v{s}} and Nov{\'a}k, Vojt{\v{e}}ch and Senjean, Bruno and Beseda, Martin},
  journal={arXiv preprint arXiv:2510.08727},
  year={2025},
  doi={10.48550/arXiv.2510.08727},
  url={https://doi.org/10.48550/arXiv.2510.08727}
}

@article{novak2025quantum,
  title={Quantum Neural Networks for Propensity Score Estimation and Survival Analysis in Observational Biomedical Studies},
  author={Nov{\'a}k, Vojt{\v{e}}ch and Zelinka, Ivan and P{\v{r}}ibylov{\'a}, Lenka and Mart{\'i}nek, Lubom{\'i}r},
  journal={arXiv preprint arXiv:2506.19973},
  year={2025},
  doi={10.48550/arXiv.2506.19973},
  url={https://doi.org/10.48550/arXiv.2506.19973}
}

@article{illesova2025complementarity,
  title={On the Complementarity of Classical Convolution and Quantum Neural Networks in Image Classification},
  author={Ill{\'e}sov{\'a}, Silvie and Obeng, Emmanuel and Bezd{\v{e}}k, Tom{\'a}{\v{s}} and Nov{\'a}k, Vojt{\v{e}}ch and Beseda, Martin},
  journal={Preprints},
  year={2025},
  doi={10.20944/preprints202512.1348.v1},
  url={https://doi.org/10.20944/preprints202512.1348.v1},
  note={Preprint}
}

@article{bezdek2025classical,
  title={Classical Optimization Strategies for Variational Quantum Algorithms: A Systematic Study of Noise Effects and Parameter Efficiency},
  author={Bezd{\v{e}}k, Tom{\'a}{\v{s}} and Yuan, Haomu and Nov{\'a}k, Vojt{\v{e}}ch and Ill{\'e}sov{\'a}, Silvie and Beseda, Martin},
  journal={arXiv preprint arXiv:2511.09314},
  year={2025},
  doi={10.48550/arXiv.2511.09314},
  url={https://doi.org/10.48550/arXiv.2511.09314}
}

@article{illesova2025classical,
  title={From Classical to Hybrid: A Practical Framework for Quantum-Enhanced Learning},
  author={Ill{\'e}sov{\'a}, Silvie and Bezd{\v{e}}k, Tom{\'a}{\v{s}} and Nov{\'a}k, Vojt{\v{e}}ch and Zelinka, Ivan and Cacciatore, Stefano and Beseda, Martin},
  journal={arXiv preprint arXiv:2511.08205},
  year={2025},
  doi={10.48550/arXiv.2511.08205},
  url={https://doi.org/10.48550/arXiv.2511.08205}
}

@inproceedings{gupta2022how,
  title={How Quantum Computing-Friendly Multispectral Data can be?},
  author={Gupta, Manish K. and Beseda, Martin and Gawron, Piotr},
  booktitle={IGARSS 2022-2022 IEEE International Geoscience and Remote Sensing Symposium},
  pages={4153--4156},
  year={2022},
  organization={IEEE},
  doi={10.1109/IGARSS46834.2022.9883676},
  url={https://doi.org/10.1109/IGARSS46834.2022.9883676}
}

@article{ciaramelletti2025detecting,
  title={Detecting Quasidegenerate Ground States in Topological Models via the Variational Quantum Eigensolver},
  author={Ciaramelletti, C. and Beseda, Martin and Consiglio, M. and Lepori, L. and Apollaro, T. J. G.},
  journal={Physical Review A},
  volume={111},
  number={2},
  pages={022437},
  year={2025},
  publisher={APS},
  doi={10.1103/PhysRevA.111.022437},
  url={https://doi.org/10.1103/PhysRevA.111.022437}
}

@inproceedings{illesova2025qmetric,
  title={QMetric: Benchmarking Quantum Neural Networks Across Circuits, Features, and Training Dimensions},
  author={Ill{\'e}sov{\'a}, Silvie and Rybotycki, Tomasz and Beseda, Martin},
  booktitle={QualITA 2025: The Fourth Conference on System and Service Quality},
  series={CEUR Workshop Proceedings},
  volume={4080},
  year={2025},
  publisher={CEUR-WS.org},
  url={https://ceur-ws.org/Vol-4080/paper3.pdf}
}

@inproceedings{trovato2025preliminary,
  title={A Preliminary Investigation on the Usage of Quantum Approximate Optimization Algorithms for Test Case Selection},
  author={Trovato, Antonio and Beseda, Martin and Di Nucci, Dario},
  booktitle={Proceedings of the 2025 29th International Conference on Evaluation and Assessment in Software Engineering (EASE)},
  pages={56--60},
  year={2025},
  note={Also available as arXiv:2504.18955},
  doi={10.48550/arXiv.2504.18955}
}

@article{illesova2025importance,
  title={On the Importance of Fundamental Properties in Quantum-Classical Machine Learning Models},
  author={Ill{\'e}sov{\'a}, Silvie and Rybotycki, Tomasz and Gawron, Piotr and Beseda, Martin},
  journal={arXiv preprint arXiv:2507.10161},
  year={2025},
  doi={10.48550/arXiv.2507.10161},
  url={https://doi.org/10.48550/arXiv.2507.10161}
}

@article{lewandowska2025benchmarking,
  title={Benchmarking Gate-Based Quantum Devices via Certification of Qubit von Neumann Measurements},
  author={Lewandowska, Paulina and Beseda, Martin},
  journal={arXiv preprint arXiv:2506.03514},
  year={2025},
  doi={10.48550/arXiv.2506.03514},
  url={https://doi.org/10.48550/arXiv.2506.03514}
}

@misc{qiskit2024,
  author       = {Ali Javadi-Abhari and Matthew Treinish and Kevin Krsulich and Christopher J. Wood and James Lishman and Julien Gacon and Samuel Martiel and Paul D. Nation and Loren S. Bishop and Andrew W. Cross and Blake R. Johnson and Jay M. Gambetta},
  title        = {Quantum computing with Qiskit},
  year         = {2024},
  url          = {https://doi.org/10.5281/zenodo.2562111},
  doi = {10.5281/zenodo.2562111}
}

@article{pyscf2018,
  author       = {Sun, Qiming and Berkelbach, Timothy C. and Blunt, Nick S. and Booth, George H. and Guo, Sheng and Li, Zhendong and Liu, Junzi and McClain, James D. and Sayfutyarova, Elvira R. and Sharma, Sandeep and others},
  title        = {PySCF: the Python-based simulations of chemistry framework},
  journal      = {Wiley Interdisciplinary Reviews: Computational Molecular Science},
  year         = {2018},
  volume       = {8},
  number       = {1},
  pages        = {e1340},
  url          = {https://doi.org/10.1002/wcms.1340},
doi = {10.1002/wcms.1340}
}

@article{liu1989,
  author       = {Liu, Dong C. and Nocedal, Jorge},
  title        = {On the limited memory BFGS method for large scale optimization},
  journal      = {Mathematical Programming},
  volume       = {45},
  pages        = {503--528},
  year         = {1989},
  url          = {https://doi.org/10.1007/BF01589116},
doi = {10.1007/BF01589116}
}

@article{dai2002,
author = {Dai, Yu-Hong},
title = {Convergence Properties of the BFGS Algoritm},
journal = {SIAM Journal on Optimization},
volume = {13},
number = {3},
pages = {693-701},
year = {2002},
doi = {10.1137/S1052623401383455},
URL = {https://doi.org/10.1137/S1052623401383455},
eprint = {https://doi.org/10.1137/S1052623401383455}
}

@article{morales2002,
title = {A numerical study of limited memory BFGS methods},
journal = {Applied Mathematics Letters},
volume = {15},
number = {4},
pages = {481-487},
year = {2002},
issn = {0893-9659},
doi = {10.1016/S0893-9659(01)00162-8},
url = {https://www.sciencedirect.com/science/article/pii/S0893965901001628},
author = {Morales, J. L.}
}

@article{spall1992,
  author       = {Spall, J. C.},
  title        = {Multivariate stochastic approximation using a simultaneous perturbation gradient approximation},
  journal      = {IEEE Transactions on Automatic Control},
  volume       = {37},
  number       = {3},
  pages        = {332--341},
  year         = {1992},
  url          = {https://doi.org/10.1109/9.119632},
doi = {10.1109/9.119632}
}

@article{spall2002,
  author={Spall, J.C.},
  journal={IEEE Transactions on Aerospace and Electronic Systems}, 
  title={Implementation of the simultaneous perturbation algorithm for stochastic optimization}, 
  year={1998},
  volume={34},
  number={3},
  pages={817-823},
  doi={10.1109/7.705889},
url={https://doi.org/10.1109/7.705889}
}

@inproceedings{maryak1999,
  author={Maryak, J.L. and Chin, D.C.},
  booktitle={Proceedings of the 1999 American Control Conference (Cat. No. 99CH36251)}, 
  title={Efficient global optimization using SPSA}, 
  year={1999},
  volume={2},
  number={},
  pages={890-894 vol.2},
  doi={10.1109/ACC.1999.783168},
url={https://doi.org/10.1109/ACC.1999.783168}
}

@Inbook{Powell1994,
author="Powell, M. J. D.",
title="A Direct Search Optimization Method That Models the Objective and Constraint Functions by Linear Interpolation",
bookTitle="Advances in Optimization and Numerical Analysis",
year="1994",
publisher="Springer Netherlands",
address="Dordrecht",
pages="51--67",
isbn="978-94-015-8330-5",
doi="10.1007/978-94-015-8330-5_4",
url="https://doi.org/10.1007/978-94-015-8330-5_4"
}

@article{powell1998,
  author       = {Powell, M. J. D.},
  title        = {Direct search algorithms for optimization calculations},
  journal      = {Acta Numerica},
  volume       = {7},
  pages        = {287--336},
  year         = {1998},
  url          = {https://doi.org/10.1017/S0962492900002841}
}

@article{powell2007,
  title={A view of algorithms for optimization without derivatives},
  author={Powell, Michael JD},
  journal={Mathematics Today-Bulletin of the Institute of Mathematics and its Applications},
  volume={43},
  number={5},
  pages={170--174},
  year={2007},
  publisher={Southend-on-Sea, Essex: Institute of Mathematics and Its Applications, 1996-},
url={https://optimization-online.org/wp-content/uploads/2007/06/1680.pdf}
}

@article{kraft1988,
  title={A software package for sequential quadratic programming},
  author={Kraft, Dieter},
  journal={Forschungsbericht- Deutsche Forschungs- und Versuchsanstalt fur Luft- und Raumfahrt},
  year={1988}
}

@article{boggs1995, 
title={Sequential Quadratic Programming}, 
volume={4}, 
doi={10.1017/S0962492900002518},
url={https:/doi.org/10.1017/S0962492900002518},
journal={Acta Numerica}, 
author={Boggs, Paul T. and Tolle, Jon W.}, 
year={1995}, 
pages={1–51}
}

@article{nelder1965,
  author       = {Nelder, J. A. and Mead, R.},
  title        = {A simplex method for function minimization},
  journal      = {The Computer Journal},
  volume       = {7},
  number       = {4},
  pages        = {308--313},
  year         = {1965},
  doi = {10.1093/comjnl/7.4.308},
  url          = {https://doi.org/10.1093/comjnl/7.4.308}
}

@article{lagarias1998,
  author       = {Lagarias, Jeffrey C. and Reeds, James A. and Wright, Margaret H. and Wright, Paul E.},
  title        = {Convergence properties of the Nelder–Mead simplex method in low dimensions},
  journal      = {SIAM Journal on Optimization},
  volume       = {9},
  number       = {1},
  pages        = {112--147},
  year         = {1998},
doi = {10.1137/S1052623496303470},
  url          = {https://doi.org/10.1137/S1052623496303470}
}

@article{hansen2003,
  author       = {Hansen, Nikolaus and Müller, Sibylle D. and Koumoutsakos, Petros},
  title        = {Reducing the time complexity of the derandomized evolution strategy with covariance matrix adaptation (CMA-ES)},
  journal      = {Evolutionary Computation},
  volume       = {11},
  number       = {1},
  pages        = {1--18},
  year         = {2003},
doi = {10.1162/106365603321828970},
  url          = {https://doi.org/10.1162/106365603321828970}
}

@inproceedings{eberhart1995,
  author       = {Eberhart, Russell and Kennedy, James},
  title        = {Particle Swarm Optimization},
  booktitle    = {Proceedings of the IEEE International Conference on Neural Networks},
  year         = {1995},
  pages        = {1942--1948},
  url          = {https://doi.org/10.1109/ICNN.1995.488968},
doi = {10.1109/ICNN.1995.488968}
}

@inproceedings{shi2001,
  author       = {Shi, Yuhui},
  title        = {Particle Swarm Optimization: Developments, Applications and Resources},
  booktitle    = {Proceedings of the 2001 Congress on Evolutionary Computation (IEEE Cat. No. 01TH8546)},
  year         = {2001},
  pages        = {81--86},
  url          = {https://doi.org/10.1109/CEC.2001.934374}
}

@Article{jain2022,
AUTHOR = {Jain, Meetu and Saihjpal, Vibha and Singh, Narinder and Singh, Satya Bir},
TITLE = {An Overview of Variants and Advancements of PSO Algorithm},
JOURNAL = {Applied Sciences},
VOLUME = {12},
YEAR = {2022},
NUMBER = {17},
ARTICLE-NUMBER = {8392},
URL = {https://www.mdpi.com/2076-3417/12/17/8392},
ISSN = {2076-3417},
DOI = {10.3390/app12178392}
}

@article{liu1989limited,
  title={On the limited memory BFGS method for large scale optimization},
  author={Liu, Dong C and Nocedal, Jorge},
  journal={Mathematical programming},
  volume={45},
  number={1},
  pages={503--528},
  year={1989},
  doi={10.1007/BF01589116},
  url={https://doi.org/10.1007/BF01589116},
  publisher={Springer}
}

@article{dai2002convergence,
  title={Convergence properties of the BFGS algoritm},
  author={Dai, Yu-Hong},
  journal={SIAM Journal on Optimization},
  volume={13},
  number={3},
  pages={693--701},
  year={2002},
  doi={10.1137/S1052623401383455},
  url={https://doi.org/10.1137/S1052623401383455},
  publisher={SIAM}
}

@article{morales2002numerical,
  title={A numerical study of limited memory BFGS methods},
  author={Morales, Jos{\'e} Luis},
  journal={Applied Mathematics Letters},
  volume={15},
  number={4},
  pages={481--487},
  year={2002},
  doi={10.1016/S0893-9659(01)00162-8},
  url={https://doi.org/10.1016/S0893-9659(01)00162-8},
  publisher={Elsevier}
}

@article{hansen2003reducing,
  title={Reducing the time complexity of the derandomized evolution strategy with covariance matrix adaptation (CMA-ES)},
  author={Hansen, Nikolaus and M{\"u}ller, Sibylle D and Koumoutsakos, Petros},
  journal={Evolutionary computation},
  volume={11},
  number={1},
  pages={1--18},
  year={2003},
  doi={10.1162/106365603321828970},
  url={https://doi.org/10.1162/106365603321828970},
  publisher={MIT Press}
}

@inproceedings{varelas2018comparative,
  title={A comparative study of large-scale variants of CMA-ES},
  author={Varelas, Konstantinos and Auger, Anne and Brockhoff, Dimo and Hansen, Nikolaus and ElHara, Ouassim Ait and Semet, Yann and Kassab, Rami and Barbaresco, Fr{\'e}d{\'e}ric},
  booktitle={Parallel Problem Solving from Nature--PPSN XV: 15th International Conference, Coimbra, Portugal, September 8--12, 2018, Proceedings, Part I 15},
  pages={3--15},
  year={2018},
  doi={10.1007/978-3-319-99253-2_1},
  url={https://doi.org/10.1007/978-3-319-99253-2_1},
  organization={Springer}
}

@book{powell1994direct,
  title={A direct search optimization method that models the objective and constraint functions by linear interpolation},
  author={Powell, Michael JD},
  year={1994},
  doi={10.1007/978-94-015-8330-5_4},
  url={https://doi.org/10.1007/978-94-015-8330-5_4},
  publisher={Springer}
}

@article{powell1998direct,
  title={Direct search algorithms for optimization calculations},
  author={Powell, Michael JD},
  journal={Acta numerica},
  volume={7},
  pages={287--336},
  year={1998},
  doi={10.1017/S0962492900002841},
  url={https://doi.org/10.1017/S0962492900002841},
  publisher={Cambridge University Press}
}

@article{powell2007view,
  title={A view of algorithms for optimization without derivatives},
  author={Powell, Michael JD},
  journal={Mathematics Today-Bulletin of the Institute of Mathematics and its Applications},
  volume={43},
  number={5},
  pages={170--174},
  year={2007},
  publisher={Southend-on-Sea, Essex: Institute of Mathematics and Its Applications, 1996-}
}

@article{ruder2016overview,
  title={An overview of gradient descent optimization algorithms},
  author={Ruder, Sebastian},
  journal={arXiv preprint arXiv:1609.04747},
  doi={https://doi.org/10.48550/arXiv.1609.04747},
  url={https://doi.org/10.48550/arXiv.1609.04747},
  year={2016}
}

@article{amari1993backpropagation,
  title={Backpropagation and stochastic gradient descent method},
  author={Amari, Shun-ichi},
  journal={Neurocomputing},
  volume={5},
  number={4-5},
  pages={185--196},
  year={1993},
  doi={10.1016/0925-2312(93)90006-O},
  url={https://doi.org/10.1016/0925-2312(93)90006-O},
  publisher={Elsevier}
}

@article{nelder1965simplex,
  title={A simplex method for function minimization},
  author={Nelder, John A and Mead, Roger},
  journal={The computer journal},
  volume={7},
  number={4},
  pages={308--313},
  year={1965},
  doi={10.1093/comjnl/7.4.308},
  url={https://doi.org/10.1093/comjnl/7.4.308},
  publisher={The British Computer Society}
}

@article{lagarias1998convergence,
  title={Convergence properties of the Nelder--Mead simplex method in low dimensions},
  author={Lagarias, Jeffrey C and Reeds, James A and Wright, Margaret H and Wright, Paul E},
  journal={SIAM Journal on optimization},
  volume={9},
  number={1},
  pages={112--147},
  year={1998},
  doi={10.1137/S1052623496303470},
  url={https://doi.org/10.1137/S1052623496303470},
  publisher={SIAM}
}

@inproceedings{eberhart1995particle,
  title={Particle swarm optimization},
  author={Eberhart, Russell and Kennedy, James},
  booktitle={Proceedings of the IEEE international conference on neural networks},
  volume={4},
  pages={1942--1948},
  year={1995},
  doi={10.1109/ICNN.1995.488968},
  url={https://doi.org/10.1109/ICNN.1995.488968},
  organization={Citeseer}
}

@inproceedings{shi2001particle,
  title={Particle swarm optimization: developments, applications and resources},
  author={Shi, Yuhui and others},
  booktitle={Proceedings of the 2001 congress on evolutionary computation (IEEE Cat. No. 01TH8546)},
  volume={1},
  pages={81--86},
  year={2001},
  doi={10.1109/CEC.2001.934374},
  url={https://doi.org/10.1109/CEC.2001.934374},
  organization={IEEE}
}

@article{jain2022overview,
  title={An overview of variants and advancements of PSO algorithm},
  author={Jain, Meetu and Saihjpal, Vibha and Singh, Narinder and Singh, Satya Bir},
  journal={Applied Sciences},
  volume={12},
  number={17},
  pages={8392},
  year={2022},
  doi={10.3390/app12178392},
  url={https://doi.org/10.3390/app12178392},
  publisher={MDPI}
}

@article{kraft1988software,
  title={A software package for sequential quadratic programming},
  author={Kraft, Dieter},
  journal={Forschungsbericht- Deutsche Forschungs- und Versuchsanstalt fur Luft- und Raumfahrt},
  year={1988}
}

@article{boggs1995sequential,
  title={Sequential quadratic programming},
  author={Boggs, Paul T and Tolle, Jon W},
  journal={Acta numerica},
  volume={4},
  pages={1--51},
  year={1995},
  doi={10.1017/S0962492900002518},
  url={https://doi.org/10.1017/S0962492900002518},
  publisher={Cambridge University Press}
}

@article{spall1992multivariate,
  title={Multivariate stochastic approximation using a simultaneous perturbation gradient approximation},
  author={Spall, James C},
  journal={IEEE transactions on automatic control},
  volume={37},
  number={3},
  pages={332--341},
  year={1992},
  doi={10.1109/9.119632},
  url={https://doi.org/10.1109/9.119632},
  publisher={IEEE}
}

@article{spall2002implementation,
  title={Implementation of the simultaneous perturbation algorithm for stochastic optimization},
  author={Spall, James C},
  journal={IEEE Transactions on aerospace and electronic systems},
  volume={34},
  number={3},
  pages={817--823},
  doi={10.1109/7.705889},
  url={https://doi.org/10.1109/7.705889},
  year={2002},
  publisher={IEEE}
}

@inproceedings{maryak1999efficient,
  title={Efficient global optimization using SPSA},
  author={Maryak, John L and Chin, Daniel C},
  booktitle={Proceedings of the 1999 American Control Conference (Cat. No. 99CH36251)},
  volume={2},
  pages={890--894},
  year={1999},
  doi={10.1109/ACC.1999.783168},
  url={https://doi.org/10.1109/ACC.1999.783168},
  organization={IEEE}
}

@misc{beseda2025results,
  author       = {Martin Beseda and Silvie Illésová and Clemens Possel and Tomáš Bezděk and Vojtěch Novák},
  title        = {Numerical Optimization Strategies for the Variational Hamiltonian Ansatz in Noisy Quantum Environments: Results},
  year         = 2025,
  publisher    = {Zenodo},
  doi          = {10.5281/zenodo.15526170},
  url          = {https://doi.org/10.5281/zenodo.15526170}
}

@article{beseda2024state,
  title={State-Averaged Orbital-Optimized VQE: A quantum algorithm for the democratic description of ground and excited electronic states},
  author={Beseda, Martin and Ill{\'e}sov{\'a}, Silvie and Yalouz, Saad and Senjean, Bruno},
  journal={Journal of Open Source Software},
  volume={9},
  number={101},
  pages={6036},
  year={2024}
}

@article{illesova2025transformation,
  title={Transformation-free generation of a quasi-diabatic representation from the state-average orbital-optimized variational quantum eigensolver},
  author={Ill{\'e}sov{\'a}, Silvie and Beseda, Martin and Yalouz, Saad and Lasorne, Benjamin and Senjean, Bruno},
  journal={Journal of Chemical Theory and Computation},
  publisher={ACS Publications},
  year={2025}
}

@article{nakanishi2019subspace,
  title={Subspace-search variational quantum eigensolver for excited states},
  author={Nakanishi, Ken M and Mitarai, Kosuke and Fujii, Keisuke},
  journal={Physical Review Research},
  volume={1},
  number={3},
  pages={033062},
  year={2019},
  publisher={APS}
}

@misc{possel2025truncatedvariationalhamiltonianansatz,
      title={Truncated Variational Hamiltonian Ansatz: efficient quantum circuit design for quantum chemistry and material science}, 
      author={Clemens Possel and Walter Hahn and Reza Shirazi and Marina Walt and Peter Pinski and Frank K. Wilhelm and Dmitry Bagrets},
      year={2025},
      eprint={2505.19772},
      archivePrefix={arXiv},
      primaryClass={quant-ph},
      url={https://arxiv.org/abs/2505.19772}, 
}

@article{xu2024quantum,
  title={Quantum convolutional long short-term memory based on variational quantum algorithms in the era of NISQ},
  author={Xu, Zeyu and Yu, Wenbin and Zhang, Chengjun and Chen, Yadang},
  journal={Information},
  volume={15},
  number={4},
  pages={175},
  year={2024},
  publisher={MDPI}
}

@article{du2022quantum,
  title={Quantum circuit architecture search for variational quantum algorithms},
  author={Du, Yuxuan and Huang, Tao and You, Shan and Hsieh, Min-Hsiu and Tao, Dacheng},
  journal={npj Quantum Information},
  volume={8},
  number={1},
  pages={62},
  year={2022},
  publisher={Nature Publishing Group UK London}
}

@article{niu2020hardware,
  title={A hardware-aware heuristic for the qubit mapping problem in the nisq era},
  author={Niu, Siyuan and Suau, Adrien and Staffelbach, Gabriel and Todri-Sanial, Aida},
  journal={IEEE Transactions on Quantum Engineering},
  volume={1},
  pages={1--14},
  year={2020},
  publisher={IEEE}
}

@article{cerezo2021variational,
  title={Variational quantum algorithms},
  author={Cerezo, Marco and Arrasmith, Andrew and Babbush, Ryan and Benjamin, Simon C and Endo, Suguru and Fujii, Keisuke and McClean, Jarrod R and Mitarai, Kosuke and Yuan, Xiao and Cincio, Lukasz and others},
  journal={Nature Reviews Physics},
  volume={3},
  number={9},
  pages={625--644},
  year={2021},
  publisher={Nature Publishing Group UK London}
}

@article{callison2022hybrid,
  title={Hybrid quantum-classical algorithms in the noisy intermediate-scale quantum era and beyond},
  author={Callison, Adam and Chancellor, Nicholas},
  journal={Physical Review A},
  volume={106},
  number={1},
  pages={010101},
  year={2022},
  publisher={APS}
}

@article{qi2024variational,
  title={Variational quantum algorithms: fundamental concepts, applications and challenges.},
  author={Qi, Han and Xiao, Sihui and Liu, Zhuo and Gong, Changqing and Gani, Abdullah},
  journal={Quantum Information Processing},
  volume={23},
  number={6},
  year={2024}
}

@article{stilck2021limitations,
  title={Limitations of optimization algorithms on noisy quantum devices},
  author={Stilck Fran{\c{c}}a, Daniel and Garcia-Patron, Raul},
  journal={Nature Physics},
  volume={17},
  number={11},
  pages={1221--1227},
  year={2021},
  publisher={Nature Publishing Group UK London}
}

@article{anand2025hamiltonian,
  title={Hamiltonian-based graph-state ansatz for variational quantum algorithms},
  author={Anand, Abhinav and Brown, Kenneth R},
  journal={Physical Review A},
  volume={111},
  number={1},
  pages={012437},
  year={2025},
  publisher={APS}
}

\end{document}